\documentclass[iop]{emulateapj-rtx4} 
\shortauthors{Sekanina \& Kracht}
\shorttitle{Outbursts and Disintegration of Comet C/2017 S3}
\slugcomment{Version 3 of Paper First Posted on December 17, 2018}

\newcommand{\Hsun}{$H_{\mbox{\boldmath ${\scriptstyle \odot}$}}$}
\newcommand{\Hssun}{$\scriptstyle H_{\mbox{\boldmath ${\:\!\!\scriptscriptstyle \odot}$}}$}

\begin{document}
\title{PREPERIHELION OUTBURSTS AND DISINTEGRATION OF COMET C/2017 S3 (PAN-STARRS)}
\author{Zdenek Sekanina$^1$ \& Rainer Kracht$^2$}
\affil{$^1$Jet Propulsion Laboratory, California Institute of Technology,
  4800 Oak Grove Drive, Pasadena, CA 91109, U.S.A. \\
 $^2$Ostlandring 53, D-25335 Elmshorn, Schleswig-Holstein, Germany}
\email{Zdenek.Sekanina@jpl.nasa.gov,\\
 {\hspace*{2.59cm}}R.Kracht@t-online.de{\vspace{-0.1cm}}}

\begin{abstract}
A sequence of events, dominated by two outbursts and ending with the preperihelion
disintegration~of comet C/2017~S3, is examined.  The onset times of the outbursts
are determined with high accuracy from the light curve of the nuclear condensation
before it disappeared following~the second outburst.  While the brightness of the
condensation was declining precipitously, the total brightness continued~to grow
in the STEREO-A's HI1 images until two days before perihelion.
The red magnitudes measured in these images refer to a uniform cloud of
nuclear fragments, 2200 km$^2$ in projected area, that began to expand at a rate
of 76~m~s$^{-1}$ at the time of the second outburst.  A tail extension, detected
in~some STEREO-A images, consisted of dust released far from the Sun.  Orbital
analysis~of~the~ground-based observations shows that the comet had arrived from
the Oort Cloud in a gravitational orbit.  Treating positional residuals as offsets
of a companion of a split comet, we confirm the existence of the cloud~of
radiation-pressure driven millimeter-sized dust grains emanating from the nucleus
during the second outburst.  We detect a similar, but compact and much fainter
cloud (or a sizable fluffy dust aggregate fragment) released at the time of the
{\vspace{-0.05cm}}first outburst. --- The debris would make a sphere of
140~m~across and its kinetic energy is equivalent to the heat of crystallization
liberated by 10$^8$\,g of amorphous water ice.  Ramifications for short-lived
companions of the split comets and for 1I/`Oumuamua are discussed.
%
%
\end{abstract}


\section{Comet's Discovery and Early Behavior}
The discovery of comet C/2017 S3 was reported jointly by R.~J.~Wainscoat and
R.~Weryk to result from systematically surveying the sky with the Pan-STARRS
1 180-cm f/4.4 Ritchey-Chr\'etien telescope at Haleakal\={a}, Hawaii, on 2017
September 23; prediscovery images~of the comet were subsequently identified in
several exposures taken on August 17 and September~7~(Green 2017).  At the time of
discovery, the object was of magnitude 21, not stellar in appearance, and showing
possible asymmetry to the east.  The preliminary orbit (Williams 2017a, 2017b)
was still too uncertain to reveal the object's origin, but a subsequent orbit
determination that linked about 50 astrometric observations over a period of
three months, from August 17 to November 18\,(Nakano 2018a), already left no doubt
that C/2017~S3 was a dynamically new comet that had arrived from the Oort Cloud.
These computations also implied that the comet was near 5~AU from both the Sun and
the Earth when discovered, that it was on its way to perihelion at 0.21~AU from
the Sun, to arrive shortly before 2018 August 16.0 UT, and that the orbital plane
had an inclination of 99$^\circ$ to the plane of the ecliptic.

The comet was an unimpressive object at helio\-centric distances greater than
1.3~AU before perihelion, and few physical observations are available from that
period of time.  Between 2.3~AU and 1.4~AU from the Sun, the comet's total
brightness corrected for the phase effect, normalized to 1~AU from the Earth,
and after personal and instrumental corrections have been applied, was found
to have varied with heliocentric distance, $r$, according to a power law
$r^{-n}$, where \mbox{$n = 3.0$\,$\pm$\,0.5}, close to Whipple's (1978)
average for Oort Cloud comets before perihelion; the ``absolute'' magnitude
(normalized to 1~AU from the Sun) amounted to \mbox{$H_0 = 10.7$\,$\pm$\,0.3},
which made the comet a likely candidate for imminent disintegration (Section
4.2).

As late as the second half of June 2018, only 7--8 weeks before perihelion, the
comet looked like a faint speck of light,\footnote{See a set of images taken by
E.~Bryssinck with a 40-cm f/3.8 astrograph at Brixiis Observatory (Code B96) on
2018 June 22--30 at {\tt http://www.astronomie.be/erik.bryssinck/c2017s3.html}.}
never reported brighter than magnitude 13.  Moreover, scattered values of a
dust-production rate proxy parameter {\it Af$\rho$} from late June (see footnote
1) clustered in a range of 50--55~cm, implying an object depleted in dust, in line
with the essentially spherical coma, hardly any dust tail at all, and additional
evidence that is presented below.  In summary, a lackluster performance.

And then it happened: As June was making way for July, the comet exploded
dramatically, rapidly developing a sharp, brilliant nuclear condensation and
finding itself all of a sudden at the life's crossroads.  Its fate was about
to be sealed in the next few weeks.

\section{Phenomenon of Cometary Outburst}
With the advent of space exploration, cometary activ\-ity has been recognized to
consist of contributions from a number of discrete sources on and beneath the
surface of the nucleus, whose emission rates vary with time, depending on their
composition, morphology, dimensions, and the solar energy input received.  As a
function of the orbit, the nucleus' shape, rotation, and other properties, the
comet's light curve is always complex, with frequent short-term ups and downs.
From time to time, a major subsurface reservoir of ices is activated, leading
to a sudden surge in brightness --- an event that is referred to as an {\it
outburst\/}.

With no universally accepted definition, we follow here the description proposed
by Sekanina (2010), according to which an outburst is {\it any sudden, prominent,
and unexpected brightening, caused by an abrupt short-term injection of massive
amounts of volatile material from the nucleus into the atmosphere\/}.  The
fundamental parameters that describe an outburst in a comet's light curve
are:\ (i)~the time of onset, (ii)~the brightness at the peak, (iii)~the
rise time (between the onset and the peak), and (iv)~the amplitude (the
increase in brightness from the onset to the peak).  Also critital is the
degree of asymmetry between the event's rising and subsiding branches.  The
rise time can be as short as a fraction of a day and very seldom exceeds a
few days.  The amplitude should equal at least 0.8--1.0 mag, equivalent to
a flux increase by a factor of 2 to 2.5, but is usually a few magnitudes.
The more appropriate parameters from the standpoint of outburst physics are
described elsewhere~(Sekanina 2017), together with analysis of the photometric
data~for 20 selected events experienced by 10 comets.

Cometary outbursts are frequent and very diverse phenomena, which can be
categorized from various points of view.  In regard of C/2017~S3, we mention
two major criteria:\ one is the overall temporal profile, which separates {\it
gas-dominated\/} from {\it dust-dominated\/} outbursts; the other criterion
is the degree of repercussions for the comet's post-event evolution, which
divides the outbursts into {\it innocuous\/} and {\it ominous\/} (or {\it
portentous\/}) and subdivides the latter into {\it nonfatal\/} and {\it
cataclysmic\/}.

The main difference between the gas-dominated and dust-dominated outbursts is
the degree of asymmetry between the rising and subsiding branches.  Gas-dominated
events are nearly symmetric relative to the peak, particularly at smaller
heliocentric distances, because the subsiding branch reflects the fairly short
photodissociation lifetime of the molecular species, primarily diatomic carbon,
observed in the light curve.  The subsiding branch of dust-dominated events is
much more extended because it is determined by the residence time of dust grains
in the coma, which, for larger particles in particular, is considerably longer
than the dissociation lifetime of molecules.  Accordingly, dust-dominated
outbursts are highly asymmetric with respect to the peak.

Having experienced an innocuous outburst, the comet does not subsequently exhibit
any anomalous changes~in its behavior, the post-event activity pattern resembling
the pre-event one.~By contrast, a nonfatal outburst~leaves a mark on the comet's
activity and/or appearance.  Following the event, the comet may stay intrinsically
either much brighter or much fainter over extended periods of time.  A more vigorous
effect of a nonfatal outburst~is~its intimate association with --- or, rather, a
{\it de facto\/} manifestation of --- a splitting of the nucleus into two or more
massive fragments.  Some time after the outburst subsides, the comet's nucleus
begins to appear double or multiple, with the separation increasing with time.
Its further evolution varies from case to case, but typically only the primary
fragment survives with no signs of deteriorating health.  In a cataclysmic
outburst the comet's existence is terminated by complete --- whether rapid or more
gradual ---  disintegration into refractory debris; following the peak, the
brightness of the comet's near-nucleus region plunges precipitously never to
recover again.  Accordingly, this type of outburst could also be referred to
as {\it terminal\/}.

Outbursts are not necessarily isolated events, as they sometimes come in pairs or
larger lineups, separated by relatively short periods of time.  The individual
events~in a group of outbursts could be either of the same category in terms of the
repercussions (e.g., two consecutive nonfatal outbursts accompanying two nucleus
fragmentation events) or of different categories (e.g., an innocuous outburst
followed by a cataclysmic outburst).  Regrettably, one cannot distinguish whether
an outburst is innocuous or ominous until the repercussions become evident.
Accordingly, an unfortunate property of outbursts is that they {\it cannot serve
as ground for prognosticating the comet's future health\/}.

\section{Continuing Saga of Comet C/2017 S3}
The comet's explosion, referred to at the end of Section~1, was the outset of a
{\it bona fide\/} outburst, as defined in Section~2.  The event was first reported
by M.~J\"{a}ger, who noticed it in an image that he had obtained with a 30-cm f/4
{\vspace{-0.04cm}}telescope at Stixendorf, Austria (Code A71), on July~1.98
UT.\footnote{See a website {\tt https:/$\!$/groups.yahoo.com/neo/groups/comets-
ml/conversations/messages/27103}.}  The comet was then 3~mag brighter than the
previous day.  From the temporal variations in the {\it Af}$\rho$ parameter and
in the nuclear magnitudes of the comet, E.~Bryssinck pointed out that the event
began on June 30.\footnote{This and following information is extracted from the
website mentioned in the footnote 2, from the messages Nos.\ 27106, 27126, 27135,
27139, 27149, 27151, 27153, 27164, 27165, 27168, 27176, and 27179.}  On July 4.0
UT, J\"{a}ger detected a 10$^\prime$ long ion tail.  The nuclear condensation
was brightening according to him until July 4, which is consistent with
A.~Novichonok's estimates of the total magnitude.  The brightness then began to
subside and by July 13, the nuclear condensation became indistinct to the
extent that J\"{a}ger suspected the comet's disintegration in progress.

However, on July 15.0 UT a second outburst was in full bloom, the comet described
by J\"{a}ger as strongly condensed and an ion tail apparent.  The brightness in an
aperture of 40$^{\prime\prime}$ across was seven times higher than in an aperture
of 12$^{\prime\prime}$, suggesting a relatively flat distribution of the surface
brightness, decreasing with the distance $\rho$ from the center only as
$\rho^{-0.4}$.  M.~Meyer noticed on July 20 that the comet was losing the
condensation, a development reminiscent of the first outburst.  The appearance
of the ion tail was intermittent; for example, it was prominent and nearly
3$^\circ$ long in an image taken by G.~Rhemann, Eichgraben, Austria (Code C14),
on July 20.0 UT, but was completely missing in his image taken 23~hours later.
J.-F.~Soulier, observing with a 30-cm f/3.8 reflector at Maisoncelles, France
(Code C10), said that in an image taken on July 27.06 UT the comet appeared to
be ``in agony'', while J\"{a}ger's animation using his images taken on August~1
showed the comet's head diffuse and clearly elongated.  The last ground-based
observation was made by Soulier on August 3, when the comet's elongation from
the Sun was 27$^\circ\!$.3; it dropped to 25$^\circ$ in the next 24~hours.

\begin{figure*}
\vspace{-3.45cm}
\hspace{-0.95cm}
\centerline{
\scalebox{0.7}{ 
\includegraphics{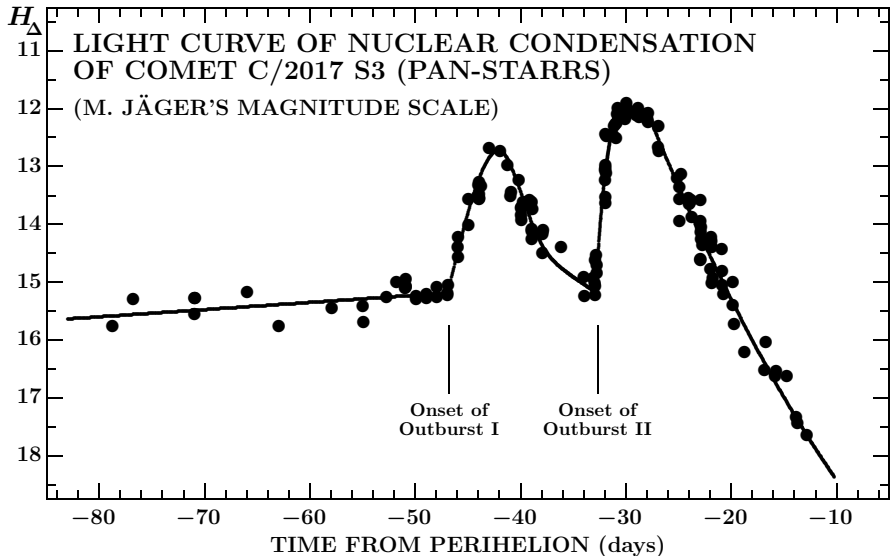}}}
\vspace{-9.95cm}
\caption{Light curve of the nuclear condensation of comet C/2017 S3, based on
129 averages of 525 individual nuclear-magnitude~determinations by 15 observers.
The data were normalized to 1~AU from the Earth and a 4$^{\prime\prime}$ aperture
and corrected for instrumental effects and for the phase effect with the Marcus
(2007) law for dust-poor comets.  After a gradual brightening, Outburst I began
46.8~days before perihelion at 1.25~AU from the Sun, Outburst II 32.6~days
before perihelion at 0.96~AU from the Sun.  The rise time and amplitude equal,
respectively, 4~days and 2.5~mag for Outburst I and 1.5~days and 3.2~mag for
Outburst~II.  The prominent, rapid post-peak fading is typical of gas-dominated
outbursts.  Note the steep terminal rate of decline at $\sim$0.3 mag~per~day.  The
perihelion occurred on 2018 August~15.95~TT.{\vspace{0.54cm}}}
\end{figure*}

It was fortunate that on July 31 the comet entered~the field of view of the HI1
imager of the STEREO-A spacecraft,\footnote{See {\tt http://stereoftp.nascom.nasa.gov}.} in which it stayed until August~14.~The comet~was easily seen in the level-2
images, which we examined.

More than a week after perihelion, the comet began to transit the field of view
of the C3 coronagraph on board the SOHO spacecraft.  Our inspection of these
images failed to show any trace of the comet.

In the light of this, a report of visual detection of the comet's debris
2--3 months after perihelion was rather unexpected.  The recovery remained
unconfirmed, as independent searches failed to corroborate the report.

The comet's apparent demise soon after the two outbursts poses questions on
their possible impact, such as:\ Which outburst was more damaging to the
comet's nucleus?  Or:\ Would the effects of the second event be less severe
in the absence of the first?  We address these and related issues (i)~by
studying the changes in the brightness and appearance of the comet with time,
and (ii)~by investigating its orbital motion.

\section{The Light Curve}
This section is divided into three parts to accentuate the differences in the
photomteric behavior of the nuclear condensation and the comet as a whole, as
well as to underline the apparent correspondence between the ground-based and
STEREO-A light curves.

\subsection{CCD Nuclear Magnitudes}
It was proposed in Sekanina (2010) that, if properly analyzed, CCD nuclear
magnitudes routinely reported with astrometric observations in the publications
of the {\it IAU Minor Planet Center\/} (MPC) can be used to constrain,
often very tightly, the onset time of outbursts.  Because of equipment
differences among the observers (especially in the size of the scanning
aperture and the filter used), appropriate corrections should be applied
before the data sets are combined and the method allowed to work.

Employing this technique, we utilized 129 averages of 525 individual
rapid-succession observations of the nuclear magnitude (marked by N), as well
as color data --- G, R or V --- if comparable to N, obtained at 15~observing
sites\footnote{See {\tt http://www.minorplanetcenter.net/db\_search}.} (Codes
970, A71, A77, B96, C10, C23, C47, D35, G40, J01, J22, J95, K02, L12, Z80).
Very few data were discarded because of their sizable deviations from the
remarkably consistent curve presented in Figure~1, in which the nuclear
magnitudes are normalized to 1~AU from the Earth and are corrected for the phase
effect using the Marcus (2007) law for dust-poor comets.  Rather arbitrarily,
the instrumental corrections were applied to reduce the data to J\"{a}ger's
magnitude scale for a scanning aperture of 4$^{\prime\prime}$ in radius.

\begin{figure*}
\vspace{-2.75cm}
\hspace{-0.8cm}
\centerline{
\scalebox{0.67}{  
\includegraphics{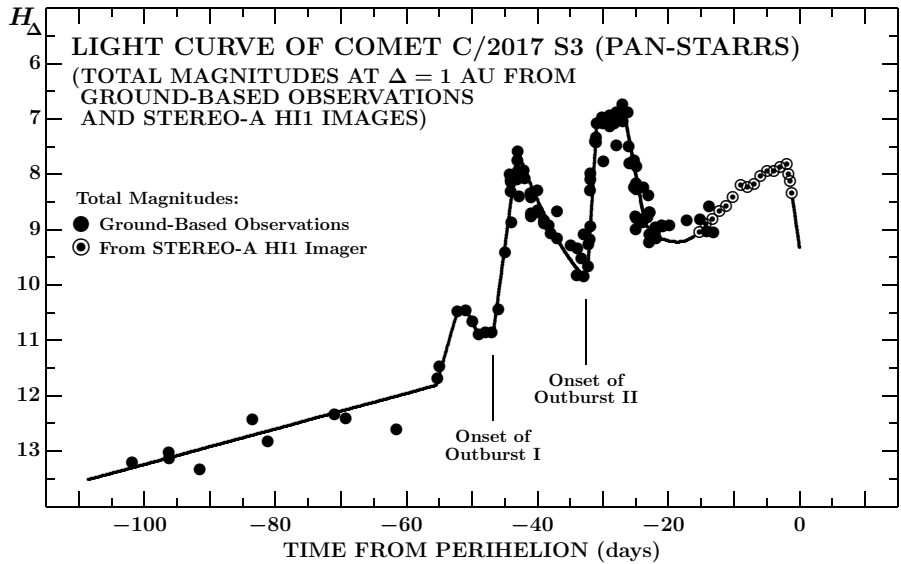}}}
\vspace{-9.7cm}
\caption{Light curve of comet C/2017 S3 based on the total visual and CCD
observations from the ground (solid circles) and measured by the second author
from level-2 images taken with the HI1 imager on board STEREO-A (circled dots).
The plotted magnitudes $H_\Delta$ were normalized and corrected in the same
fashion as the nuclear magnitudes in Figure~1.  While both light curves show
the two outbursts, they differ from one another dramatically after Outburst~II
terminated, when the total brightness began to increase again.  There is also
possible evidence for a precursor flare-up prior to Outburst~I.  The measured
R magnitudes of the STEREO-A images were converted from the R to V magnitudes
by applying an approximate color index of +0.4 mag.  The perihelion occurred
on 2019 August 15.95 TT.{\vspace{0.51cm}}}
\end{figure*}

The outbursts differ from one another in several respects.  Both the rising
and subsiding branches for the second outburst are distinctly steeper and its
amplitude is higher (3.2~mag against 2.5~mag) than for the first outburst.
The peak of the second outburst appears to be distinctly wider, suggesting
perhaps that the event consisted of several explosions in rapid succession.
The steep subsiding branches are strong evidence that both outbursts were
unquestionably gas-dominated events.  And the steeper slope of the second
outburst may imply an effect of the lifetime of molecules, which varies
with the square of heliocentric distance.  The rate of fading after the
second outburst was brutal, $\sim$0.3~mag per day, an indication that
the {\it nuclear region\/} was being very thoroughly vacated by gas and 
photometrically effective dust.  However, one cannot rule out that sizable
{\it inactive\/} fragments, which are difficult to detect, still persisted
near the location of the parent nucleus.

The first outburst must have begun many hours before July~1.0~UT, because
three observers reported the comet to have already been anomalously bright,
displaying a brilliant, starlike nuclear condensation, around the UT midnight
from the 30th to the 1st.  These and other observations made between July~1.0
and 3.0~UT line up in the plot of the light curve along
a slope implying that the event commenced close to June~30.0~UT, except that
the images by A.~Diepvens with a 20-cm refractor at Olmen, Belgium (Code C23),
preclude an onset time before June~30.1~UT.  These constraints provide for the
nominal time of onset for this Outburst I an estimate of June~30.2\,$\pm$\,0.1 UT.

Similarly, the second outburst could not commence~after July~14.9 UT because
it was already in progress~three quarters of an hour before the UT midnight
of July~15, when Soulier took the first image of the night.~The~flare-up
was confirmed a half an hour later, still~before the midnight, by P.~Carson's
observations at the Eastwood Observatory (Code K02), Leigh-on-Sea, Essex,
United Kingdom, with a 32-cm reflector and f/5.4 focal reducer; and within
minutes of the midnight by other observers, including G.~Dangl at Nonndorf,
Austria (Code C47), G.~Vandenbulcke at Koksijde, Belgium (Code L12), as well
as Bryssinck and J\"{a}ger.~On~the other hand, the observations by
B.~L\"{u}tkenh\"{o}ner et al.\ at the Slooh Observatory on Mt.~Teide,
Tenerife, Canary Islands (Code G40), and by K. Hills, who worked with a
50-cm f/2.9 astrograph at the Tacande Observatory, La Palma, Canary Islands
(Code J22), rule out an onset before July~14.3~UT.  We adopt July~14.4\,$\pm$\,0.1
UT as the start of Outburst II.

\subsection{$\!$Total~Magnitudes~from~Ground-Based~Observations, and the Comet's
Appearance}
We were able to collect ground-based observations of the comet's total brightness
made by 20 observers.  Most of the data come from 14 visual observers, who
reported their results either to the Crni Vrh Observatory's {\it COBS
Database\/},\footnote{See {\tt http://www.cobs.sl/analysis}.{\vspace{0.035cm}}}
or to the {\it International Comet Quarterly\/},\footnote{See {\tt
http://www.icq.eps.harvard.edu/CometMags.html}.{\vspace{0.035cm}}} the two
sources we consulted to collect the data for analysis.  However, because the
comet had been very faint before undergoing the first ourburst, it was
necessary to supplement the visual observations with a set of total CCD
magnitudes.  Most of these data were obtained from the {\it Minor Planet
Center\/}'s database~of~astro\-metric observations {\vspace{-0.06cm}}(see
footnote~5), the source that also provided the large set of nuclear
magnitudes.\footnote{The data reported by observers as the CCD total magnitudes
(with no filter) are marked by T to distinguish them from the CCD nuclear
magnitudes and various color magnitudes.{\vspace{0.035cm}}}  Additional total
CCD magnitudes were found in the two sources of visual magnitudes.

Although combining visual and CCD magnitudes carries risks of their questionable
compatibility, we used the method, expounded in some detail elsewhere
(Sekanina 2017), that applies appropriate personal and instrumental corrections
to minimize these risks.  Since information on the comet's brightness at very
large heliocentric distances was too fragmentary, we limited our analysis to
an orbital arc of less than $\sim$100~days from perihelion.  The compatibility
tests were passed by the data sets reported by only seven CCD
observers.\footnote{One observer provided both visual and CCD magnitudes,
making the sum of visual and CCD observers exceed the total number of
observers.}

\begin{figure*}
\vspace{-2.7cm}
\hspace{-0.8cm}
\centerline{
\scalebox{0.66}{
\includegraphics{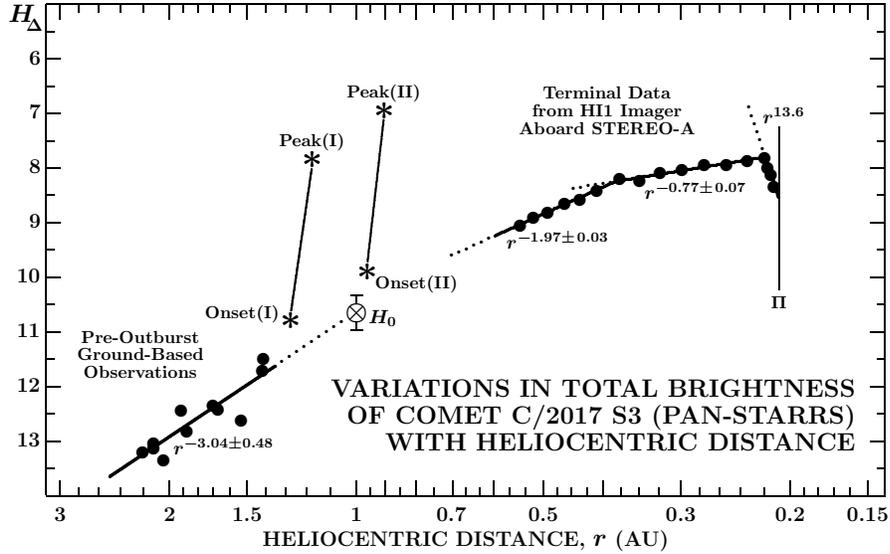}}}
\vspace{-9.6cm}
\caption{Variations in the total normalized brightness of the comet C/2017 S3
as a function of heliocentric distance.  The magnitudes $H_\Delta$ are identical
with those in Figure~2.  The comet's pre-outburst gradual brightening at distances
exceeding 1.4~AU from the Sun preperihelion is shown at the lower left.  The
extrapolated absolute magnitude $H_0$ and the onset points, Onset(I), Onset(II),
and the peak points, Peak(I), Peak(II), of the two outbursts are depicted.
Selected total magnitudes from the level-2 images by the HI1 camera on board
STEREO-A, displayed to the upper right, are the same data as in Figure~2.  The
comet's perihelion distance of 0.208~AU is marked by $\Pi$.{\vspace{0.5cm}}}
\end{figure*}

The total number of ground-based observations, either visual or CCD, employed
in our analysis and plotted in Figure~2, is 109.  All 11 data points beyond
1.4~AU from the Sun are CCD magnitudes.  The scatter among the total brightness
entries is greater than that among the nuclear magnitudes and it is estimated at
$\pm$0.3~mag on the average.  The main problem with the total CCD magnitudes is
is that they often exclude the outermost coma and require large corrections to
account for this deficit.

The light curve in Figure 2 shows a very gradual brightening of the comet when
it was more than 55 days before perihelion or $>$1.4~AU from the Sun.  Plotted
in Figure~3 against heliocetric distance $r$ rather than time, the normalized
brightness varies as $r^{-3.04 \pm 0.48}$, a rate that is rather typical of
Oort Cloud comets, as already noted in Section~1.  The visual absolute magnitude,
which characterizes the comet's stamina, is --- when extrapolated from the
pre-outburst orbital arc between 2.2~AU and 1.4~AU from the Sun --- equal to
\mbox{$H_0 = 10.66 \pm 0.32$}, more than 2~mag below Bortle's (1991)
perihelion-survival limit (of 8.2 for this object) that identifies objects prone to
early disintegration.  The subsequent evolution of C/2017~S3, while unpredictable
in detail, was not entirely surprising.  One unexpected minor feature in the light
curve in Figure~2 is a possible precursor flare-up, with an amplitude slightly
exceeding 1~mag, which, if genuine, began about 54 days before perihelion, on
June~23.  This feature does not show up in Figure~1.

The profiles of the two outbursts, including their rise times and the plateau
of Outburst~II, look rather alike in Figures~1 and 2.  The only slight disparity
for Outburst~I is its higher amplitude, equaling 3.5 mag (and rivaling Outburst~II)
in the total light.

The stunning difference between the light curves based on the nuclear and total
brightness data is apparent at the end of Outburst II:\ while the nuclear
condensation was fading dramatically, the comet's total brightness
stopped subsiding once the flare-up died down.  Unfortunately, the object
was by then less than 30$^\circ$ from the Sun and ground-based observations
terminated.  The subsequent developments could luckily be followed in the
images taken by the HI1 camera on board STEREO-A, as described below.

A characteristic property of C/2017 S3 was its prominent green color,
repeatedly commented on by observers.  The color was undoubtedly a corollary of
the comet's visible spectrum being dominated by the \mbox{$d^3\Pi_g\!\rightarrow
\!a^3\Pi_u$} transition of the diatomic carbon molecule, whose strongest (0--0)
band is near 517~nm.

\begin{figure}[b]
\vspace{0.6cm}
\hspace{-0.16cm}
\centerline{
\scalebox{0.465}{
\includegraphics{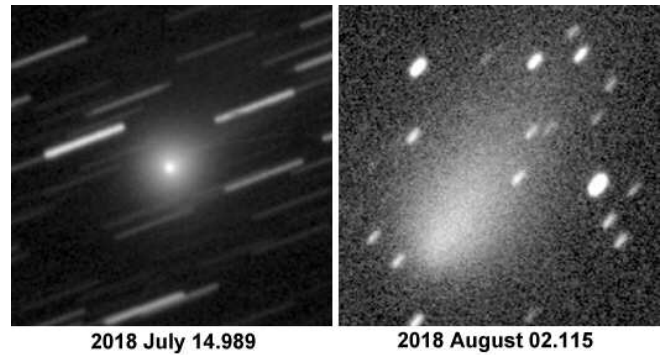}}} 
\vspace{-0.1cm}
\caption{Comparison of the appearance of C/2017 S3:\ highly condensed in an
early stage of Outburst~II (left) and after it permanently lost the nuclear
condensation (right).  The images were obtained with a 30-cm f/3.8 reflector + CCD
(no filter) and each is 4$^{\prime\!}$.6 on a side. (Image credit:\ J.-F.\
Soulier, Maisoncelles, France.){\vspace{-0.1cm}}}
\end{figure}

The appearance of the comet was correlated with its brightness.  In the early
stage of Outburst~I the most striking feature was a bright, starlike nuclear
condensation gleaming through the green coma, with ion tail reported on
a few occasions, but not continuously.  As the outburst proceeded, the
condensation was becoming progressively less distinct and more diffuse over
a period of several days.  At the outset of Outburst~II, this cycle of
morphological changes repeated itself, but as the event was subsiding, the
observers noticed that the condensation continued to expand and fade to the
point of disappearance, with the flat surface-brightness distribution in the
inner coma somewhat elongated in the antisolar direction, thereby degrading
the quality of astrometry.
\begin{figure*}[t]
\vspace{0.15cm}
\hspace{-0.19cm}
\centerline{
\scalebox{1.18}{
\includegraphics{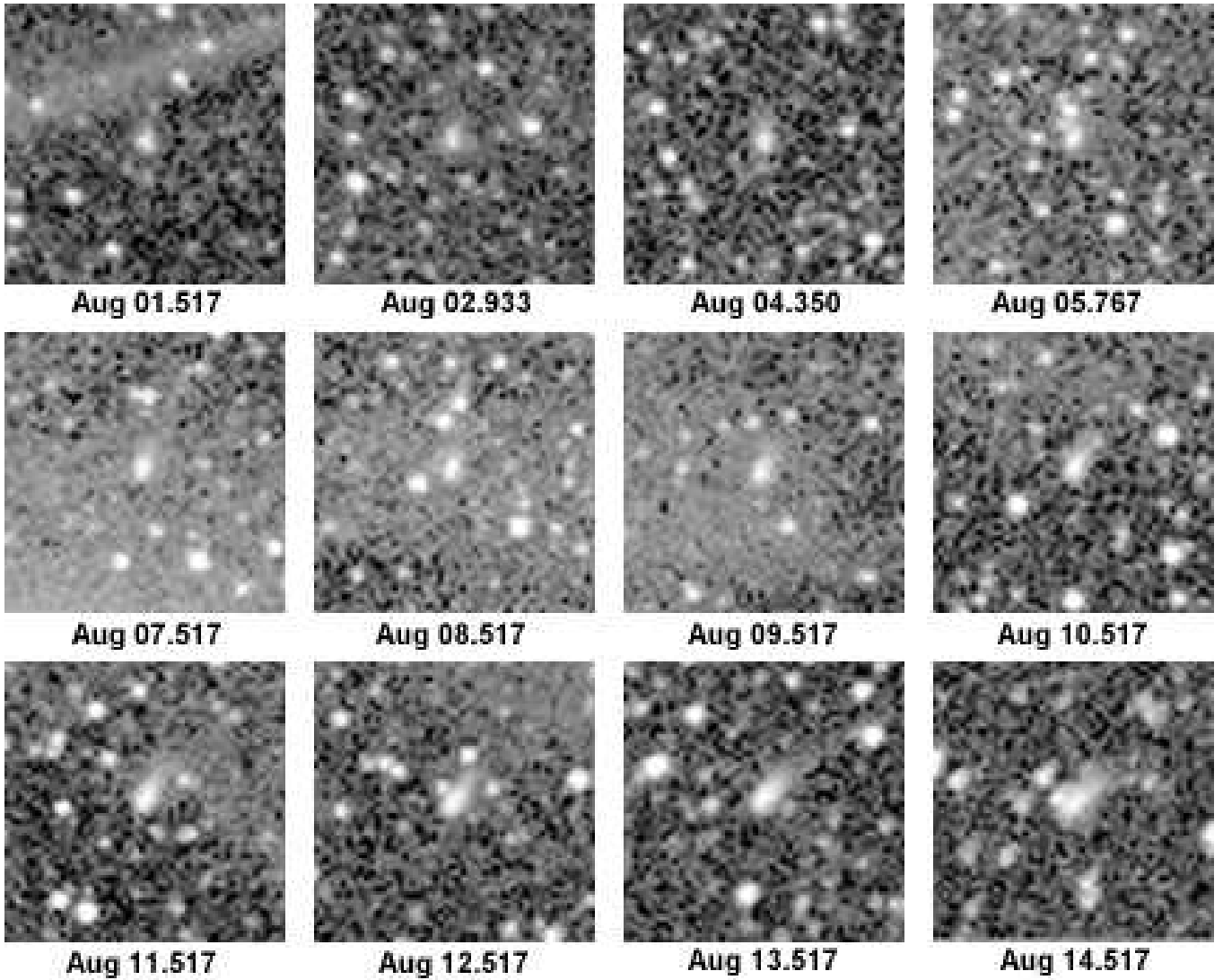}}} 
\vspace{-0.22cm}
\caption{Examples of the level-2 images of C/2017 S3 taken by the HI1 camera
on board the STEREO-A spacecraft on 2018 August~\mbox{1--14}, shortly before
perihelion.  The cloud of nuclear debris is located in each frame's center.
From August 7 on, a tail is seen to point to the upper right.  Each frame is
1$^\circ$ on a side, the position angle of the direction up is $\sim$335$^\circ$.
(Image credit:\ NASA/SECCHI consortium.){\vspace{0.5cm}}}
\end{figure*}
Figure 4 illustrates the contrast between the comet's appearance soon after the
onset of Outburst II and following the disappearance of the nuclear condensaton
less than three weeks later.

\subsection{Total CCD Magnitudes of the Comet's Debris\\from Images Taken by
STEREO-A}
By the time the comet entered the field of view of the HI1-A imager, it was
clear that the loss of the nuclear condensation was permanent and that the
integrity of the comet's nucleus was compromised, its mass subjected to severe
fragmentation.  We provide evidence for this conclusion by investigating the
comet's orbital motion in Sections~7--8, but the ground-based imaging from the
last days of July and the first days of August leaves no doubt that in late
July the object's ability to function as an {\it active comet was paralyzed\/}
and that a cloud of some sort of {\it debris\/} now occupied the
\mbox{position}~of~the~\mbox{former}~\mbox{nucleus}.  To describe the debris is a
major goal of this study. We suspect that\,{\it Outburst II was a cataclysmic
event\/} and in the following we search for more supporting evidence.

The comet entered the field of view of the HI1 imager on board STEREO-A
on July 31 and left it on August 14.  The second author used the {\it
Astrometrica\/} interactive software tool to determine magnitudes of the
comet's debris clearly visible in the level-2 images, examples of which are
displayed in Figure~5.  The measurements were made with a scanning aperture
of 2~pixels in radius.  With a pixel size of $\sim$72$^{\prime\prime}$ and
the detector's spectral bandpass of $\sim$600~nm to $\sim$750~nm, the measured
data represent {\it total red magnitudes\/}.  Their mean error is about
$\pm$0.1~mag, substantially better than the uncertainty of the ground-based
brightness estimates.  Converted to the visual magnitudes by applying an
approximate correction of +0.4~mag, they are plotted in Figures~2 and 3.
Figure~2 exhibits their excellent compatibility with the ground-based
observations; demonstrates, from about July~25 on, an enormous disparity between
the light curves derived from, respectively, the nuclear and total brightness
(already alluded to in Section 4.2); and indicates that the comet's debris
in the aperture continued to brighten until August 14.0 UT, or 2~days before
perihelion. Only at that point did a fading set in.

The most significant piece of information on the brightness evolution of the
cloud of debris in Figure~3 is the brightness-variation law of $r^{-2}$ between
July 31 and August 7, implying {\it no loss in the projected cross-sectional
area\/} of the cloud of fragments in this time span!  Regardless of the extent
of damage inflicted upon the nucleus, it is obvious that in this period of time
the field of debris was still traveling in an organized manner.  Excluding an
unlikely scenario in which the loss rate of a cross-sectional area is always
compensated by the exactly same rate of debris fragmentation, the $r^{-2}$ law
suggests that during the week-long period the {\it scanning aperture contained
the whole volume of the debris cloud\/}.

Between {\vspace{-0.05cm}}August 7.0 and 14.0 UT, the comet's debris still
brightened, but at a much slower rate, as $\sim\!\!r^{-\frac{3}{4}}$.  In relation
to the previous period, this trend can be interpreted to mean that the dimensions
of the debris cloud now began to spill outside the field covered by the scanning
aperture, with an ever decreasing fraction of the cloud detected.  The peak on
August 14.0 UT is sharp, the brightness then starting to drop precipitously.
This event is deemed to display the last gasp of life, resulting apparently in
a rapidly accelerated expansion of a second generation of debris and signaling
the imminent, ultimate demise of the comet.

\section{Interpretation of the STEREO-A Light Curve}
In an effort to model the brightness variations in the STEREO-A images, we
formulate a simple hypothesis:\ at time $t_{\rm frg}$ the comet's nucleus suddenly
disintegrated into a cloud of fragments of equal dimensions, which was optically
thin and expanding isotropically with a uniform radial velocity $v_{\rm exp}$,
reaching at time $t$ a radius
\begin{equation}
\rho(t) = v_{\rm exp} (t \!-\! t_{\rm frg}).
\end{equation}
Let the spatial density of the fragments be independent of the position in the
cloud and decreasing with time as $\rho^{-3}$.  Centered on a circular scanning
aperture $n$ pixels in radius, the cloud stays within the aperture's bounds as
long as
\begin{equation}
\rho \leq a,
\end{equation}
where $a$ is the radius of the aperture at the comet's distance $\Delta$ from
the STEREO-A spacecraft,
\begin{equation}
a = 725 n p \Delta,
\end{equation}
with $p$ being a pixel size in arcsec, while $\Delta$ is in AU; $a$ is then
in km.

As the cloud of debris keeps expanding, its radius is sooner or later bound
to exceed the radius of the scanning aperture, \mbox{$\rho > a$}, the volume
that stays confined to the aperture is given by the intersection of a cylinder
of radius $a$ and infinite length (closely approximating at the comet the
scanning cone whose vertex is at the spacecraft) with a sphere, of radius
$\rho$, whose center is located on the cylinder's axis.  This confined volume
of space equals the sum of the volume of the cylinder of radius $a$ and length
2$b$, where
\begin{equation}
b = \sqrt{\rho^2 \!-\! a^2},
\end{equation}
plus the volume occupied by two spherical caps, each having a height $\rho - b$
and a base radius $a$.  The volume of the expanding spherical cloud of debris
that is confined to the aperture equals
\begin{equation}
U_{\rm apert} = 2 \pi a^2 b + \frac{2 \pi}{3} (\rho \!-\! b)^2 (2 \rho + b).
\end{equation}
{\vspace{0.25cm}}

{\vspace{-0.43cm}}
{\noindent}Since the total volume of the cloud equals \mbox{$U_{\rm cloud} =
\frac{4}{3} \pi \rho^3$}, the volume fraction in the scanning aperture amounts
to (for $\rho > a$)
\begin{equation}
{\cal A} = \frac{U_{\rm apert}}{U_{\rm cloud}} = 1 - \! \left[1 - \! \left(
 \frac{a}{\rho} \right)^{\!\!2} \right]^{\!\frac{3}{2}} \!\!\! < 1,
\end{equation}
which is, on our assumptions, also the fraction of the total cross-sectional
area of the debris cloud in the aperture's field.  When \mbox{$\rho \leq a$},
the fraction is \mbox{${\cal A} = 1$}.

For the images taken after August 7.0 UT, the expression (6) is to be compared
for each imaging time $t$ with the fraction of the cross-sectional area of the
debris in the aperture, computed from the absolute magnitude $H_0(t)$,
\begin{equation}
{\cal A} = 10^{0.4(\widehat{H}_0 - H_0)},
\end{equation}
where $\widehat{H}_0$ is the constant absolute magnitude derived from the images
taken between July 31 and August 7, representing the total cross-sectional area
of the debris in the cloud.

From Equation (6) the radius of the debris cloud equals
\begin{equation}
\rho = a \left[ 1 - (1 \!-\! {\cal A})^{\frac{2}{3}} \right]^{-\frac{1}{2}}
\end{equation}
and the hypothesis of a uniform isotropic expansion of the debris cloud is tested
by its basic condition, expressed by Equation~(1).  Written as
\begin{equation}
t = t_{\rm frg} + \frac{1}{v_{\rm exp}} \rho,
\end{equation}
the two parameters of the hypothesis, the fragmentation time of the debris,
$t_{\rm frg}$, and the expansion velocity, $v_{\rm exp}$, are given as the
ordinate and the reciprocal of the slope, respectively.  The degree, with
which the relation approximates a straight line, measures how successful our
hypothesis is.  Also, if our suspicion that Outburst~II was the event that
doomed the comet is correct, there should be a correspondence between the
fragmentation time $t_{\rm frg}$ and the time of Outburst~II.

The measurements of the comet's apparent red magnitudes in the HI1-A level-2 
images were made with a circular aperture of 2~pixels.  With a pixel size of
71$^{\prime\prime\!}$.94, the aperture radius (in km) at the comet's distance
$\Delta$ (in AU) from STEREO-A{\vspace{-0.03cm}} becomes according to
Equation~(3) \mbox{$a = 10.43 \times \!10^4 \Delta$}.

Table 1 presents the apparent magnitude measurements (in column~3) as a function
of time, together with the comet's distances from STEREO-A and the Sun, the
phase angle, and the derived quantities:\ the absolute red magnitude $H_0$,
the aperture radius $a$, the fraction of the cross-sectional area of the cloud
of debris in the aperture, $\cal A$, derived from Equation~(7), and the debris
cloud's radius $\rho$, calculated from Equation~(8).

\begin{table*}[t]
\vspace{-4.2cm}
\hspace{-0.6cm}
\centerline{
\scalebox{1}{
\includegraphics{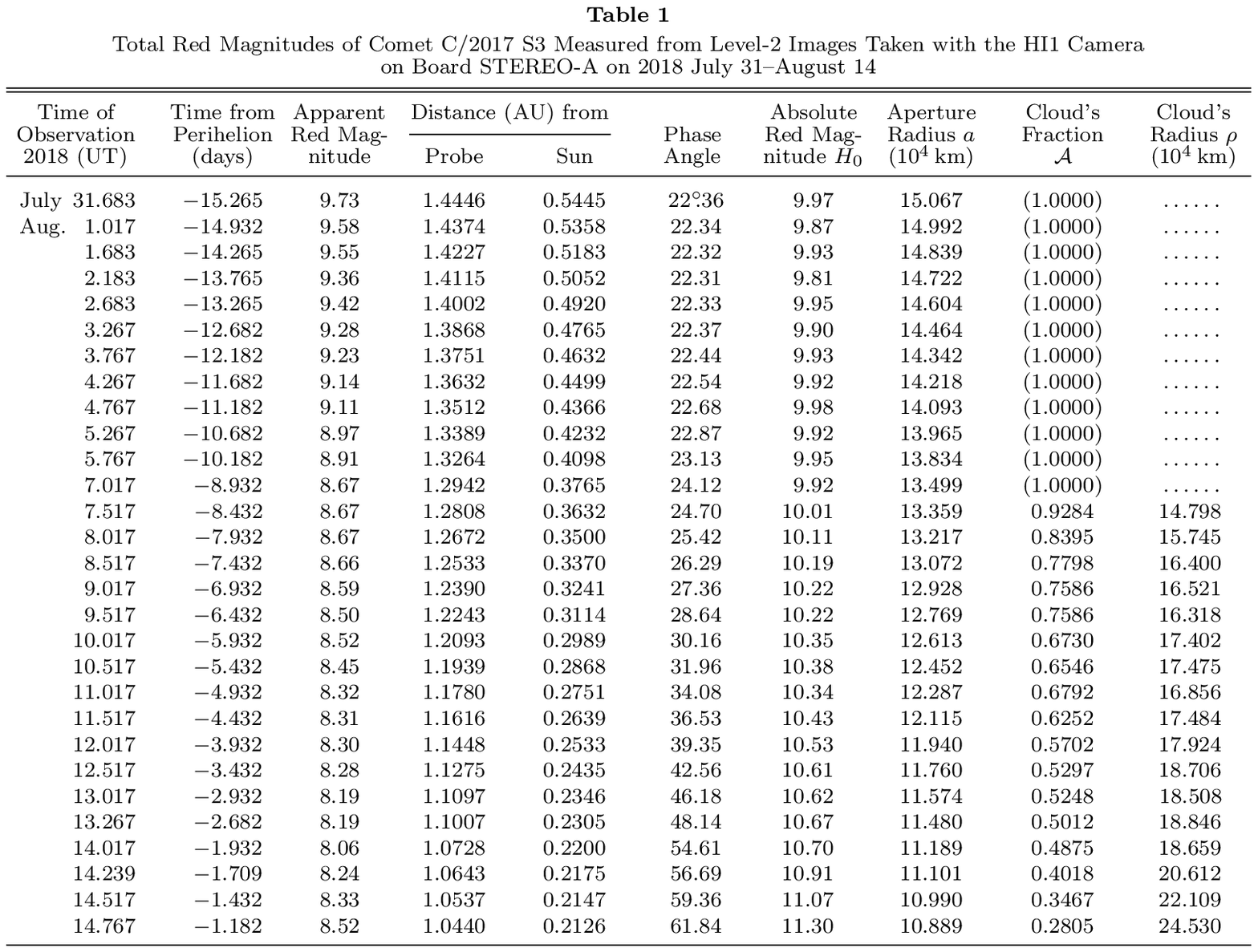}}}
\vspace{-12.06cm}
\end{table*}

The most striking feature of the first part of the table, July~31.7 to
August~7.0 UT, is the essentially constant value of the absolute red magnitude,
which averages
\begin{equation}
\widehat{H}_0 = 9.92 \pm 0.05.
\end{equation}
This absolute magnitude reflects the $r^{-2}$ dependence of the normalized
magnitude in Figure~3 and implies a constant cross-sectional area of the
fragmented nucleus in the scanning aperture (as noted in Section 4.3), which
equals the total cross-sectional area of the cloud, $X_{\rm frg}$.  Thus, 
{\vspace{-0.07cm}}besides its use in computing the fraction ${\cal A}$ in
Table~1 from Equation~(7), $\widehat{H}_0$ is employed to determine $X_{\rm
frg}$ from
\begin{equation}
X_{\rm frg} = \frac{7.03}{p_{\rm R}} 10^{16 + 0.4(\mbox{\Hssun} - \widehat{H}_0)},
\end{equation}
where $p_{\rm R}$ is the geometric albedo of the debris in the read part of
the spectrum and {\Hsun} is the Sun's red magnitude.  Taking \mbox{$p_{\rm
R} = 0.05$} and \mbox{{\Hsun} = $-$27.10}, we find
\begin{equation}
X_{\rm frg} = 2200 \pm 100 \:{\rm km}^2.
\end{equation}

The hypothesis of a uniform isotropic expansion of the debris cloud is tested
in Figure~6, in which the time of observation is plotted as a function of the
derived radius of the cloud once it exceeded the radius of the aperture.  A
least-squares fit to the data in the interval of time August 7.5--13.3 UT
yields a solution
\begin{eqnarray}
t\!-\!t_\pi & = & -31.5 + 1.52 \rho, \\[-0.1cm]
& & \hspace{0.17cm} \pm 2.5\hspace{0.06cm} \pm \!0.15 \nonumber
\end{eqnarray}
where time is in days, $\rho$ is in 10$^4$\,km, and $t_\pi$ is the comet's
perihelion time derived in Section~7.  Figure~6 clearly demonstrates the
presence of a linear relationship between $\rho$ and $t$ and submits for
the fragmentation time
\begin{equation}
t_{\rm frg} \!-\! t_\pi \!=\! -31.5 \pm 2.5 \;{\rm days} \hspace{0.3cm}(t_{\rm
 frg} \!=\! {\rm July \, 15.5}\pm{\rm 2.5\: UT)},
\end{equation}
in excellent conformity with the timing of Outburst~II, which according to
Figure~1 began 32.6 days before perihelion (July~14.4 UT) and reached a
peak 1.5~days later, 31.1~days before perihelion.  {\small \bf The cataclysmic
nature of Outburst~II is hereby strongly supported.}  From the slope in
Equation~(13) one gets for the cloud's expansion velocity before August~14
\begin{equation}
v_{\rm exp} = 76 \pm 7 \;\,{\rm m\,s}^{-1}.
\end{equation}

The last four points of Table 1, referring to the measured terminal fading
over a period of August~14.0-14.8 UT in Figures 2 and 3, do not fit
Equation~(13).  Instead, they follow a very different straight line of a
considerably flatter slope,
\begin{eqnarray}
t \!-\! t_\pi & = & -4.4 + 0.131 \rho, \\[-0.1cm]
 & & \hspace{-0.01cm} \pm 0.2 \hspace{0.06cm} \pm \! 0.010 \nonumber
\end{eqnarray}
implying an expansion velocity of 880\,$\pm$\,70 m\,s$^{-1}$.  We offer no
conclusive interpretation of this terminal event, but suggest that it may
signal a rapid rate of sublimation of the debris in the cloud, a process
that would be consistent with the high expansion velocity.

\begin{figure}[t]
\vspace{-0.25cm}
\hspace{1.22cm}
\centerline{
\scalebox{0.7}{
\includegraphics{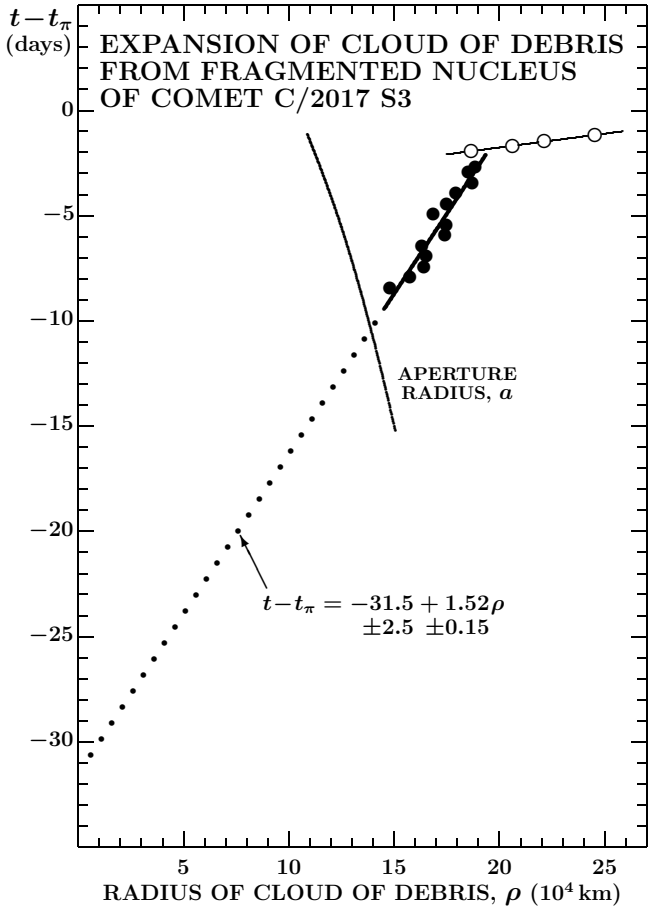}}}
\vspace{-8.8cm}
\caption{Expansion of the cloud of debris after the fragmentation of the nucleus
of comet C/2017 S3.  Solid circles refer to the times between August 7.0 and 14.0
UT, when the expansion followed Equation (13).  The open circles refer to the last
four entries in Table 1.  The aperture radius is also plotted.{\vspace{0.55cm}}}
\end{figure}

The fit to the cross-sectional area of the fragmented nucleus confined to the
scanning aperture, ${\cal A}$, is provided by Equation~(6) via Equation~(13).
It is plotted as a function of time in Figure~7.  The fit actually suggests that
the scanning aperture was already filled with the fragments by August~5.56~UT,
or 10.39~days before perihelion, when the debris cloud expanded to
\mbox{$13.89 \times \!10^4$\,km} in radius, which was at the time also the
radius of the aperture at the comet's distance from STEREO-A.

In an effort to further strengthen the case for the debris cloud's expansion,
an obvious avenue was to search for additional evidence among the ground-based
imaging observations.  Unfortunately, we were able to find no data of this kind.
The expansion is surely documented qualitatively beyond a shadow of a doubt by,
for example, comparing the two images in Figure~4, but quantitative data are
lacking.  This may in part be due to the fact that the boundary of the expanding
cloud was buried in the coma.  The radius $\rho$ in our model thus~remains~a
derived, not measured, quantity.  Yet, its introduction in our formulation was
a convenient mathematical tool, which was justified by strong evidence from
the STEREO-A light curve of the fragmented comet and which allowed us to examine
and eventually establish the relationship between the debris and Outburst~II.

\begin{figure}[t]
\vspace{-4.11cm}
\hspace{1.08cm}
\centerline{
\scalebox{0.74}{
\includegraphics{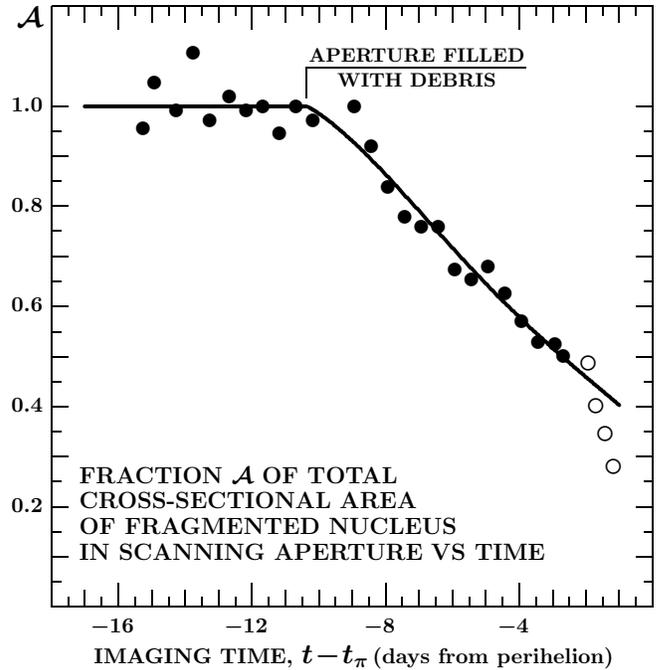}}}
\vspace{-9.25cm}
\caption{Interpretation of the STEREO-A light curve of comet C/2017 S3 in terms
of a temporal dependence of the fraction ${\cal A}$ of the cross-sectional area
of the fragmented nucleus, detected in the scanning aperture 2$^{\prime\!}$.4
in radius.  The solid circles are the data from Table~1 between July~31.7 and
August 13.3 UT, the open circles are the last four data points, as in Figure~6,
Note that the fit suggests that the cloud of debris began to fill the aperture
as early as August 5.56 UT, or \mbox{$t\!-\!t_\pi = -10.39$ days}.{\vspace{0.4cm}}}
\end{figure}

\section{Tail-Like Extension of Fragmented Comet\\in STEREO-A Images}
The images displayed in Figure 5 reveal, at least from August~7 on, an extension
from the cloud to the upper right, slowly rotating clockwise.  If this feature was
related to the comet's fragmentation, it would imply that a fraction of the debris'
mass was contained in dust particles much smaller in size than the fragments in
the cloud, in fact small enough to be subjected to solar radiation-pressure
accelerations high enough to show up outside the cloud on a time scale of only
a few weeks after the outburst.  If confirmed, this would mean that our assumption
of the debris cloud's isotropic expansion did not apply fully, even though the
extension is much fainter than the cloud itself.

\begin{table*}[t]
\vspace{-4.2cm}
\hspace{-0.53cm}
\centerline{
\scalebox{1}{
\includegraphics{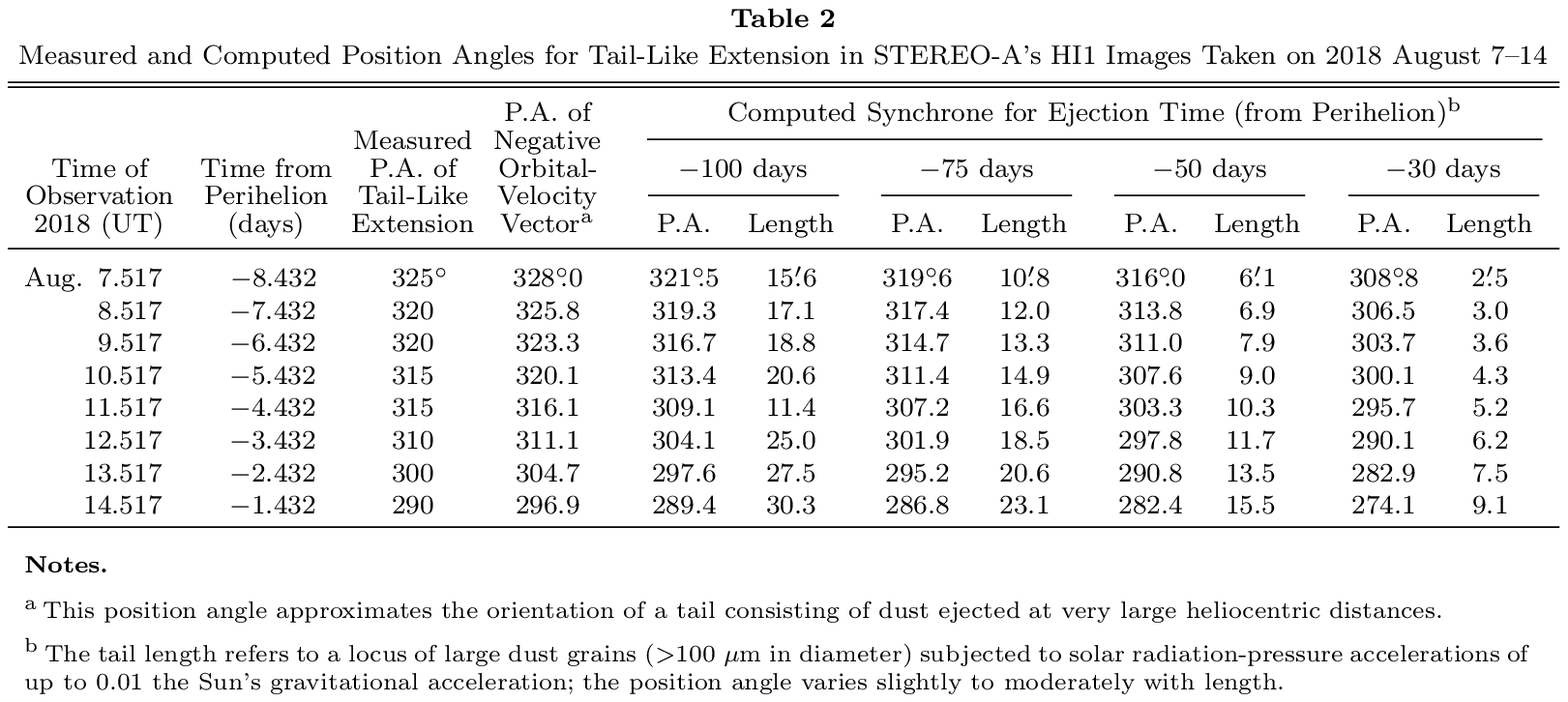}}}
\vspace{-17.48cm}
\end{table*}

The extension appears to be fairly narrow, suggesting that an approximation
by a synchrone should provide at least a crude estimate of the age of the dust
that makes the feature up.  Measurement of the position angle was extremely
difficult because of the low surface brightness of the feature, the crowded
star fields, and the large pixel size of the detector.  In Table~2 we compare
our measurements, which could at best be accomplished with a precision of
$\pm$5$^\circ$, with the expected position angles for four different assumed
ejection (i.e., fragmentation) times and with the negative orbital-velocity vector,
which is a limiting direction for early dust ejecta at large heliocentric distance.
Comets arriving from the Oort Cloud are known to release copious amounts of sizable,
submillimeter-sized and larger dust far from the Sun on approach to perihelion
(e.g., Sekanina 1978; Meech et al.\ 2009).

Comparison of the measured and computed position angles clearly suggests that the
tail-like extension in the STEREO-A's HI1 images was a product of dust emission
at early times, more than 100 days before perihelion, and that it could not be
associated with Outburst II and the comet's disintegration because of systematic
differences in the orientation of more than 10$^\circ$.  The feature does not fit
potential dust ejecta from Outburst~I either.  In summary, the feature does not
contradict our hypothesis of an isotropic, uniform expansion of the comet's
fragmented nucleus (Section 5).

The question that remains to be answered is why this important conclusion requires
the low-resolution \mbox{HI1-A} imaging and does not appear to be supported by
ground-based imaging observations of higher quality.  Here two constraints --- one
physical, the other orbital --- conspire that make in this particular respect the
\mbox{STEREO-A} imaging superior in spite of its low resolution power.  The
physical constraint is the comet's dust-poor nature:\ as long as the object was
active --- until Outburst~II was over --- the dust features were outperformed by
the more prominent gas features.  It was only after the comet's activity ceased ---
the time approximately coinciding with the end of the ground-based and the
beginning of the \mbox{STEREO-A} observations --- that the dust features began
to dominate the comet's appearance.

The orbital constraint is even more important.  Along an essentially parabolic
orbit, the preperihelion dust tail is always extremely narrow until a distance
from the Sun that does not exceed the perihelion distance by more than a factor
of two or so.  This is an effect of angular momentum, which in practice means
that a dust tail is restricted to a sector between the negative orbital-velocity
vector and the radius vector.  The angle subtended by this sector before
perihelion is extremely narrow regardless of the geometry in the Sun-comet-Earth
configuration and is especially constraining for comets with small perihelion
distances such as C/2017~S3.  Indeed, for ground-based observations of this
comet the sector's width never exceeded 40$^\circ$ from the time of Outburst~I
on and was merely 35$^\circ$ at the time of the last observation on August~3.
By contrast, for \mbox{STEREO-A} the sector was 71$^\circ$ wide on
August~7 and 106$^\circ$ wide on August 14, allowing thus a considerably
better angular resolution of dust features.  Hand in hand with this effect
went the features' apparent length.  No dust extensions could at all be
detected in late June and early July, when they were pointing almost exactly
away from the Earth, and they would generally be quite short on ground-based
images in later times as well.  STEREO-A was much better than Earth positioned
for the detection of dust ejecta (especially the early ones) and the images
taken shortly before perihelion suited this purpose nearly perfectly both
timewise and locationwise.

\section{Orbit Determination, and Investigation of the Motion of Fragmented
 Nucleus}
In a quest for information on the role of the two outbursts in the disintegration
of comet C/2017 S3, we provided, in Section 5, compelling evidence of the
cataclysmic nature of Outburst~II.  So far, however, we have been unable to
detect any effect on the comet by Outburst~I, a circumstance that would
support the notion that it apparently was an innocuous event.  In the following
we investigate whether the comet's orbital motion was in any way impacted by
Outburst~I and whether the conclusions from Section~5 on Outburst~II and the
nucleus' disintegration could further be corroborated.

\begin{table*}[t]
\vspace{-3.77cm}
\hspace{-0.53cm}
\centerline{
\scalebox{1}{
\includegraphics{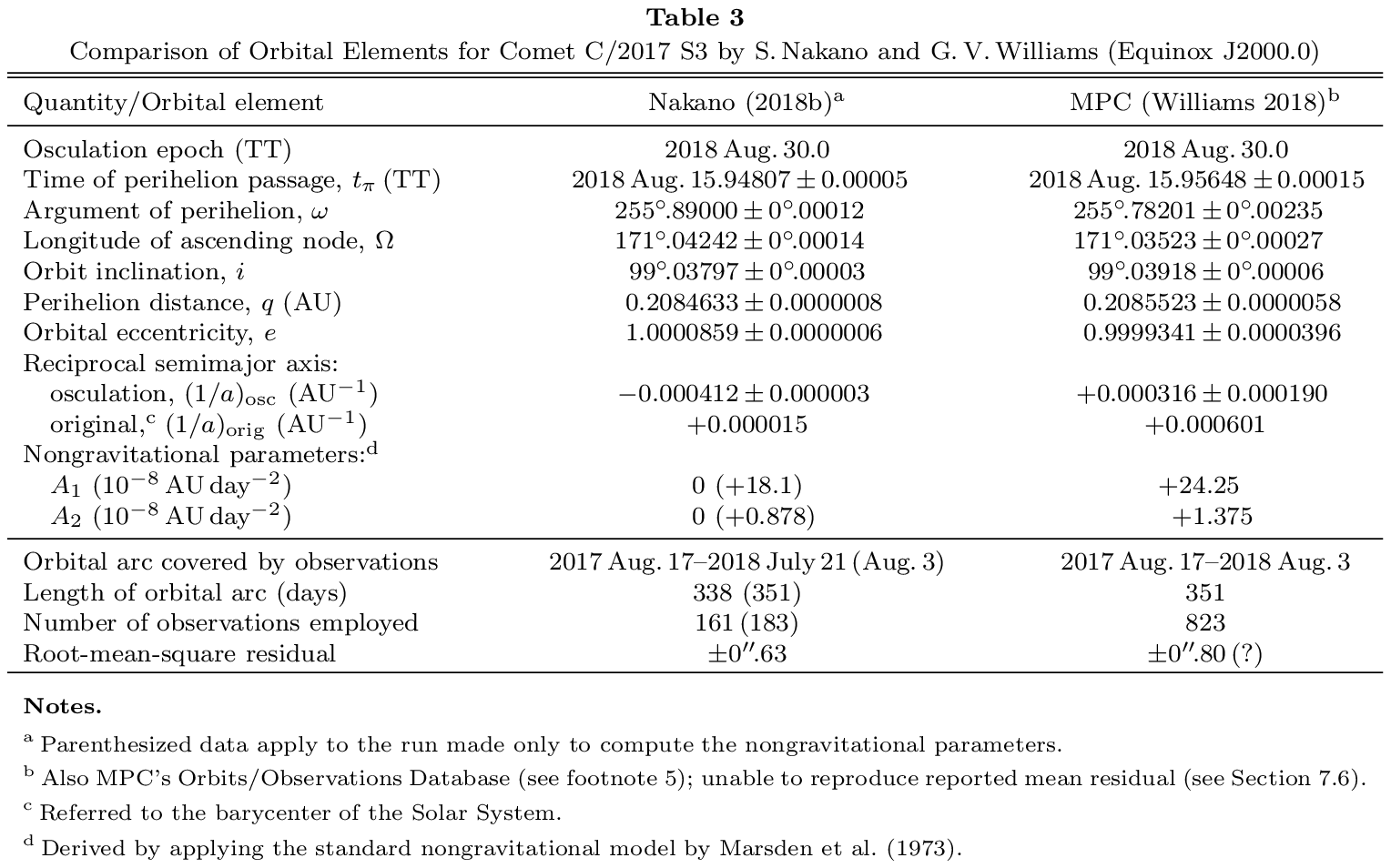}}}
\vspace{-15.31cm}
\end{table*}

\subsection{Current Status of Comet's Orbit Investigation}
Thanks to the Pan-STARRS pre-discovery images, the orbital arc covered by the
ground-based observations was extended to 351 days, from 2017 August 17 to 2018
August 3. To our knowledge, two independent orbital solutions are available at
the time of this writing that link positions from this entire period of time:\
one by Nakano (2018b) and the other, also referred to as the MPC orbit, by
Williams (2018).  They are compared in Table~3.

Nakano completed his computations shortly before a massive amount of astrometric
data was released by the MPC on August 23, which, with several additional data
issued on September~21,{\vspace{-0.04cm}}\footnote{See MPEC 2018-Q62 and MPEC
2018-S50, respectively.{\vspace{-0.08cm}}} brought their total to a very
respectable number of 1034, but did not extend the covered orbital arc.  Nakano's
gravitational solution provides a useful starting point for a project aimed
at a definitive orbit determination.  He found that until 2018~July~21 the 161
astrometric positions used could be fitted with a mean residual of
$\pm$0$^{\prime\prime\!}$.63, but that all observations made after July~21 ---
specifically between July~23 and August~3 --- deviated from the gravitational
orbit systematically and increasingly with time, with the residuals of up to
30$^{\prime\prime}$, negative in right ascension and positive in declination.
Thus, whatever remained of the comet's nucleus after Outburst~II, it was located
to the northwest of the expected position.  To describe the magnitude of the
anomalous residuals, Nakano provided a second, nongravitational solution, in
which the observations from the period of July~23--August~3 were included and
which resulted in the radial component of the nongravitational acceleration at
1~AU from the Sun amounting to \mbox{$+18.1 \times \!10^{-8}$\,AU day$^{-2}$},
comparable to the effect in the motion of comet C/1993~A1 Mueller (Nakano 1994)
and equaling 0.06 percent of the Sun's gravitational acceleration.  Nakano also
detected a much smaller transverse component of the force (see Table~3).

An insight into the quality of Nakano's computations is facilitated because he
offers a table of residuals for all the observations that he collected, both
employed in the solution and rejected ones.  The table demonstrates that the
temporal distribution of residuals up to July~21 was generally satisfactory,
rarely with greater than sub-arcsec systematic trends detectable over fairly
short periods of time.  For example, on July~14--18 all residuals were positive,
up to 2$^{\prime\prime}$, in right ascension and negative (and lower) in
declination.  Similarly, all residuals between 2017 December 13 and 2018 March~24
were negative and up to more than 1$^{\prime\prime\!}$.5 in right ascension.

We note that Nakano's gravitational solution included observations made as late
as one week after the onset of Outburst II, and it is unclear whether the minor
systematic trends in the residuals were a corollary of this late cutoff.
Nonetheless, his orbit shows that C/2017~S3 was clearly a dynamically new comet,
arriving from the Oort Cloud, as shown in Table~3.

By contrast, Williams presented a solution that linked 823 observations from the
entire 351-days long orbital arc.  He applied the nongravitational terms and
obtained the radial- and transverse-components' parameters that were a little
higher than the nongravitational parameters obtained by Nakano when he
incorporated the post-July~21 observations.  As expected, the orbital
elements by Williams differ from Nakano's set rather significantly, much more
than the mean errors suggest.  In particular, the MPC orbit implies that
the comet did {\it not\/} arrive~from the Oort Cloud!

In his presentation, Williams provides no information on the distribution of the
residuals, so it is not possible to examine the quality of fit, including the
presence of long-term systematic trends.  However, the mean residual significantly
higher than Nakano's is worrisome.  We return to this issue in Section~7.6.

\subsection{Strategy and Methodology of the Present\\Orbital Investigation}
The enormous, systematic positional residuals that Nakano (2018b) obtained from
all the observations made in late July and early August represent independent
evidence that after Outburst~II the comet's nucleus was in shambles and that
powerful nongravitational forces were at work.  Since positional offsets are
the second integral of nongravitational perturbations, an inertia causes a delay
before the latter show up to a degree that the scientist computing the orbit
can no longer tolerate.  Nakano unquestionably experimented with the orbital
fit before he selected July~21 as the limit for the astrometric positions with
still acceptable residuals from his orbital solution.  However, the
date of July~21 is not associated with any milestone in the comet's physical
behavior.  The light curves in Figures~2 and 3 show that both the nuclear and
total intrinsic brightness were on this date already in rather steep decline.
Potentially correlated with this brightness behavior was a nongravitational
acceleration, of \mbox{$\sim \!\! 18 \times \! 10^{-8}$\,AU day$^{-2}$} at
1~AU from the Sun, which made the center of the debris cloud move, by
July~21, some 900~km away from the expected position of the nucleus, which
corresponded to an angular deviation of about 1$^{\prime\prime\!}$.1 after
accounting for the projection foreshortening.  This is about a half of the
maximum {\vspace{-0.04cm}}residual allowed by Nakano (2018b) in either
coordinate,\footnote{In his investigation of C/2017~S3, Nakano accepted in
the orbital solution an astrometric position that left a residual as high as
2$^{\prime\prime\!}$.1 in one coordinate, but he rejected a position that
left a residual of 2$^{\prime\prime\!}$.5.{\vspace{-0.25cm}}} nearing his
rejection cutoff.

This outline of Nakano's work leads us to a conclusion that the {\small
\bf proper procedure for computing an orbit~that is unaffected by perturbations
of the comet's motion exerted in the course of an outburst is by employing
only astrometric observations that were made before the outburst had begun}.
Accordingly, we focus in the following on two classes of orbital solution:

(i) {\small \bf Orbits A}, derived by linking accurate observations made between
2017 August~17 and 2018 June~30.2~UT, the onset of Outburst~I, thus eliminating
any effects on the comet's motion by Outbursts~I and II; and

(ii) {\small \bf Orbits B}, derived by linking accurate observations made between
2017 August~17 and 2018 July~14.4~UT, the onset of Outburst~II, thus eliminating
any effects~by Outburst~II.  Comparison with Orbits A isolates and measures the
effects by Outburst~I.

Nakano (2018b) listed 22 observations from the time span of July~23 through
August~3.  At present the number is more than seven times as large.  There are
also well over 300 observations from the time between July~14.4 and July~23.  Given
that the fragmented nucleus consisted of essentially inert, refractory material,
the strong trends in the residuals from the gravitational orbit can be used to
provide important information on its properties.  The premise of a dominant size
among the fragments --- like in the model formulated in Section~5 --- allows one
to {\small \bf treat the residuals as offsets} of the center of the debris cloud
from the positions that the nucleus would occupy if it did not disintegrate.
Mathematically, we deal with a problem equivalent to that of the relative motion
of a companion fragment departing from the primary fragment of a split comet
(Sekanina 1977, 1982).  In the absence of activity, the nongravitational force
acting on the debris is identified as solar radiation pressure, which requires
that the acceleration vary as an inverse square of heliocentric distance.  On
the other hand, the isotropic expansion of the debris cloud implies a zero
impulse at fragmentation, in which case the solution to the problem has only two
parameters.  One is the nucleus'~fragmentation time, $t_{\rm frg}$ (equivalent
to the companion's separation time); the other is the fragments' deceleration
(i.e., acceleration in the antisolar direction) normalized to 1~AU from the
Sun, $\gamma$.  As a measure of solar radiation pressure, this deceleration,
expressed in units of the Sun's gravitational acceleration at 1~AU (equal to
\mbox{$2.96 \times \!10^{-4}$\,AU day$^{-2}$}), is related to the mean diameter,
$D_{\rm frg}$, of the fragments in the cloud (in cm) by
\begin{equation}
D_{\rm frg} = \frac{1.148 \, Q_{\rm pr}}{\gamma \delta} \! \times \!10^{-4},
\end{equation}
where $\delta$ is the bulk density of the fragments (in g\,cm$^{-3}$) and
$Q_{\rm pr}$ is a dimensionless efficiency factor for radiation pressure,
which is close to unity for all fragments larger than several microns in
diameter.

The methodology of orbital analysis of C/2017~S3 was motivated by the goals
of this investigation, primarily the understanding of the comet's fragmented
nucleus.  An {\it EXORB8\/} orbit-determination code, written and updated
by A.~Vitagliano, was employed by the second author to carry out the
computations.  The code integrates the comet's orbital motion using a
variable step and accounts for the perturbations by the eight planets,
by Pluto, and by the three most massive asteroids, as well as for the
relativistic effect.  The nongravitational terms are directly incorporated
into the equations of motion, following the standard Style II model by
Marsden et al.\ (1973); modified nongravitational solutions with an arbitrary
scaling distance $r_0$ (e.g., Sekanina \& Kracht 2015) are readily accommodated
(Section~7.5).  The orbital elements are computed by applying a least-squares
differential-correction optimization procedure.  The standard JPL DE421 ephemeris
is used and the precision of our computations is 17 decimal places.

An early task was to examine Nakano's (2018b) finding on the absence of a
nongravitational acceleration until days after Outburst~II.  Given that this was
an intrinsically faint comet (Section 4.2), with a presumably small nucleus,
we felt that his result was rather surprising.

Our work on this and related problems proceeded in four steps:\ we started
by examining the comet's orbital motion in the pre-outburst period of time
(i.e., Orbits~A), which terminated three weeks before the cutoff date that Nakano
chose for his gravitational solution.  Subsequently, in an effort to isolate
potential orbital effects by Outburst~I, we investigated the orbital motion in
an extended period of time (Orbits~B).

Next, we addressed the issue of feasibility to accommodate all observations
into one solution, examined the resulting distribution of residuals, and
compared it with the distributions derived in the first two steps.  Finally,
we focused on a simulation of the orbital motion of the fragmented nucleus,
attempting to match the distribution of residuals left by the cloud of debris
in terms of an effect by solar radiation pressure.  This task was accomplished,
as explained above, by applying a standard model for split comets.

\begin{table*}
\vspace{-4.2cm}
\hspace{-0.5cm}
\centerline{
\scalebox{1}{
\includegraphics{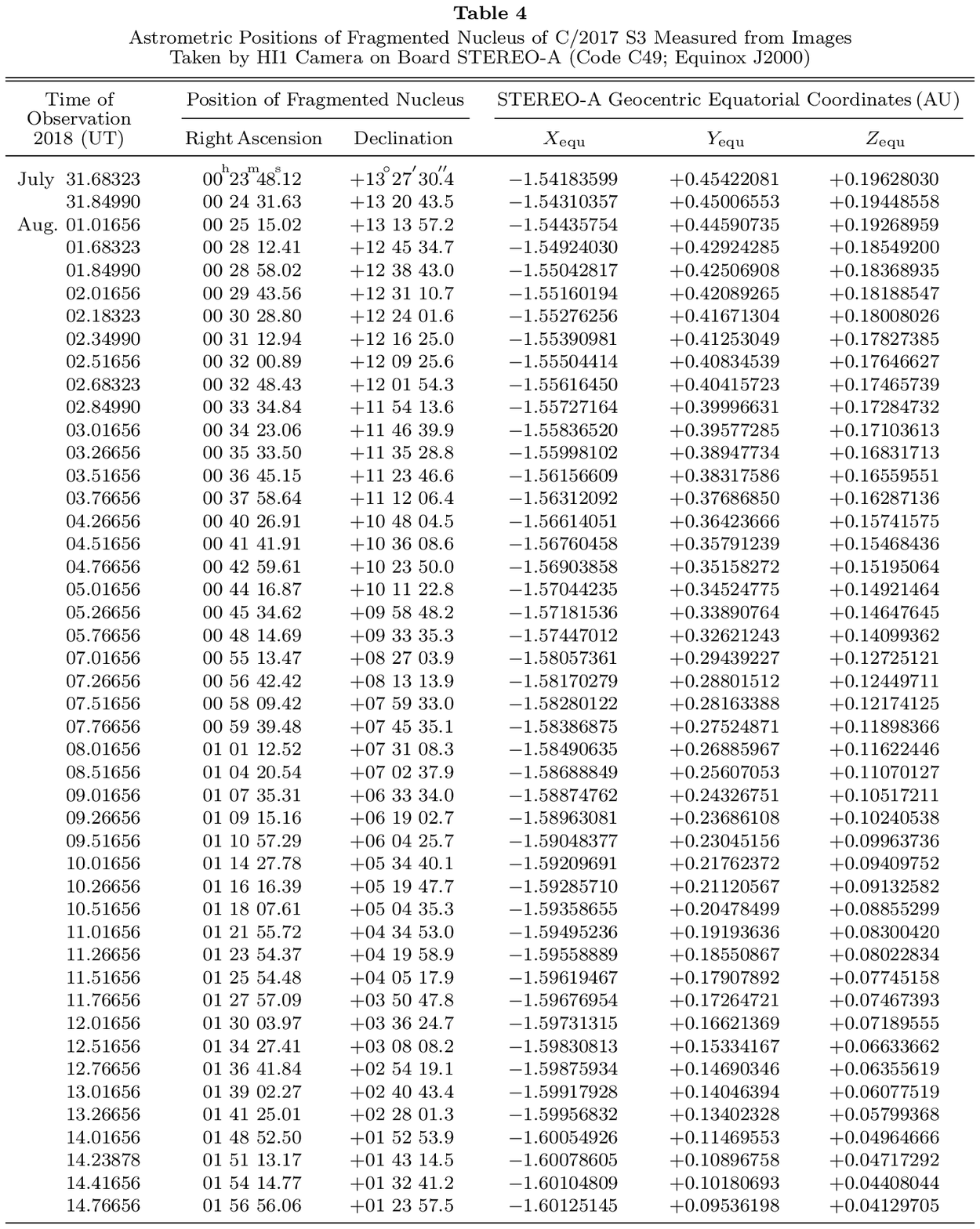}}}
\vspace{-6.42cm}
\end{table*}

The orbital solutions presented in Section 7.3 and~beyond were derived using the
set of 1034 \mbox{ground-based} observations available from the MPC (see footnote
5).  The data's merit was extensively tested, as described in the Appendix.
Additional astrometric data were obtained by the second author, who measured
46~images of the comet taken by the HI1 camera~on~board~\mbox{STEREO-A}.  Listed
in Table~4, these positions have limited accuracy on account of the detector's
large pixel size.

\begin{figure*}[t]
\vspace{-1.68cm}
\hspace{-0.1cm}
\centerline{
\scalebox{0.8}{
\includegraphics{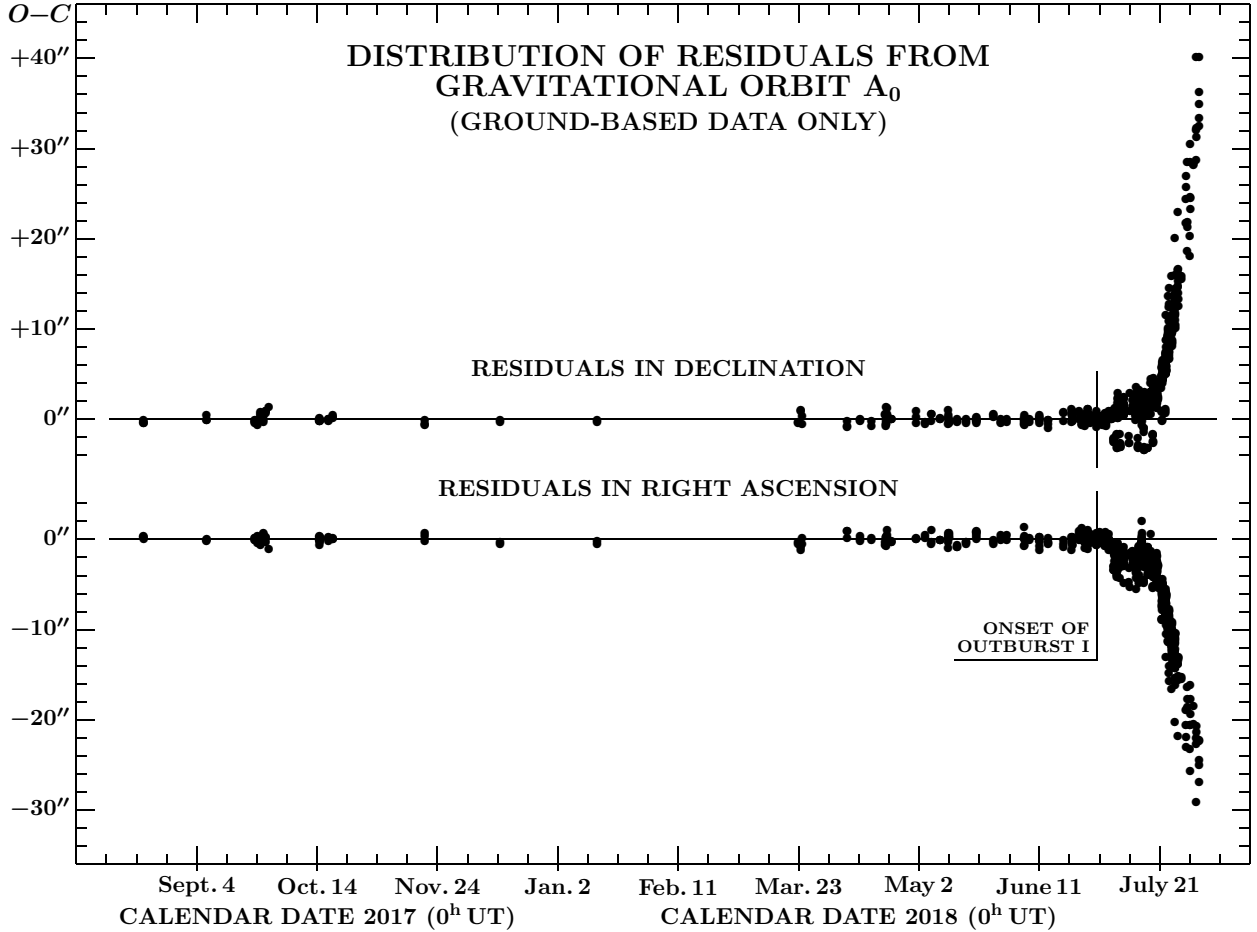}}}
\vspace{-10cm}
\caption{Distribution of residuals $O\!-\!C$ from the gravitational Orbit A$_0$
left by 219 ground-based observations from between 2017 August 17 and 2018
June~30.2 UT that were used in the solution, and by 807 ones from 2018 June~30.2
through August 3.1 UT that were not used.{\vspace{0.5cm}}}
\end{figure*}
\subsection{Orbits A: Solutions Terminating at the\\Onset of Outburst~I}

\vspace{0.15cm}
We began with 227 ground-based observations available for the Orbits~A class
of solutions.  Because of a high quality of an overwhelming majority of the
data, we used only those leaving in either coordinate a residual not exceeding
$\pm$1$^{\prime\prime\!}$.5 (see the Appendix).

\begin{table}[b]
\vspace{-3.65cm}
\hspace{4.2cm}
\centerline{
\scalebox{1}{
\includegraphics{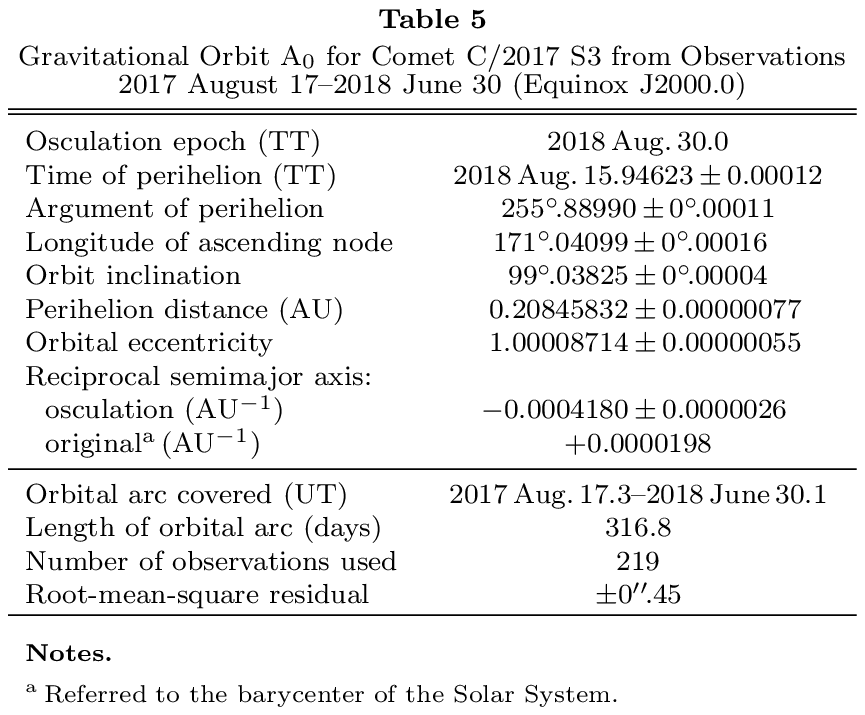}}}
\vspace{-18.95cm}
\end{table}

We first fitted a gravitational solution, which is from now on referred to
as Orbit A$_0$.  It turned out that only 8 observations, less than 4~percent
of the total, failed to satisfy the strict rejection cutoff.  The 219 data points
covered an orbital arc of 317~days and, as is illustrated by the distribution
of residuals in Figure~8, the fit --- within the limits of the orbital arc
used in the computation and terminating at the onset of Outburst~I --- appears
to be perfect in either coordinate, with the mean residual amounting to
$\pm$0$^{\prime\prime\!}$.45.  However, within a few days of the termination
date, the residuals begin to exhibits systematic trends, which are particularly
strong in right ascension.  The effect is displayed prominently in Figure~9,
a close-up of Figure~8 for the period of 45~days, from June~19 through August~3,
which includes 883 ground-based observations.  Yet, for two weeks after the
onset of Outburst~I, the systematic residuals did not exceed several arcsec in
either coordinate, until the onset of Outburst~II, at which time the disparity
exploded exponentially, reaching 40$^{\prime\prime}$ in declination by August~3
and being confirmed by the residuals from the STEREO-A~astrometry presented in
Figure~10.

\begin{figure}[t]
\vspace{-1.7cm}
\hspace{3.1cm}
\centerline{
\scalebox{0.82}{
\includegraphics{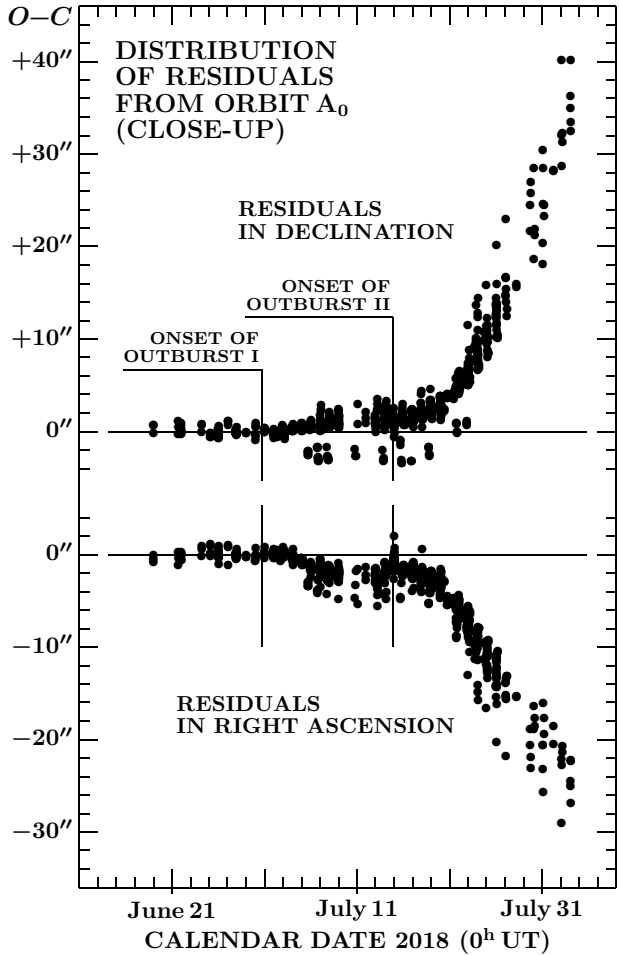}}}
\vspace{-10.27cm}
\caption{Close-up, from Figure~8, of the distribution of residuals \mbox{$O\!-\!C$}
from Orbit~A$_0$, left by 883 ground-based observations from a period of 45 days,
between June~19 and August~3.  Note that the 76 observations between June 19
and June 30 that were used in the solution show no systematic trends in the
residuals.{\vspace{0.6cm}}}
\end{figure}

As a means of further testing the quality of Orbit~A$_0$, we also computed a
standard Style II nongravitational solution (see Section~7.2 for a reference)
that rested on the same 219 ground-based observations.  This solution, referred
to as Orbit~A$_1$, provided, in addition to the orbital elements, the parameter
$A_1$ of the radial component of the nongravitational acceleration.  These orbital
elements and the associated mean residual were practically identical with those
for Orbit~A$_0$ and the residuals never differed by more than a few hundredths of
an arcsec.  For the nongrvitational parameter we obtained \mbox{$A_1 \!=\! (+0.9
\pm 1.9) \! \times \! 10^{-8}$\,AU day$^{-2}$}, thus confirming that {\small \bf
the comet's motion between 2017 August 17 and~2018 June 30 was unaffected by
nongravitational forces (of measurable magnitude) and is adequately described by
Orbit A{\boldmath $_0$}} presented in Table~5.  Orbit A$_0$ also leaves no doubt
that the comet has indeed arrived from the Oort Cloud.

\subsection{Orbits B:\ Solutions Terminating at the\\Onset of Outburst~II}
The moderate systematic trends in the residuals from Orbit A$_0$ over the
extrapolated interval of time between the onset of Outburst~I and the onset
of Outburst~II, clearly seen in Figure~9, suggest that Outburst~I may have
detectably affected the comet's orbital motion.  To gain a greater insight
into this problem, we derived new solutions by linking the nearly
300 ground-based observations made between 2018 June 30.2 and July 14.4 UT
with the observations used to compute Orbit A$_0$.  These class B solutions
should be sensitive to potential effects by Outburst~I but not Outburst~II.

We started with a gravitational solution, referred to as Orbit~B$_0$, that
linked 443~ground-based observations; 70, or nearly 14~percent of a total of
513~observations available, were removed because their residuals exceeded the
1$^{\prime\prime\!}$.5~rejection cutoff in at least one coordinate (see the
Appendix).  The quality of the distribution of residuals left by the observations
made before mid-June 2018 was as high as that from Orbit A$_0$.  Figure~11
displays the residuals left by the observations between June~19 and July~14.4~UT
as well as by the ignored observations made following the onset of Outburst~II.
The fit before~July~14 is~not~quite perfect, but it is better in declination.
The value of Orbit~B$_0$, whose elements are in Table~6, is that its residuals
provide us with a fairly authentic record of the comet's genuine orbital motion
with respect to the hypothetical, purely-gravitational motion of the original,
intact nucleus at the times after Outburst~II had begun.

\begin{figure}[t]
\vspace{-1.7cm}
\hspace{3.1cm}
\centerline{
\scalebox{0.82}{
\includegraphics{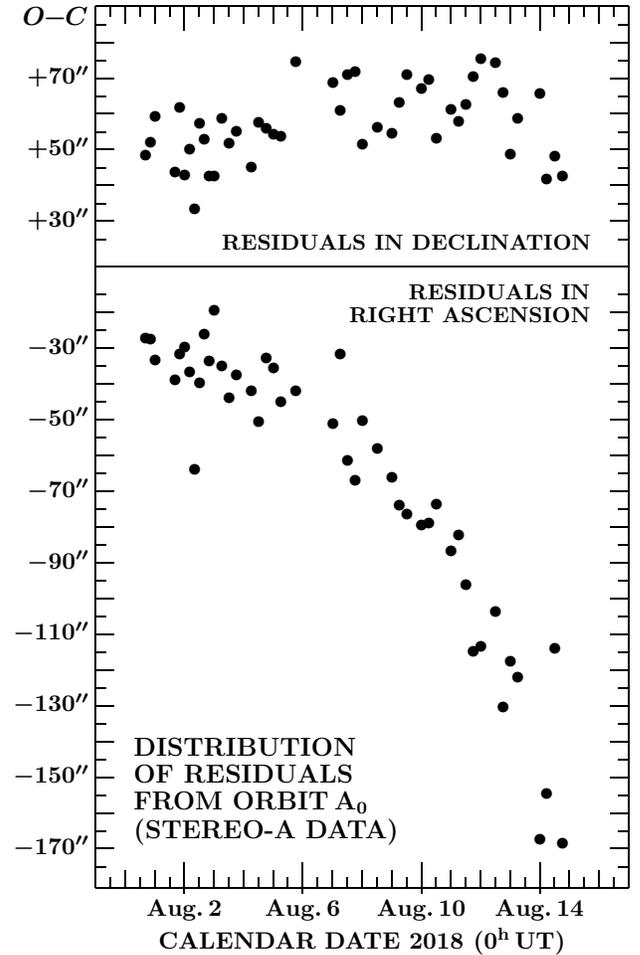}}}
\vspace{-10.3cm}
\caption{Distribution of residuals \mbox{$O\!-\!C$} from Orbit A$_0$ left by 46
astrometric positions from July~31--August 14, measured from the STEREO-A images
and listed in Table~4.{\vspace{0.5cm}}}
\end{figure}

Before getting to the next phase of orbital analysis,~we make two comments.
One, an important aspect of Figure~11 is the absence of any progressive
deviation from the gravitational motion until about July~20.  This date is
nearly identical with the end point of Nakano's (2018b) gravitational solution.
A minor trend in the distribution of residuals in right ascension, starting
already in late June, offers by itself enough evidence for a slight~nudge
to the nucleus, which originated with Outburst~I.  Granting that Nakano's
result may still stand as a fair approximation, we will return to this point
in Section~8.

Two, comparison of Figures~9 and 11 suggests that the distribution of residuals
left by the post-Outburst~II ground-based observations does not depend strongly
on which of the two gravitational orbits, A$_0$ or B$_0$, was~used to fit the
pre-Outburst~II observations.  The same likewise applies to the STEREO-A data.
Similarly, the mean residual of the fit by Orbit~B$_0$ to the observations from
2017 August 17 through 2018 July 14, which amounts to $\pm$0$^{\prime\prime\!}$.53
(Table~6), is only moderately higher than the mean residual of
$\pm$0$^{\prime\prime\!}$.45 (Table~5) describing the fit by Orbit~A$_0$ to the
shorter arc.  Yet, in order to improve the solution over the orbital arc ending
with the onset~of Outburst~II, the introduction of a nongravitational acceleration
became desirable.
\begin{figure}[t]
\vspace{-3.1cm}
\hspace{3.15cm}
\centerline{
\scalebox{0.82}{
\includegraphics{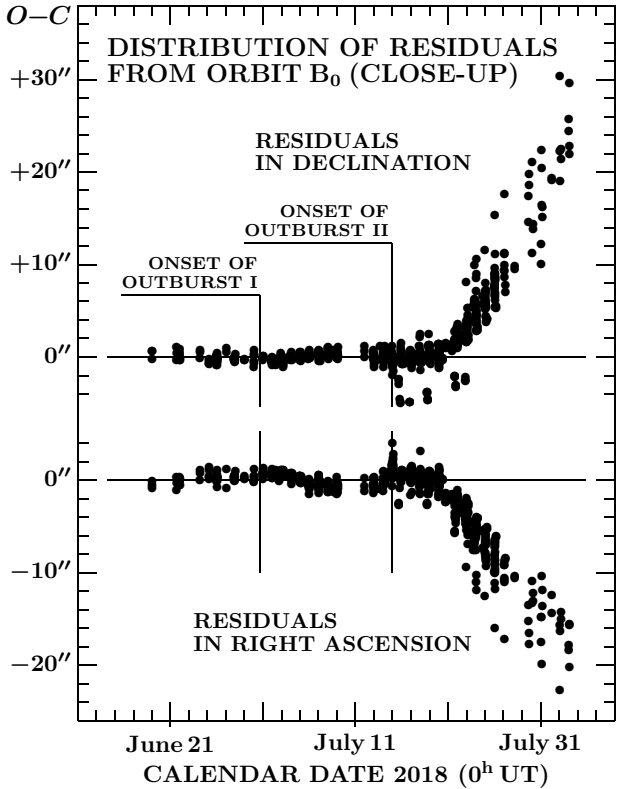}}}
\vspace{-11.1cm}
\caption{Distribution of residuals \mbox{$O \!-\! C$} from Orbit~B$_0$ left by
824 ground-based observations from a time span of June~19 to August~3.  Note
that the peak residuals from the early August days are smaller than the same
residuals from Orbit~A$_0$ in Figure~9.{\vspace{0.55cm}}}
\end{figure}

We began by including the radial component of the nongravitational acceleration
into the equations of motion of the standard Style~II model (Section~7.2) in order
to determine, besides the orbital elements, the parameter $A_1$.  Referred to as
Orbit~B$_1$, this solution was based on 454 ground-based observations, thus
allowing us to incorporate 11 additional observations that satisfied the rejection
threshold of 1$^{\prime\prime\!}$.5.  At the same time, the mean residual was
brought down to $\pm$0$^{\prime\prime\!}$.50.  Unlike in the case of Orbit~A$_1$
(Section~7.3), {\vspace{-0.02cm}}the nongravitational parameter was now well
defined, \mbox{$A_1 \!=\! (+10.68 \pm 0.65) \!\times\! 10^{-8}$\,AU day$^{-2}$},
with a signal-to-noise ratio exceeding 16. The observations before June 19 were
fitted by Orbit~B$_1$ equally well as by Orbit~B$_0$, so there is no need to plot
this early part of the distribution of residuals.  A close-up of the critical
period of time, arranged in the same fashion as Figure~11 for Orbit~B$_0$, is
displayed in Figure~12. 

Comparison of the two figures suggests that the improvement in the quality
of fit between Orbits~B$_0$ and B$_1$ in the period of time between
June~19 and July~14.4~UT is at best marginal; at the beginning of Outburst~II
it is in fact Orbit B$_0$ that provides a somewhat better fit, especially in
right ascension.  This conclusion implies that the search for an improved
orbital~solution is to continue.
\begin{figure}[t]
\vspace{-3.33cm}
\hspace{3.15cm}
\centerline{
\scalebox{0.82}{
\includegraphics{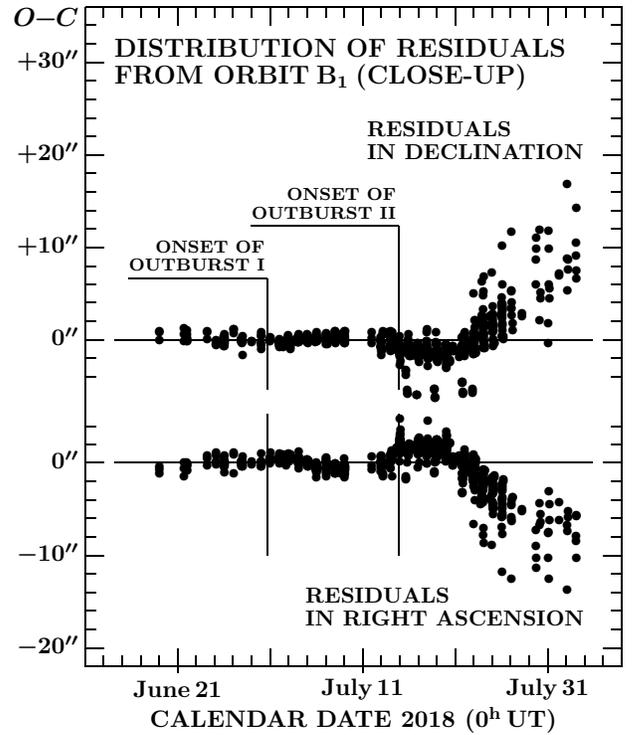}}}
\vspace{-11.57cm}
\caption{Distribution of residuals \mbox{$O \!-\! C$} from Orbit~B$_1$, left~by
832 ground-based observations from a period of June 19--August~3.  Note that
{\vspace{-0.01cm}}the peak residuals from the early August days are nearly
$\frac{1}{2}$ the magnitude of the same residuals from Orbit~B$_0$ in
Figure~11.{\vspace{0.58cm}}}
\end{figure}

Although we looked skeptically at the chance that the incorporation of a
transverse component (or, for that matter, a normal component) of the
nongravitational acceleration could appreciably improve the fit, we tested
this option briefly by computing Orbit~B$_2$, a standard Style II
nongravitational solution with the parameter $A_2$ of the transverse
component added to $A_1$.
%
%
Orbit B$_2$ did further reduce the mean residual, but only insignificantly, to
$\pm$0$^{\prime\prime\!}$.49, and the signal-to-noise of the $A_2$ determination
amounted to only about 3; we consider the improvement over the quality of
Orbit~B$_1$ marginal and unimpressive, given that an additional parameter was
solved for.  Hence, our effort to find a refined fit to the astrometric
observations made at the time between the two outbursts should take yet another
turn.

\begin{table}[b]
\vspace{-3.43cm}
\hspace{4.22cm}
\centerline{
\scalebox{1}{
\includegraphics{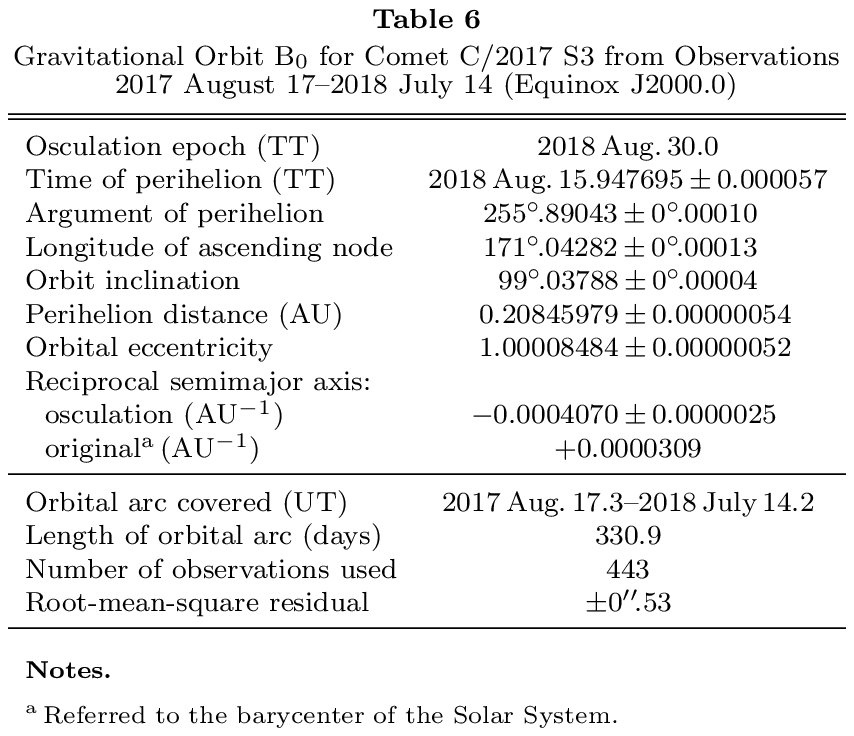}}}
\vspace{-18.68cm}
\end{table}
\begin{table*}[t]
\vspace{-4.2cm}
\hspace{-0.53cm}
\centerline{
\scalebox{1}{
\includegraphics{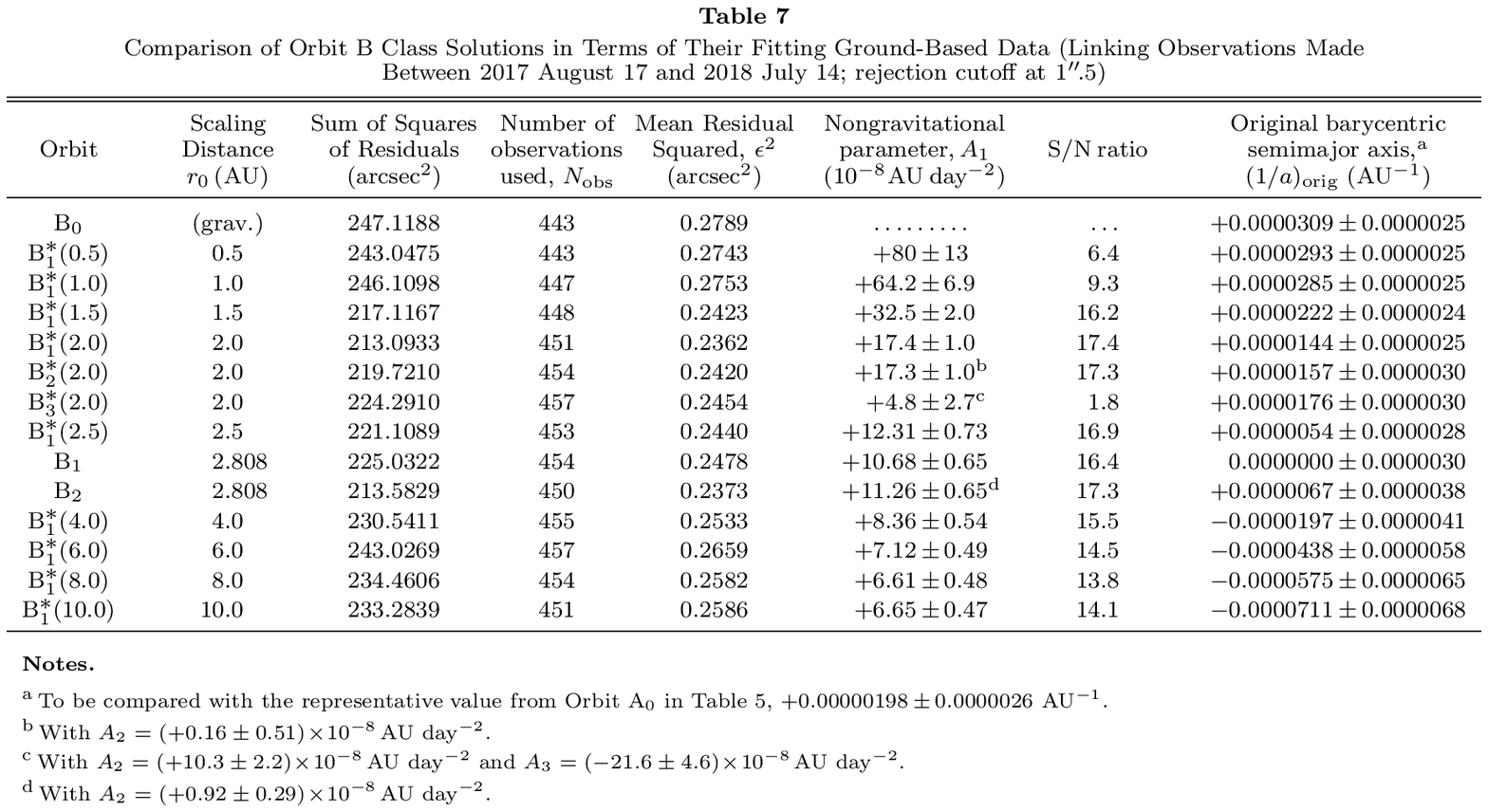}}}
\vspace{-15.02cm}
\end{table*}

\subsection{Orbits B Subclass:$\!$ Modified Nongravitational Laws}
The standard Style II nongravitational model of Marsden et al.\ (1973), mentioned
in Section 7.2, is described by a scaling distance of \mbox{$r_0 = 2.808$ AU}.  
This model~was over the past four decades tested extensively on a large sample
of comets, short-period ones in particular, and found to work satisfactorily in
the majority of cases.  The standard nongravitational law mimicks the outgassing
curve of water ice sublimating from a comet's spherical isothermal nucleus.

More recently, however, a broad variety of alternative, {\small \bf modified
nongravitational laws} proved more successful than the standard model in
solving some specific problems associated with cometary motions.  A good
example is a work by the present authors (Sekanina \& Kracht 2015) on the
strong erosion-driven nongravitational acceleration experienced by the Kreutz
sungrazing system's dwarf comets at extremely small heliocentric distances.
A remarkable property of the empirical nongravitational law employed is that
it does not vary with a heliocentric distance $r$, but with a ratio of $r/r_0$,
in which the scaling distance $r_0$ is in principle a measure of the latent heat
of sublimation, ${\cal L}$, of the ice that dominates the comet's outgassing
activity.  Since \mbox{$r_0 \!\sim\! {\cal L}^{-2}$}, highly refractory
material with excessive values of the sublimation heat, which sublimates only
near the Sun, requires extremely low scaling distances.  In our study of the
dwarf Kreutz comets we derived scaling distances as low as $\sim$0.01~AU; this
is consistent with the sublimation of silicates, as for example for forsterite
we found \mbox{$r_0 = 0.015$ AU}~(\mbox{Sekanina} \& Kracht 2015).  At the other
extreme, for highly volatile ices, whose sublimation heat ${\cal L}$ is very low,
the scaling distance \mbox{$r_0 \gg 2.8$ AU} and the nongravitational
acceleration varies essentially as $r^{-2}$, as recently found by{\nopagebreak}
Micheli{\nopagebreak} et al.\ (2018) for 1I/2017~U1 ('Oumuamua).

Being unsure of an appropriate scaling distance for the comet C/2017~S3, we ran
a number of orbital solutions~in a broad range of $r_0$, from 0.5~AU to 10~AU.
The quality of fit to a set of $N_{\rm obs}$ ground-based observations used in a
given solution is measured by the mean residual squared, $\epsilon_{\rm rms}^2$,
expressed as
\begin{equation}
\epsilon_{\rm rms}^2 = (2 N_{\rm obs})^{-1} \sum_{i=1}^{N_{\rm obs}} \left[
 (O\!-\!C)_{\rm RA}^2 + (O\!-\!C)_{\rm Decl}^2 \right]_i,
\end{equation}
where \mbox{$(O \!-\! C)_{\rm RA}$} and \mbox{$(O \!-\! C)_{\rm Decl}$} are the
respective residuals in right ascension and declination left by each individual
observation.  In line with the designation introduced in Section 7.4 for the
Orbits~B class standard Style II nongravitational solutions, we use an index
\mbox{$k \!=\! 1$} for a modified solution that provides the radial-component
parameter $A_1$; \mbox{$k \!=\! 2$} for a modified solution that, next~to $A_1$,
also provides the transverse-component parameter~$A_2$; and extend the designation
to \mbox{$k \!=\!3$} for a modified solution that provides, in addition, the
normal-component parameter $A_3$.  The modified solutions, now referred to as
Orbits~B$_k^{\textstyle \ast}(r_0)$, are compared in Table~7 with the class B
gravitational solution, B$_0$, and the two class B standard nongravitational
solutions.  All the tabulated orbits offer a good match to the observations
made before June~19.{\hspace{0.3cm}}

The best solution, B$_1^{\textstyle \ast}$(2.0), indicates that the scaling
distance is very close to 2~AU and confirms that the radial component of the
nongravitational acceleration accounts for the entire effect.  Inclusion of the
transverse component, in Orbit B$_2^{\textstyle \ast}$(2.0), shows its
contribution to be essentially zero.  Inclusion of the transverse and normal
components, in Orbit B$_3^{\textstyle \ast}$(2.0), leads to a spurious solution
because the radial component becomes virtually indeterminate.  Also, the match
to the observations by both B$_2^{\textstyle \ast}$(2.0) and B$_3^{\textstyle
\ast}$(2.0) is worse than by B$_1^{\textstyle \ast}$(2.0), in spite of the
greater number of free parameters.  All tested solutions with a scaling distance
greater than that of the standard model show a less satisfactory match to the
observations and, in contrast to the solutions with a scaling distance near 2~AU,
imply an unlikely interstellar origin of C/2017~S3.

\begin{figure}[t]
\vspace{-4.05cm}
\hspace{3.15cm}
\centerline{
\scalebox{0.82}{
\includegraphics{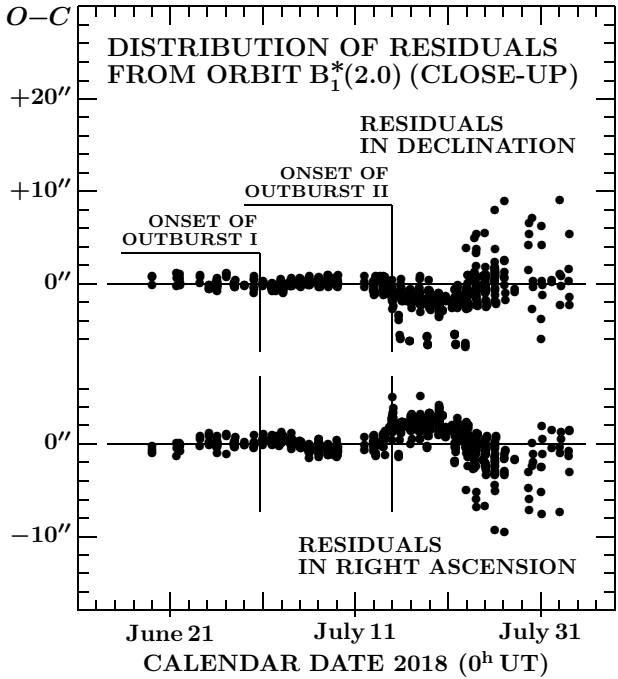}}}
\vspace{-11.6cm}
\caption{Distribution of residuals \mbox{$O \!-\! C$} from Orbit B$_1^{\textstyle
\ast}$(2.0) left by 829 ground-based observartions made between June 19 and
July 14.4~UT.  Unlike in case of Orbit B$_1$ (Figure~12), all observations made
after July 14, not used in the solution, leave residuals
$<$10$^{\prime\prime}$.{\vspace{0.45cm}}}
\end{figure}

Orbit B$_1^{\textstyle \ast}$(2.0) has another, rather remarkable property,
which is demonstrated in Figure~13.  Although we were fitting only the
observations made {\it prior to\/} July 14.4 UT, {\it all\/} ground-based
observations made {\it after\/} this time leave the residuals that are less
than 10$^{\prime\prime}$ in either coordinate, and many of them are much
smaller.  However, the figure shows the presence in the post-Outburst~II
period of time of prominent systematic trends in either coordinate with
amplitudes of $>$5$^{\prime\prime}$ and, in right ascension at least, with
the hint of a surprisingly short period on the order of perhaps three weeks
or so. In addition, Figure~14 shows that Orbit B$_1^{\textstyle \ast}$(2.0)
fails to fit the positions of the fragmented nucleus derived from the STEREO-A
images (Table~4), even though their residuals are smaller than from other
solutions (see Figure~10 for comparison).

\begin{figure}[t]
\vspace{-1.67cm}
\hspace{3.15cm}
\centerline{
\scalebox{0.82}{
\includegraphics{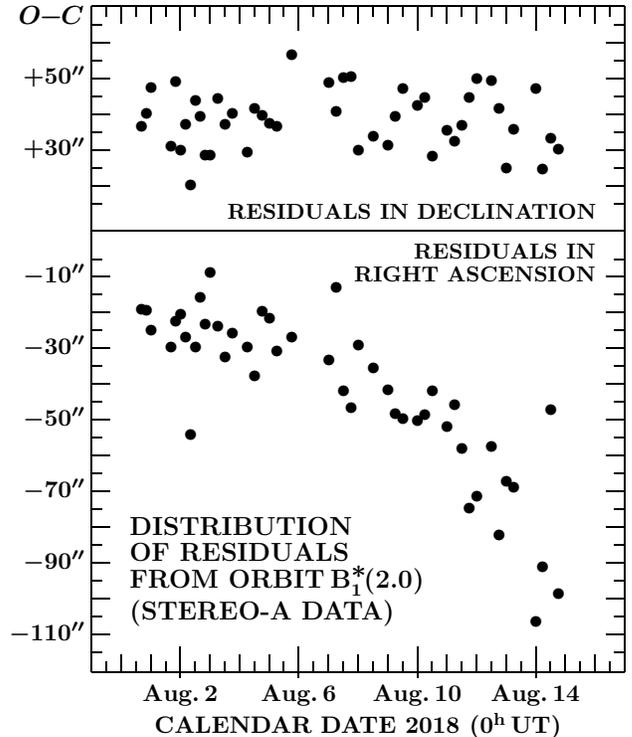}}}
\vspace{-13.2cm}
\caption{Distribution of residuals \mbox{$O \!-\! C$} from Orbit B$_1^{\textstyle
\ast}$(2.0) left by 46 astrometric positions from July 31--August 14, obtained
by measuring the STEREO-A images.{\vspace{0.45cm}}}
\end{figure}

We conclude this section by stating that Orbit~B$_1^{\textstyle \ast}$(2.0),
presented in Table~8 and incorporating a modified nongravitational law with a
scaling distance of 2.0~AU, is helpful in that it provides a fair (but by no
means fully satisfactory) fit to the ground-based observations made between
2017 August~17 and 2018 July~14.  Moreover, when extrapolated, it unexpectedly
well approximates the positions of the fragmented nucleus over the extended
period of time between mid-July and August~3 (when the ground-based observations
terminated), but it does not fit the positions determined from the STEREO-A
images.  The presence of minor nonrandom trends in the residuals as early as
the beginning of July leaves no doubt that the orbital motion of the comet was
indeed affected by Outburst~I, but only insignificantly.  Comparison of Orbits
B$_0$, B$_1$, and B$_1^{\textstyle \ast}$(2.0) in terms of the distribution
of residuals left by the observations made prior to July 14.4~UT shows that
they are in fact quite similar, the major differences taking place only among
the extrapolated residuals in late July and early August.

\begin{table}[b]
\vspace{-3.55cm}
\hspace{4.22cm}
\centerline{
\scalebox{1}{
\includegraphics{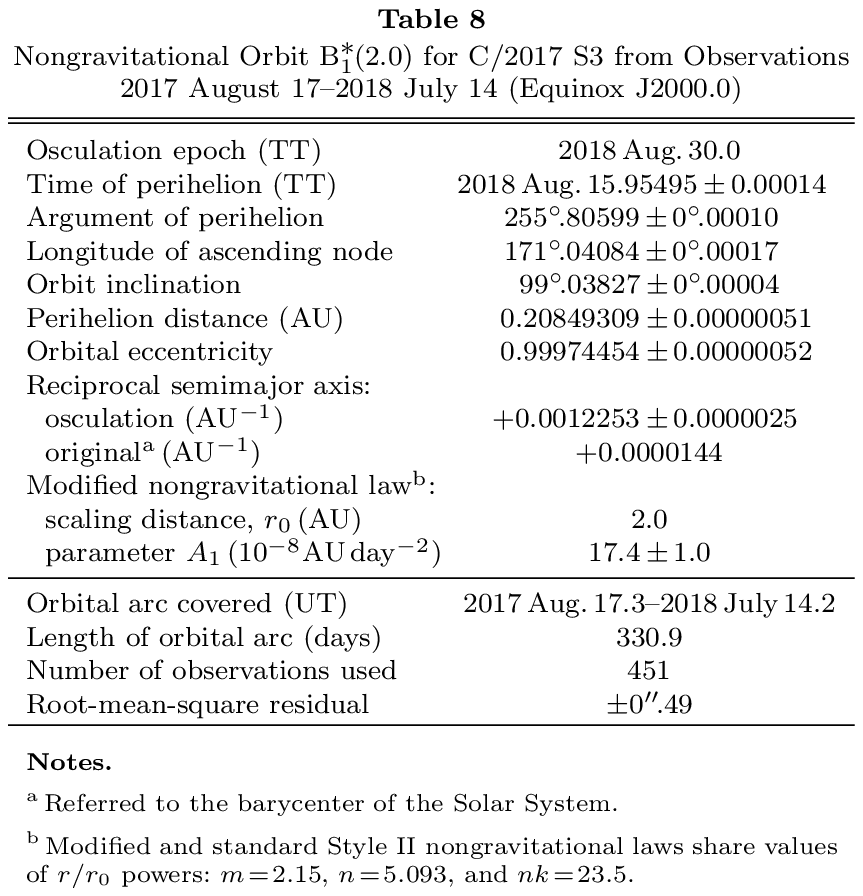}}}
\vspace{-17.06cm}
\end{table}
\begin{table*}[t]
\vspace{-4.2cm}
\hspace{-0.53cm}
\centerline{
\scalebox{1}{
\includegraphics{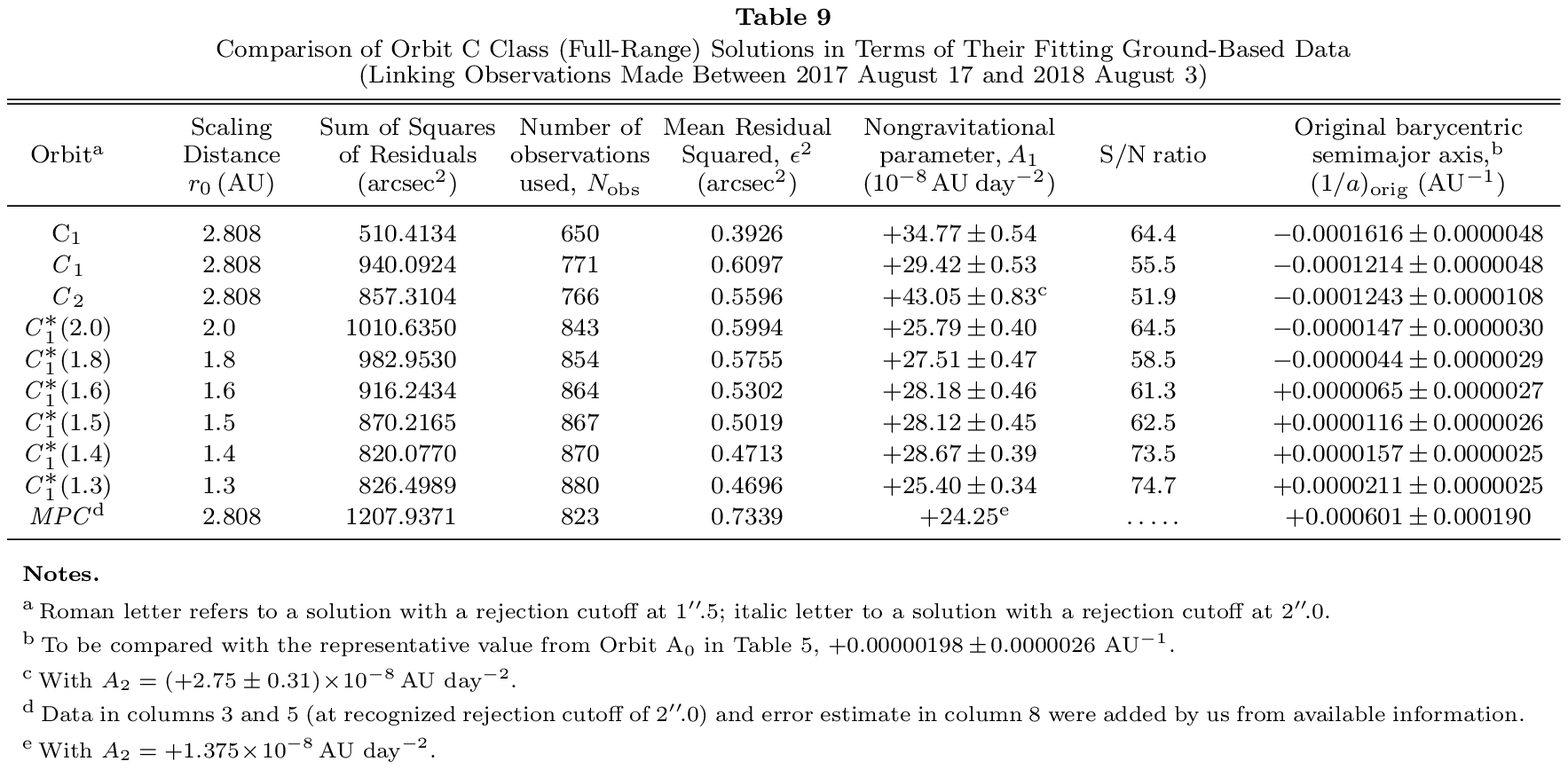}}}
\vspace{-16.1cm}
\end{table*}
\subsection{Are~There~Orbital~Solutions~That~Match~Full~Range\\[0.04cm]of This Comet's
 Ground-Based Observations?}

\vspace{0.08cm}
Given the fair degree of success of Orbit B$_1^{\textstyle \ast}$(2.0), the
question addressed in this section is whether it is at all possible to
formulate an orbital solution that could satisfactorily link the motion of the
original, intact comet before Outburst~I with the motion of the cloud of debris
observed following Outburst~II.  For this purpose we introduce a new {\small
\bf Orbits C} class of solutions derived by including all accurate ground-based
observations made between 2017 August~17 and 2018 August~3 (referred to below
as the full-range solutions); they are extending the classification proposed in
Section~7.2. 

%
%
To gain an insight into the matter, we considered the standard Style~II
nongravitational law, a radial component only (the parameter $A_1$), and a
1$^{\prime\prime\!}$.5 rejection cutoff and tried to link the full range of the
observations.  The result, Orbit C$_1$, was a disappointment because no more than
650~observations could be linked; the remaining 384 observations --- fully 37
percent of the total --- left residuals~in excess of the imposed rejection cutoff
and were discarded.  Surprisingly, the fraction of rejected observations from the
period following the outset of Outburst~II was almost exactly the same, 195 out
of 521.  This implied that the degree of success of this orbital solution was no
better prior to Outburst~II; indeed, it turned out that {\it all\/}~79~observations
made between the beginning of November 2017 and the end of May 2018 --- long
before Outburst~I --- had to be discarded because their residuals in right
ascension consistently exceeded 1$^{\prime\prime\!}$.5, reaching a peak of
6$^{\prime\prime}$(!) in early April.  This is an excellent example of
residuals caused by improper modeling of the orbital motion (see the
Appendix).  Although the nonrandom deviations in declination were less dramatic,
with an amplitude of 2$^{\prime\prime}$ in November 2017, the overall effect of
the systematic trends in the residuals was clearly severe and the orbital
solution unacceptable.{\hspace{0.3cm}}

%
%
Our response to this failure was to raise the rejection limit to 2$^{\prime\prime
\!}$.0 and to repeat the exercise in order to obtain Orbit\,{\it C\/}$_1$\,(the
italics signaling the increased rejection cutoff). The result was by no means
encouraging:\ the systematic trends in the residuals in right ascension remained,
peaking again in early April, even though the amplitude dropped from
6$^{\prime\prime}$ to less than 5$^{\prime\prime}$.  The new cutoff required
that 53 of 56 observations made between the beginning of January and mid-May
be rejected.  Yet, we were able to accommodate 771 ground-based observa\-tions
with a mean residual of $\pm$0$^{\prime\prime\!}$.78.

%
%
Next, we computed Orbit\,{\it C\/}$_2$ by solving, besides $A_1$, also for the
parameter $A_2$ of the transverse component of the nongravitational acceleration.
With no changes to the law or rejection cutoff, we found that the distribution
of residuals from this solution was in fact worse than for Orbit\,{\it C\/}$_1$.
Although we were able to link nearly the same number of observations, 766, and
to bring the mean residual from $\pm$0$^{\prime\prime\!}$.78 down to
$\pm$0$^{\prime\prime\!}$.75, the amplitude of the persisting systematic trends
in the residuals in right ascension, peaking in early April, grew to fully
7$^{\prime\prime\!}$.  This highly unsatisfactory orbital solution, parameterized
in the same fashion as the orbit by Williams (2018), showed that increasing the
number of unknowns to solve for is not the avenue to pursue any further.

%
%
Our search for an acceptable full-range solution was eventually at least partially
rewarded when we tested several modified nongravitational solutions, having been
encouraged by the fair success of Orbit B$_1^{\textstyle \ast}$(2.0).  We kept
solving for $A_1$ only, holding the rejection cutoff at 2$^{\prime\prime}$.  As
shown in Table~9, we continued to decrease the scaling distance from 2.0~AU down
to 1.3~AU and thereby succeeded in accommodating an ever greater number of
ground-based observations, from 843 up to 880, and simultaneously decreasing the
mean residual from $\pm$0$^{\prime\prime\!}$.77 down to $\pm$0$^{\prime\prime\!}$.68
and completely eliminating the systematic trends in the residuals over the period
of time before Outburst~I.

\begin{figure}[t]
\vspace{-4.38cm}
\hspace{2.9cm}
\centerline{
\scalebox{0.785}{
\includegraphics{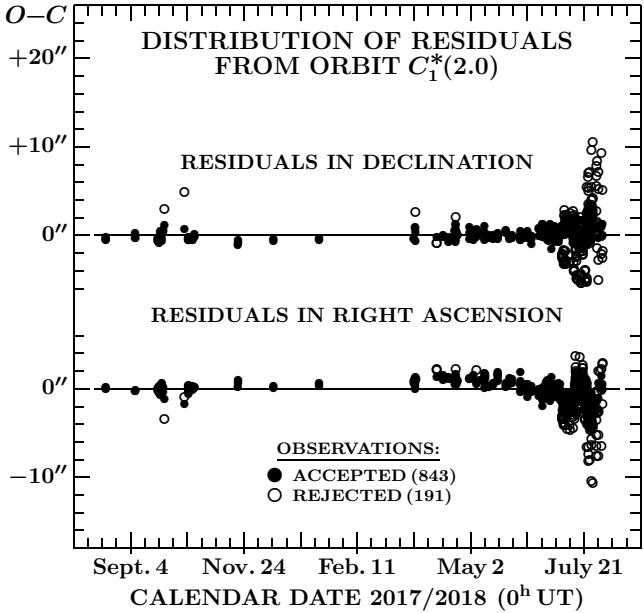}}}
\vspace{-11.12cm}
\caption{Distribution of residuals \mbox{$O \!-\! C$} of 1034 ground-based
observations from Orbit\,{\it C\/}$_1^{\textstyle \ast}$(2.0) (rejection cutoff
at 2$^{\prime\prime\!}$.0). Note that the systematic trends in right ascension
reach a peak of more than the rejection cutoff in April, but the overall range
of the rejected residuals after Outburst~II, in July--August, does not exceed
16$^{\prime\prime}$ in either coordinate.{\vspace{0.55cm}}}
\end{figure}
\begin{table}[b]
\vspace{-3.5cm}
\hspace{4.22cm}
\centerline{
\scalebox{1}{
\includegraphics{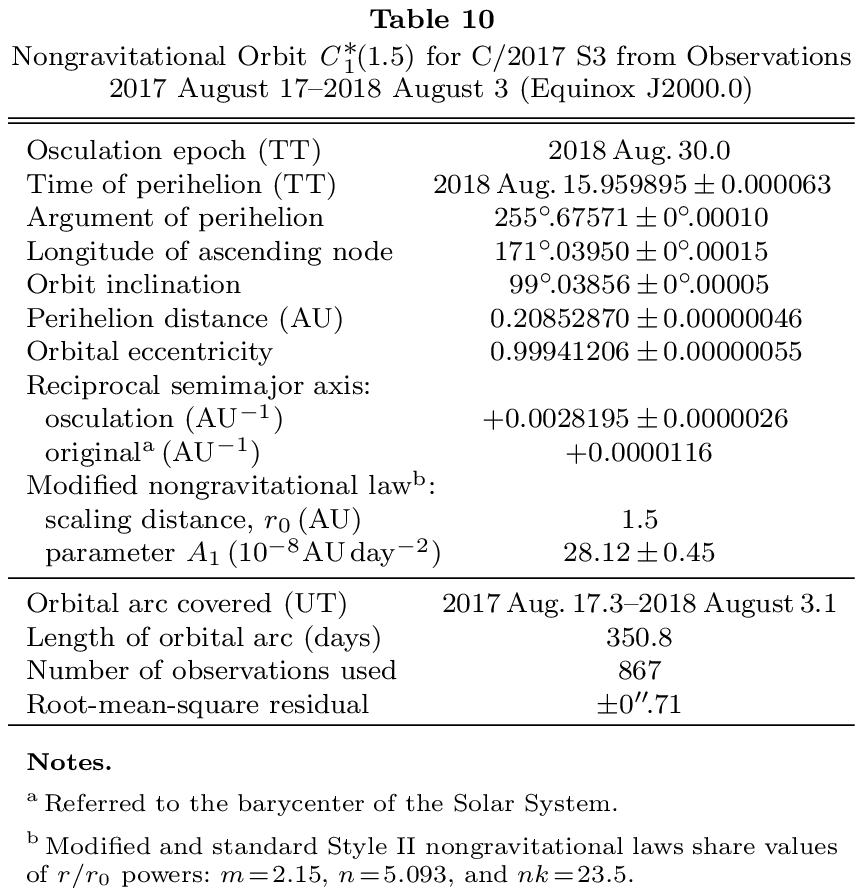}}}
\vspace{-17.06cm}
\end{table}
\begin{figure}[t]
\vspace{-3.6cm}
\hspace{2.88cm}
\centerline{
\scalebox{0.785}{
\includegraphics{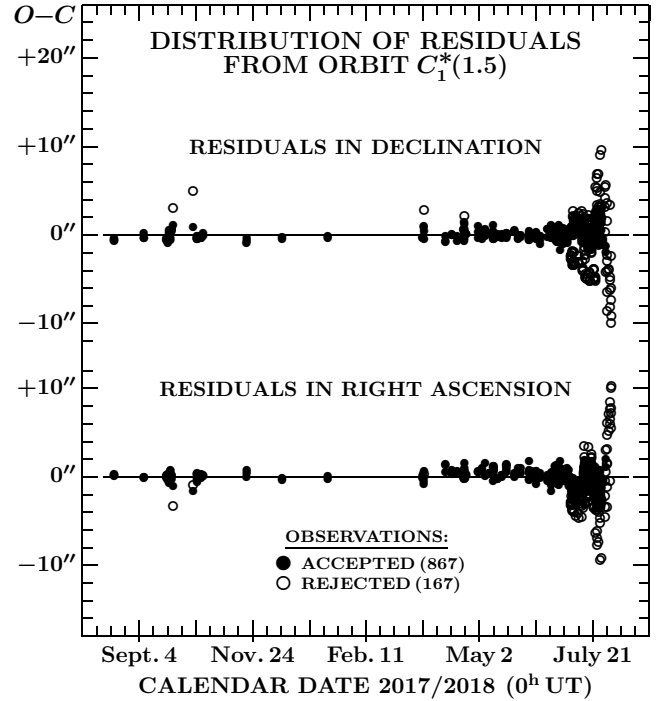}}}
\vspace{-10.73cm}
\caption{Distribution of residuals \mbox{$O \!-\! C$} of 1034 ground-based
observations from Orbit\,{\it C\/}$_1^{\textstyle \ast}$(1.5) (rejection cutoff
at 2$^{\prime\prime\!}$.0).  Note the marginal systematic trends in right
ascension peaking in April and the overall range of the rejected residuals
after Outburst~II, in July--August, reaching $\sim$20$^{\prime\prime}$ in
either coordinate.{\vspace{0.6cm}}}
\end{figure}

Although we could have continued to explore modified solutions with still lower
scaling distances, we stopped~at \mbox{$r_0 = 1.3$ AU} because of an increase to
17$^{\prime\prime}$ in the magni\-tude of the residuals, in both coordinates, of
rejected observations from early August 2018, even though the systematic trend in
the residuals in right ascension, peaking in April 2018, vanished completely.
A gradual removal of this trend with decreasing scaling distance is clearly seen
from comparsion of Orbit~{\it C\/}$_1^{\textstyle \ast}$(2.0) in Figure~15 with
Orbit~{\it C\/}$_1^{\textstyle \ast}$(1.5) in Figure~16.{\vspace{-0.05cm}}
Other solutions between Orbits {\it C\/}$_1^{\textstyle \ast}$(2.0) and {\it
C\/}$_1^{\textstyle \ast}$(1.3) have intermediate properties.  The orbits with
the scaling distance of \mbox{1.3--1.5 AU} also offer the best approximations to
the original semimajor axis (Table~9) and leave the smallest systematic deviations
from the STEREO-A positional data.  On balance, we slightly prefer Orbit~{\it
C\/}$_1^{\textstyle \ast}$(1.5) to {\it C\/}$_1^{\textstyle \ast}$(1.3), because
the former leaves the residuals of rejected observations in early August
substantially lower, not exceeding $\sim$10$^{\prime\prime}$; the elements are 
presented in Table~10.

For comparsion we also computed the distribution of residuals from the MPC orbit
by Williams (2018) (reproduced in Table~3); it is displayed in Figure~17.  Even
though the MPC solution has one more free parameter than the modified-law solutions
in Table~9, the systematic trend in the residuals in right ascension that peak in
April 2018 is much more prominent, having an amplitude of 3$^{\prime\prime}$.
In addition, the MPC orbit accommodates some 50 fewer observations (at the
same rejection level of 2$^{\prime\prime\!}$.0) than the modified-law solutions
with \mbox{$r_0 \leq 1.5$ AU} and its mean residual is substantially
higher.\footnote{We were unable to reproduce the reported mean residual of
$\pm$0$^{\prime\prime\!}$.80 for the MPC solution; instead, our computation of
the distribution of residuals left by the 823~linked observations yields a mean
residual of $\pm$0$^{\prime\prime\!}$.86.} The MPC orbit also leaves asymmetric,
systematic trends of up to 12$^{\prime\prime}$ in the residuals from the
rejected July--August ground-based observations and, just as the modified-law
solutions, it fails to fit the STEREO-A astrometric observations from Table~4.
The poor fit may also have led to the strongly elliptical original orbit, which
is evidently inconsistent with both our and Nakano's results.

\begin{figure}[t]
\vspace{-4.35cm}
\hspace{2.88cm}
\centerline{
\scalebox{0.785}{
\includegraphics{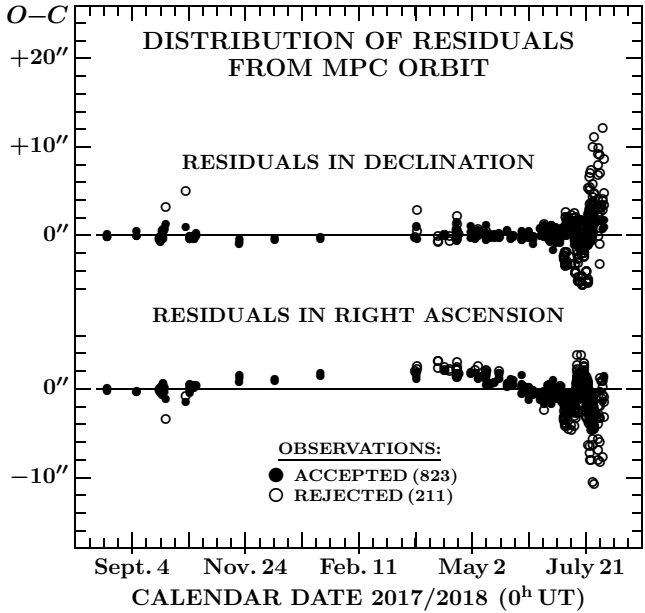}}}
\vspace{-11.13cm}
\caption{Distribution of residuals of 1034 ground-based observations from the
MPC orbit by Williams (2018), based on the standard nongravitational law and
823 data points that satisfy a rejection cutoff at 2$^{\prime\prime\!}$.0.  The
solution includes the parameters of the radial and transverse components of the
nongravitational acceleration.  Note that the systematic trend in the residuals in
right ascension stretching from November 2017 through May 2018 is quite formidable,
amounting to at least 3$^{\prime\prime}$ in April.  Besides, the residuals of
the rejected ground-based observations in late July and early August 2018 are
asymmetric and extend to $\sim$12$^{\prime\prime}$ in declination.{\vspace{0.63cm}}}
\end{figure}

While the modified-law orbits with the scaling distance of \mbox{1.3--1.5 AU} are
superior to the standard-law solutions because they are more successful in
simulating the apparent absence of nongravitational effects in the comet's motion
before Outburst~I (which began at a heliocentric distance of 1.25~AU), we feel
that the effort aimed~at accommodating the full range of ground-based observations,
from 2017 August 17 to 2018 August 3, by a single set of orbital elements is too
ambitious given the perceived anomalies in the comet's motion especially during
Outburst~II.  These nongravitational perturbations were not only substantial in
magnitude but also tightly constrained in time.  They cannot be fully accounted
for by methods that are designed to describe an essentially continuous action of
nongravitational forces.  In the following, we employ analogy with cometary
splitting in our quest to gain a greater insight into the enigmatic orbital
behavior of comet C/2017~S3.

%
\section{Nucleus' Fragmentation:\ Detection of\\Two Independent Clouds of
 Debris,\\and Their Physical Properties}
In Section 7.2 we pointed out that the strong systematic trends in the fragmented
nucleus' residuals from the (extrapolated) orbital motion of the intact comet
mimicked the motion of a companion fragment, after its separation from the parent
comet, relative to the primary fragment --- the problem of a split comet.  Because
the motion of the intact nucleus of C/2017~S3 was unaffected by a nongravitational
acceleration, it is imperative that the residuals from a {\it gravitational\/}
orbital solution be employed in this exercise.

As with any other motions driven mainly by a differential deceleration (rather
than a relative velocity),~the separation between the fragments first increases
very slowly, because the deceleration needs time to build up the velocity, which
in turn needs time to build up the separation distance.  However, in the advanced
stages of fragment separation the relative motion increases at rates that increase
rapidly.  The introduction of the deceleration as the dominant factor in this
process always results in a {\it much earlier\/} time of fragmentation relative
to the time determined by the models that attribute the effect to an (exaggerated)
separation velocity.  The needed fragmentation-time corrections{\vspace{-0.03cm}}
can in some instances become enormous.\footnote{Neglect of a deceleration in the
motion of the companion to comet C/1956~F1 (Wirtanen) offers an example of such
extreme errors.
Fitting the apparent gradual increase in the separation distance between the
comet's two nuclei over a period of more than two years, from 1957 May~1 through
1959 September~2, Roemer (1962, 1963) determined that the parent comet had split,
with an uncertainty of a few days, on 1957 January 1, the separation velocity
projected on the sky having reached $\sim$1.6~m~s$^{-1}$.  She went on to use
this result in her determination of the comet's mass.  A subsequent rigorous
analysis of the motions of the two fragments showed that the breakup had in fact
occurred in September 1954, more than 2~years (!) earlier, with an uncertainty of
{\vspace{-0.04cm}}some 2~months, and that the total separation velocity had been
merely 0.26~m~s$^{-1}$ (Sekanina 1978).  Although the comet was not seen double
when discovered and observed in 1956, the predicted separation at the time was
only $\sim$2$^{\prime\prime}$, too minute (and the companion probably too faint)
to detect. --- Another, a far less dramatic case is C/1947~X1, for which Guigay
(1955) {\vspace{-0.04cm}}presented three fragmentation scenarios, the one with
the least separation velocity, of 4.8~m~s$^{-1}$, implying a fragmentation time
of 1947 December~8.0 UT, 5.4~days after perihelion.  Yet, the best rigorous
solution offers a separation velocity of only 1.9~m~s$^{-1}$ and a fragmentation
time of November~30.5~UT, or 2.1~days {\it before\/} perihelion (Sekanina 1978).
Numerous other~examples could be cited.}

This major fragmentation-time correction and the slow buildup of the separation
distance between the fragments implies that they cannot be spatially resolved
for a fairly long time after the event took place, and that therefore only the
resulting {\small \bf duplicity or multiplicity of a comet is detected by
ground-based observers, not the splitting itself}, despite frequent claims to
the contrary.

Application of the model by Sekanina (1977, 1982) requires that the fragmentation
time of the cloud of nuclear debris of C/2017~S3 be equated with the companion's
time of separation and the effect of solar radiation pressure on the debris
particulates with the companion's nongravitational acceleration, which the model
assumes to vary as the inverse square of heliocentric distance.  The degree to
which the additional condition of an isotropic expansion of the cloud of dust
debris (Section~7.2) is in fact satisfied should be tested by checking the
absence of a separation-velocity effect.

Equipped with the outlined methodology, our primary interest was to apply the
fragmentation model to the rapidly increasing residuals from Orbit~B$_0$
days after the onset of Outburst~II, as depicted in Figure~11.  A less
prominent effect of the same kind appears to be displayed by the distribution of
residuals from Orbit~A$_0$ left by the observations made between Outburst~I and
Outburst~II, as shown in Figure~9.  For analysis of the process of fragmentation
related to Outburst~II, we chose Orbit B$_0$ as the most appropriate reference,
in part because it absorbs much of the modest nongravitational perturbation effect
associated with Outburst I.

\begin{figure*}[t]
\vspace{-2.5cm}
\hspace{-0.23cm}
\centerline{
\scalebox{0.82}{
\includegraphics{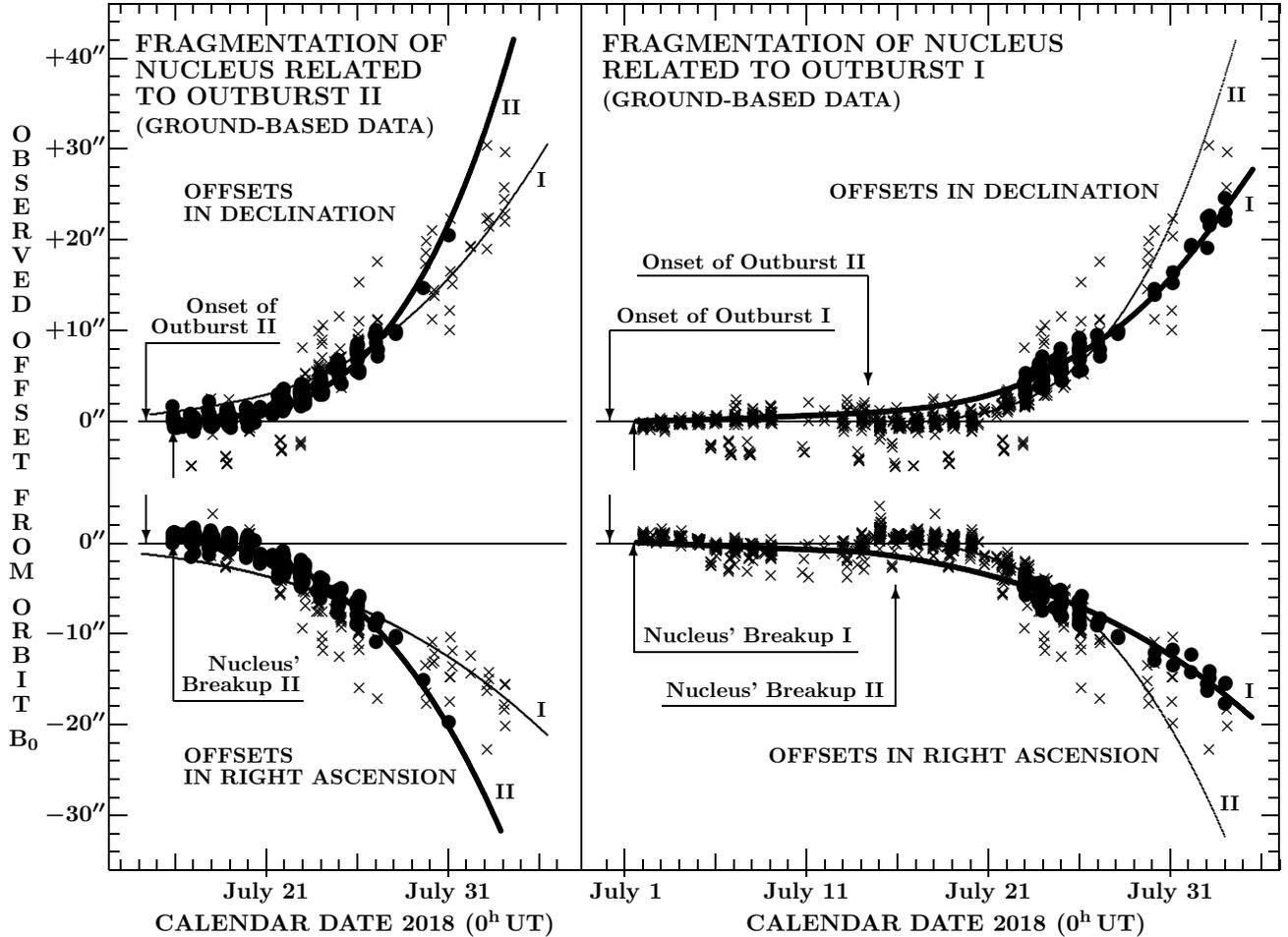}}}
\vspace{-9.15cm}
\caption{Evidence of sudden fragmentation of the comet's nucleus associated
with the two outbursts, I and II.  The thick curves fit the offsets, from
Orbit~B$_0$, of the observations used in the specified solution (solid
circles); the thin curves show the other fit (left vs right); and the crosses
are the ignored observations.  Rejection threshold at 2$^{\prime\prime\!}$.0.
{\it Left\/}:\ Fit to 367~observations made between July~16 and 31, indicating
a fragmentation time of July~15.9\,$\pm$\,0.1~UT (1.5~days after the onset
of {\vspace{-0.04cm}}Outburst~II) and a nongravitational deceleration of
216\,$\pm$\,6 units of 10$^{-5}$ the Sun's gravitational acceleration or
\mbox{(64\,$\pm$\,2)$\, \times 10^{-8}$\,AU day$^{-2}$}.  None of the 12~data
points obtained on August~1--3 and only one out of 23 (or 4~percent) obtained
from July~30 on could be accommodated, the remaining ones deviating from the
least-squares solutions by more than 2$^{\prime\prime}$ in at least one coordinate.
 {\it Right\/}:\ Fit to 86~observations made between July~23 and August~3,
indicating a fragmentation time of July~1.5\,$\pm$\,0.7~UT (1.3~days after
{\vspace{-0.04cm}}the onset of Outburst~I) and a nongravitational deceleration
of 57\,$\pm$\,2 units of 10$^{-5}$ the Sun's gravitational acceleration or
\mbox{(16.9\,$\pm$\,0.6)$\, \times 10^{-8}$\,AU day$^{-2}$}.  We were able to
accommodate 9 out of 12 data points obtained on August~1--3 and 13 out of a
total of 23 obtained from July~30 on, that is, 75 and 58 percent,
respectively.{\vspace{0.55cm}}}
\end{figure*}

\subsection{Effect of Outburst~II in Ground-Based Observations}
\vspace{-0.03cm}
Treated as offsets from Orbit B$_0$, the residuals from Figure~11 are replotted,
from July 16 on, in the left-hand side panel, and from July 1 on in the right-hand
side panel, of Figure~18.  We first focus on the left-hand side panel of the
figure and notice that, consistent with expectation, in the early days after the
onset of Outburst~II the offsets are distributed along the axis of abscisas, very
slowly building up a detectable deviation from it.  Later on, the offsets grow at
a sharply accelerating rate, both in right ascension and declination.  It is
noted that a least-squares fit to the offsets in this advanced stage only would
show that the fragmentation occurred some time on July~20 or 21.  As long as
the scientist is not concerned with the physical implications of the orbital
solution, he would incorporate the observations from up to July~20 or 21 into
his input file --- just as did Nakano (2018b).  However, fitting the offsets
with the fragmentation model shows that the breakup did indeed take place nearly
a week earlier.  Applying the model to 367 offsets between July~16 and 31 and
choosing a rejection cutoff of 2$^{\prime\prime\!}$.0, we obtain a solution with
a mean residual of $\pm$0$^{\prime\prime\!}$.77, resulting in a fragmentation
time of July~15.9\,$\pm$\,0.1~UT, which is lagging the onset of Outburst~II by
1.5~days and coinciding with the event's brightness peak (Figure~1).  The
deceleration $\gamma$, implied by the rapidly increasing offsets, amounts to
\mbox{216\,$\pm$\,6 units} of 10$^{-5}$\,the Sun's gravitational acceleration
and is equivalent~to~\mbox{(64\,$\pm\,2)\!\times\!10^{-8}$AU\,day$^{-2}$}
at 1~AU from the Sun.  At an assumed bulk density~of 0.53~g~cm$^{-3}$, the
dominant dust grains in the debris cloud were, following Equation~(17), exactly
1~mm in diameter.

This deceleration is nearly identical in magnitude to the value reported by
Sekanina \& Chodas (2012) for the sungrazing comet C/2011~W3 (Lovejoy).  From~their
examination of its spine tail, they determined for the debris, released from the
disintegrating comet shortly after perihelion, a radiation-pressure effect of
\mbox{$\gamma = 191$\,$\pm$\,42}~units of 10$^{-5}$\,the Sun's gravitational
acceleration.  Among the split comets, the short-lived companion to C/1942~X1
(Whipple-Fedtke-Tevzadze), observed over a period of only nine days by G.\,Van
Biesbroeck, was subjected to a similar deceleration, equaling 228\,$\pm$\,16~units
of 10$^{-5}$\,the Sun's gravitational acceleration (Sekanina 1979, 1982).

We also ran solutions that included, as additional parameters, the normal
and/or transverse components of the separation velocity only to find that they
both were insignificant, with no effect on the result.\footnote{The radial
component of the separation velocity could not be determined because of its
very high correlation with the fragmentation time, but it probably was negligibly
small as well.}~Accordingly, we see no reason for questioning the validity of the
two-parameter solution.

A caveat of major consequence does, however, pertain to the discarded
observations.  Employing a rejection threshold of 2$^{\prime\prime}$, we found
two groups of offsets in right ascension and/or declination that could not be
used because of their unacceptably large deviations from otherwise a very
satisfactory fit.  The first group was about two dozen offsets on July~16--26,
an overwhelming majority of which had the same observatory code, undoubtedly
an observer/instrumental problem.  The other group was also about two dozen
offsets, from July~29--August~3, at the very end of the orbital arc observed
from the ground.  All these rejected data are depicted by the crosses in the
left-hand side panel of Figure~18.  Only one offset satisfied the solution on
July~29, another one on July~31, but not a single one on August~1--3; over
90~percent were systematically off.  In the context of this discrepancy, we
recall reports by independent observers (Section~3) that in late July and
early August the comet essentially lost its nuclear condensation, aggravating
positional determination.  But Soulier, who contributed more than 25~percent of
the astrometric data reported for July~27--August~3 and 50~percent of the data
reported for August~1--3, not only complained that it was difficult to measure
the comet's position (by applying a centroid routine of the Astrometrica software
tool), but also pointed out that the comet did not occupy the expected position,
as if it were {\small \bf completely displaced}.\footnote{See Soulier's message
27180 in the Comets Mailing List website; cf.\ footnote 2.}{\vspace{-0.2cm}}

\subsection{Effect of Outburst I in Ground-Based Observations}
The remarkable, albeit entirely unexpected, property of the rejected offsets from
July~29--August~3 on the~left~of Figure~18 is their well-organized distribution
with time, which is obvious to the extent that we could not resist subjecting
these data to the same kind of treatment as the offsets with a confirmed
reference to Outburst~II.  We initiated the procedure by collecting all 23~data
reported for July~30--August~3 and by applying the two-parameter fragmentation
model.  The result of this first step was most surprising:\ we obtained
July~1.9\,$\pm$\,5.9~UT for the fragmentation time, a value that is centered
on the time lagging the onset of Outburst~I by 1.7~days; and a deceleration
\mbox{$\gamma = 61$\,$\pm$\,15}~units of 10$^{-5}$\,the Sun's gravitational
acceleration.~The mean residual was $\pm$3$^{\prime\prime\!}$.2, the maximum
residual amounting to 8$^{\prime\prime}$. This solution provides us with clear
{\small \bf evidence of a surviving cloud of debris related to Outburst~I},
a feature similar to the cloud associated with Outburst~II, but whose existence
we have as yet{\nopagebreak} been unaware~of.

The large errors are caused by a long gap between the fragmentation time and the
short period of time occupied by the offsets, as well as by our refraining from
applying any rejection cutoff in this particular case.  Next we searched
for all offsets between July~23 and August~3 consistent with the initial
solution.  Re-imposing the 2$^{\prime\prime}$ rejection threshold, we collected
86~offsets and solved for the two parameters again.  We obtained a fragmentation
time of July~1.5\,$\pm$\,0.7~UT, centered on 1.3~days after the onset of
Outburst~I.  The deceleration was \mbox{57\,$\pm$\,2 units} of 10$^{-5}$\,the
Sun's gravitational acceleration, equivalent to (16.9\,$\pm$\,0.6)$\times
10^{-8}$\,AU~day$^{-2}$ and implying, on the same bulk-density assumption as
before, the dominant presence of dust grains 3.8~mm across.  The mean
residual was $\pm$0$^{\prime\prime\!}$.98 and the run is depicted in the
right-hand side panel of Figure~18.

As with the other debris cloud, we also ran three- and four-parameter solutions
that would determine the normal and/or transverse components of the separation
velocity, but they always were very near zero.  The deceleration resulting from
the adopted two-parameter solution, is remarkably close to the value of the $A_1$
parameter of the previously derived nongravitational Orbit B$_1^{\textstyle
\ast}$(2.0), which was linking the observations covering the orbital arc up to
the beginning of Outburst~II (Table~8).  This encouraging agreement suggests
that the debris cloud associated with Outburst~I did affect the comet's
astrometry at the time between the outbursts, governing the non\-gravitational
solution.

\begin{table}[t]
\vspace{-4.21cm}
\hspace{4.2cm}
\centerline{
\scalebox{1}{
\includegraphics{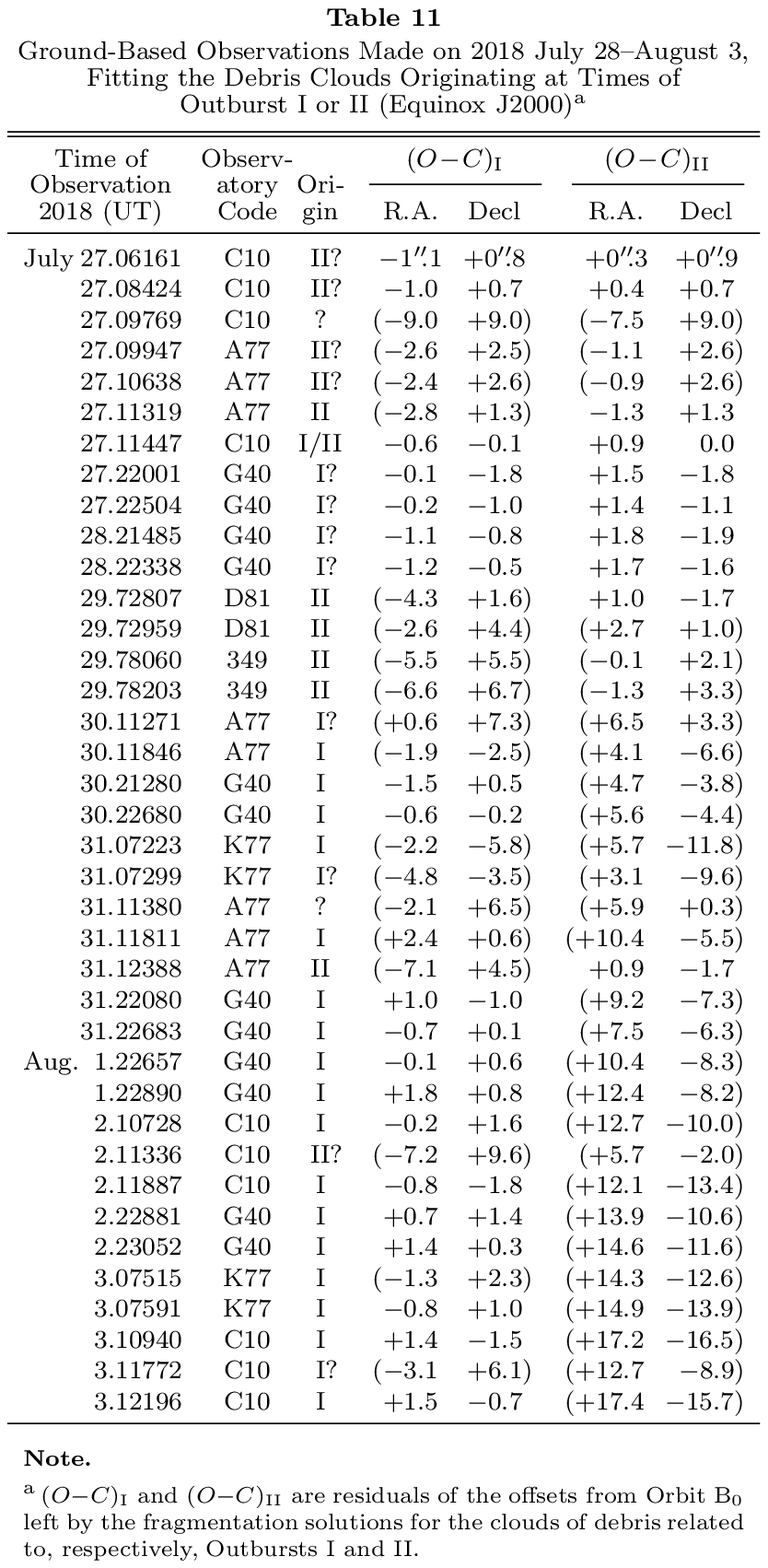}}}
\vspace{-8.05cm}
\end{table}

The residuals from the two-parameter solutions to the debris clouds related to
the two outbursts are presented in Table~11 for all ground-based astrometric
observations made between July~27 and August~3.  In the upper half of the table,
covering July 27--29, the observations refer mostly to the debris cloud associated
with Outburst~II, while its lower half, July 30--August 3, is dominated by the
observations of the debris cloud related to Outburst~I.  The same result emerges
from the summary Table~12, which lists daily totals of the observations:\ their
numbers have a tendency to increase with time for the debris cloud associated with
Outburst~I, but to decrease for the cloud related to Outburst~II.  The time at
which the numbers of the two trends equate each other is not sharply defined, but
Table~11 and Table~12 consistently suggest that this happens in general proximity
of July~29/30.  Since it is the peak surface brightness that determines the spot
the measurers bisect as the ``comet's position'', the transition from one cloud
to the other suggests that they were then of approximately equal peak surface
brightness, one fading more steeply than the other.  We show in Section~8.4
that the critical time, $t_{\rm crit}$, when the balance was achieved --- adopted
to have occurred on July~30.0~UT --- provides an important piece of information,
despite the fact that the fading trends are at first sight counterintuitive.

%
\subsection{Split-Comet Scenario}
The comet's nuclear condensation, determining the comet's astrometric position,
was measured with high accuracy as a well-behaving peak of light until the
time of Outburst~I.  This is documented by a low mean residual of
$\pm$0$^{\prime\prime\!}$.45 of Orbit A$_0$ and a very few observations that
had to be discarded because their residuals exceeded the rejection threshold
of 1$^{\prime\prime\!}$.5 (see Appendix).

Starting with Outburst I, C/2017 S3 became effectively a {\it double\/} comet.
During this event, the major part of the parent nucleus became the primary
fragment, while the less sizable part began disintegrating, presumably upon
separation, into a compact cloud of debris.  The active phase of the event
was characterized by a sharp, starlike condensation --- typical for comets
in outburst --- consisting primarily of the emitted gas and microscopic dust
ejecta, the latter being confirmed by an instant increase in the {\it
Af}$\rho$ parameter (e.g., Bryssinck, footnote~1).  The parallel surge of
total brightness was dominated by gas, as already acknowledged in Sections~1
and 4.2.  The contribution by the cloud of millimeter-sized debris remained
largely obscured for a number of days, until the activity of the primary
fragment subsided enough.  Only then did the chances of the cloud's detection
improved, especially in properly centered small apertures used in astrometric
observations, with the opportunity peaking just before the onset of Outburst~II.

\begin{table}[b]
\vspace{-3.8cm}
\hspace{4.22cm}
\centerline{
\scalebox{1}{
\includegraphics{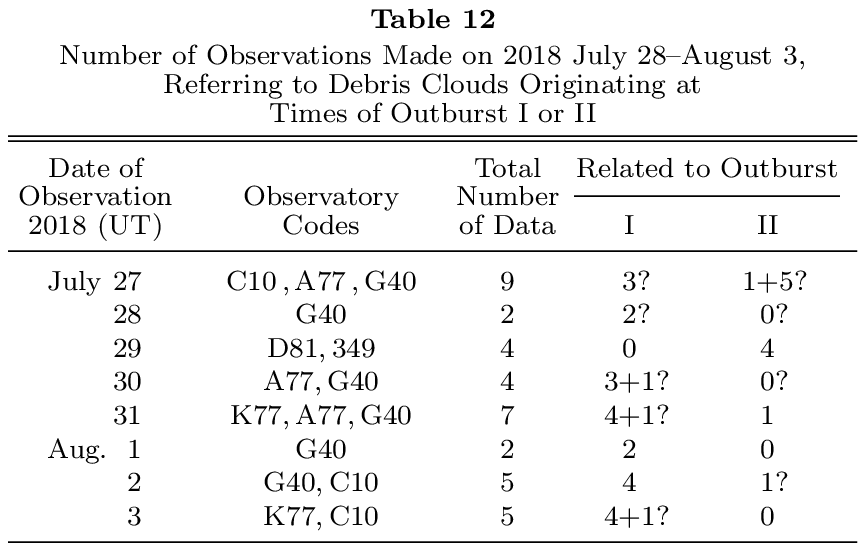}}}
\vspace{-20.22cm}
\end{table}

With the arrival of Outburst II, all traces of the previous activity got
concealed by the newly developing feature.  Within two days of the event's
onset, the primary fragment appears to have completely disintegrated into
a second cloud of debris, believed to be much more massive but also rapidly
expanding with time.  Measured astrometric positions, referring to the bright
and initially stellar nuclear condensation at the location of the disintegrating
nucleus, did not begin to deviate from the extrapolated Orbit B$_0$ until
July~20/21, suggesting that, by then, the cloud of millimeter-sized grains
associated with Outburst~II began to dominate the comet's brightness in the
small apertures. The expansion rate of this cloud, given by Equation~(15),
explains the sharp drop in the nuclear-condensation's brightness seen in
Figure~1.  If the cloud of millimeter-sized grains related to Outburst~I
was expanding, as it turned out to be the case, at a much lower rate, the
surface-brightness peak at this cloud's center
was dimming with time more slowly than the peak of the debris cloud
associated with Outburst~II; it was only a matter of time for the former
outshining~the latter.  As documented by their astrometry, this is exactly
what Soulier and other observers detected in the period of July~30 through
August~3:\ the bisected point of maximum surface brightness moved from the
debris cloud related to Outburst~II to that related to Outburst~I; the comet
looked displaced, as illustrated in Figure~18.

The existence of two nearby peaks in the coma at the time is corroborated by
Soulier's (see footnote 15) brightness measurements in two apertures on
August~3.12~UT.  Magnitude 16.2 in a 6$^{\prime\prime\!}$.5 peak-centered
aperture implies a magnitude fainter than 12.7 in a 32$^{\prime\prime\!}$.4
aperture, yet Soulier reports magnitude 11.5.  Since there was no major
condensation in the coma, the discrepancy implies that the second peak was
likely to have been located within the larger aperture's radius; Table~11
does indeed predict that the two clouds' centers were then about
22$^{\prime\prime}$ apart.   

\begin{figure}[t]
\vspace{-2.64cm}
\hspace{1.95cm}
\centerline{
\scalebox{0.72}{
\includegraphics{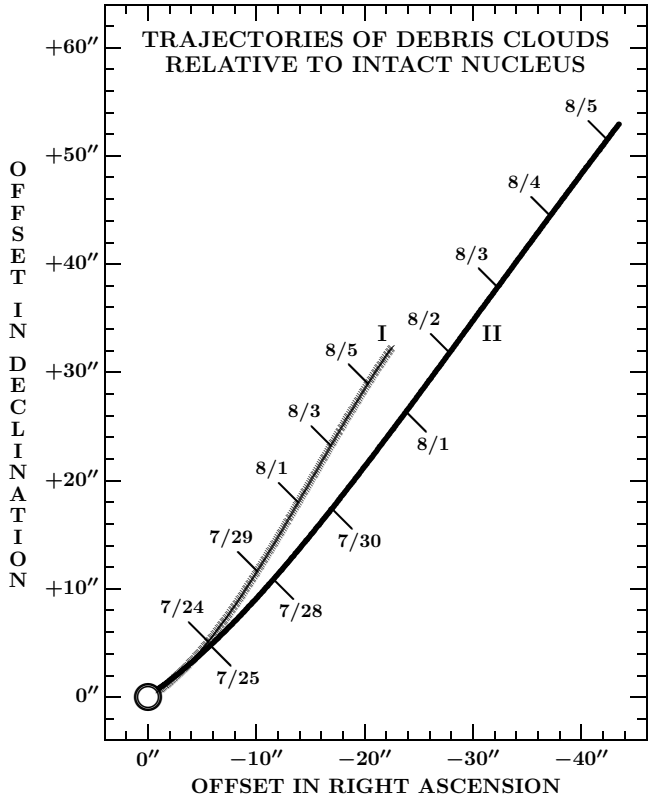}}}
\vspace{-8.4cm}
\caption{Projected trajectories of the clouds of millimeter-sized dust grains
relative to the extrapolated Orbit B$_0$.  Although the breakup occurred on
July~1.5~UT for the cloud related to Outburst~I and on July~15.9~UT for the
cloud related to Outburst~II, either cloud remained close to the position
determined by Orbit B$_0$ for an extended period of time following the
fragmentation event.  For example, the separation distance of 2$^{\prime\prime}$
was reached only two weeks after fragmentation in the first case and 5 days after
fragmentation in the second case.  There was no need to introduce a separation
velocity in order to fit the motions of the two debris clouds, either one of
which moved under the effects of a repulsive force that varied as an inverse
square of heliocentric distance and is believed to be solar radiation pressure.
The deceleration on the cloud related to Outburst II was 3.8 times higher than
on the other cloud.{\vspace{0.65cm}}}
\end{figure}

In summary, the distribution of offsets of the observed positions from Orbit
B$_0$ conforms closely to a concatenated curve representing the motions of two
distinct clouds of millimeter-sized dust grains that were released during the
two major outbursts, always some 1--2 days after their onset.  The debris cloud
associated with Outburst~I fits (i)~the offsets at times between the outbursts
(although these offsets were not used in the solution, they stayed within
2$^{\prime\prime}$ and the cloud's motion essentially coincided with the
center of the nuclear condensation that appeared after Outburst~I):
(ii)~some of the offsets between July 23 and 27; and, most importantly,
(iii)~nearly all offsets from July~30 on; but {\it not\/} between July~16 and
22.  The cloud of millimeter-sized grains related to Outburst~II fits the
offsets in the period of time between July~16 and 27, but only sporadically
those at the later time, and {\it none\/} on August~1--3.  Between July~23 and
28 the projected motions of the two clouds were close enough to each other that
it is difficult to resolve to which any particular observation referred.  The
centers of the two clouds had equal right ascension on July~25.5~UT and equal
declination on July~27.0~UT.  The separation distance between them increased
to 13$^{\prime\prime}$ on August 1 and to about 22$^{\prime\prime}$ on
August~3.  Although the disintegrating object was technically a double comet,
this status was concealed by the described configuration and brightness
relationship between the two components  The trajectories of the two debris
clouds relative to the extrapolated Orbit~B$_0$ are plotted in Figure~19.

\subsection{Condition of Equal Peak Surface Brightness, and Properties of Debris
Cloud Related to Outburst I}
The condition of equal peak surface brightness implies constraints on the
relationship between the expansion velocities, $v_{\rm exp}$, and the total
projected scattering cross-sectional areas, $X_{\rm frg}$, of the two debris
clouds.  A cloud's surface brightness is proportional to the sum of the
cross-sectional areas of the grains situated in a column of unit projected area
that extends along the line of sight.  In an isotropic and uniformly expanding,
optically thin spherical cloud of debris (see Section~5), which at time $t$
projects as a disk of radius $\rho(t)$, the surface brightness decreases
symmetrically with increasing distance $x$ from the center of the disk.  The
sum of the cross-sectional areas of the grains located inside a column of unit
projected area at distance $x$ from the disk's center equals $\sigma(t,x)$ and
varies as the column's length,
\begin{equation}
\sigma(t,x) \sim \sqrt{\rho^2 \!-\! x^2}.
\end{equation}
The surface brightness reaches a peak at the center~of~the disk, where
\mbox{$\sigma(t,0) \equiv \sigma_0(t)$}, so that
\begin{equation}
\sigma(t,x) = \sigma_0(t) \sqrt{1 \!-\! \left(\frac{x}{\rho}\right)^{\!\!2}} .
\end{equation}
A total cross-sectional area of all grains in~the~cloud is calculated by
integrating over the disk,
\begin{equation}
X_{\rm frg} = \!\!\int_{0}^{\rho} \!\! 2 \pi x \sigma(t,x)\, dx =
 {\textstyle \frac{2}{3}} \pi \sigma_0 \rho^2.
\end{equation}
{\vspace{-0.04cm}}While $\sigma_0$ and $\rho$ vary with time, $X_{\rm frg}$ is for
either cloud a constant, so \mbox{$\sigma_0 \!\sim\! \rho^{-2}$}.  Inserting for
$\rho$ from Equation~(1); referring, respectively, to the fragmentation time, the
expansion velocity, the peak columnar cross-sectional area of the grains
(determining the peak surface brightness), and the total cross-sectional area of
the cloud of debris associated with Outburst~I as $t_{\rm frg}$({\scriptsize I}),
$v_{\rm exp}$({\scriptsize I}), $\sigma_0$({\scriptsize I}), and $X_{\rm
frg}$({\scriptsize I}); referring, respectively, to those associated with
Outburst~II as $t_{\rm frg}$({{\scriptsize II}), $v_{\rm frg}$({\scriptsize II}),
$\sigma_0$({\scriptsize II}), and $X_{\rm frg}$({\scriptsize II}); and equating,
at time $t_{\rm crit}$, the two peak values of the surface brightness (i.e.,
the sums of the peak cross-sectional areas of the grains); we find for the
relationship between the two events' parameters:
\begin{equation}
\frac{X_{\rm frg}{\mbox{({\scriptsize I})}}}{v_{\rm exp}^2{\mbox{({\scriptsize
 I})}} \left[t_{\rm crit}\!-\!t_{\rm frg}{\mbox{({\scriptsize I})}}\right]^2} =
 \frac{X_{\rm frg}{\mbox{({\scriptsize II})}}}{v_{\rm exp}^2{\mbox{({\scriptsize
 II})}} \left[ t_{\rm crit} \!-\! t_{\rm frg}{\mbox{({\scriptsize II})}}\right]^2}.
\end{equation}
In this dimensionless equation we know $t_{\rm frg}$({\scriptsize I}) and
$t_{\rm frg}$({\scriptsize II}) from, respectively, Sections 8.2 and 8.1; $v_{\rm
exp}$({\scriptsize II}) from Equation~(15); $X_{\rm frg}$({\scriptsize II}) from
Equation~(12); and $t_{\rm crit}$ from Section~8.2.  Equation~(22) can be written
as a relation between the two unknowns,
\begin{equation}
X_{\rm frg}{\mbox{({\scriptsize I})}} = \eta \, v_{\rm exp}^2{\mbox{({\scriptsize
 I})}},
\end{equation}
where
\begin{equation}
\eta = \frac{X_{\rm frg}{\mbox{({\scriptsize II})}}}{v_{\rm
 exp}^2{\mbox{({\scriptsize II})}}} \left[ \frac{t_{\rm crit} \!-\! t_{\rm
 frg}{\mbox{({\scriptsize I})}}}{t_{\rm crit} \!-\! t_{\rm frg}{\mbox{({\scriptsize
 II})}}} \right]^{\!2}\!\!.
\end{equation}
Numerically, \mbox{$\eta = 1.56 \times \!10^6$\,s$^2$}.

In Section 8.3 we suggested that the contamination of the cloud of millimeter-sized
grains associated with Outburst~I should have been minimized shortly before the
onset of Outburst~II, although the question of what fraction of the comet's
brightness it accounted for remains open.  An example for testing the cloud's
signal is the observation
on July~14.03~UT by E.~Bryssinck (see footnote~1), who reported the apparent red
magnitude of 14.24 in an aperture 9$^{\prime\prime\!}$.92 in radius and of 15.56
in an aperture 4$^{\prime\prime\!}$.96 in radius.  They are equivalent to the
phase-corrected absolute magnitudes of 13.37 and 14.69, respectively.  To find
out whether at this time $t_{\rm obs}$ the debris cloud, of an expansion-velocity
dependent radius $\rho_{\rm obs}$ extended beyond the limits of an aperture of
radius $a_{\rm obs}$, we checked the validity of Equation~(6).  Denoting the
cross-sectional area of the debris cloud inside the aperture at the time of
observation as $X_{\rm obs}$, the equation can be rearranged thus:
\begin{equation}
\left[1\!-\!\frac{\xi}{v_{\rm exp}^2{\mbox{({\scriptsize I})}}} \right]^{\!2}
 \!\!=\! \left[1\!-\!\frac{\zeta}{v_{\rm exp}^2{\mbox{({\scriptsize I})}}}
 \right]^{\!3} \;\;\;\; ({\rm for} \; \rho_{\rm obs} \!>\! a_{\rm obs}),
\end{equation}
where with use of Equation (23),
\begin{equation}
\xi = \frac{X_{\rm obs}}{\eta}, \;\;\;\;\zeta = \!\left[\frac{a_{\rm obs}}{t_{\rm
 obs}\!-\! t_{\rm frg}{\mbox{({\scriptsize I})}}} \right]^{\!2} \!\!,
\end{equation}
and $X_{\rm obs}$ is expressed as a function of the absolute magnitude according
to Equation~(11).  If the flux of the cloud of millimeter-sized grains associated
with Outburst~I is contaminated in the aperture by other ejecta, one gets an upper
limit on $X_{\rm obs}$.

The condition (25) is a quadratic equation in $v_{\rm exp}^2(\mbox{\scriptsize
I})$,{\vspace{-0.04cm}} with meaningful solutions limited to
\begin{equation}
\zeta < \xi < {\textstyle \frac{3}{2}} \, \zeta
\end{equation}
and to the root with the positive sign in front of the~discriminator.  Writing
further \mbox{$\xi = \zeta (1 \!+\! \epsilon)$} (with \mbox{$0 \!<\! \epsilon
\!<\!  \frac{1}{2}$}), the solution becomes
\begin{equation}
v_{\rm exp}^2(\mbox{\scriptsize I}) = \zeta \! \left[1 \!-\! \epsilon \!\left(
 \!1\!+\!  {\textstyle \frac{1}{2}} \epsilon \!+\! \sqrt{\epsilon} \sqrt{1 \!+\!
 {\textstyle \frac{1}{4}} \epsilon} \right) \! \right]^{\!-1} \!\!.
\end{equation}

Returning now to the July 14 observation by Bryssinck, the results for the larger
aperture, 9$^{\prime\prime\!}$.92 in radius, are \mbox{$\xi = 58.5$ m$^2$~s$^{-2}$}
and \mbox{$\zeta = 66.8$ m$^2$~s$^{-2}$}, so that the condition (27) is not
satisfied.  The dimensions of the cloud related to Outburst~I are smaller than
the aperture, and its expansion velocity is lower than \mbox{$\sqrt{\xi} =
7.6$~m~s$^{-1}$}, which happens to be a tenth of the expansion velocity of the
cloud associated with Outburst~II.  This result already suggests that Outburst~I
was a far less violent event than Outburst~II.

For the smaller aperture, 4$^{\prime\prime\!}$.96 in radius, the condition (27)
is satisfied, even though only marginally.~We~derive \mbox{$\xi = 17.3$ m$^2$
s$^{-2}$} and \mbox{$\zeta = 16.7$ m$^2$ s$^{-2}$}, so that \mbox{$\epsilon =
0.0359$} and the solution (28) then implies an expansion{\vspace{-0.03cm}}
velocity of \mbox{$v_{\rm exp}({\mbox{\scriptsize I}}) = 4.18 \; {\rm m}\;{\rm
s}^{-1}$}, unless the observed signal's contamination by other ejecta (e.g.,
microscopic dust) exceeds 3--4~percent.  On the same assumption we have for
the {\vspace{-0.05cm}}cross-sectional area of the cloud from Equation~(23)
\mbox{$X_{\rm frg}({\mbox{\scriptsize I}}) = 27 \; {\rm km}^2$}, only a little
more than 1~percent of the cross-sectional area of the cloud related to
Outburst~II.{\nopagebreak}

In order to test the validity of the assumption~on~the lack of a major
contamination of the measured signal, we confront the result based on the July~14
\mbox{observation} with the last ground-based observations, made by Soulier on
August 3.11--3.12~UT, when he used an aperture of 6$^{\prime\prime\!}$.5 in radius
to measure an average apparent CCD magnitude of 16.4 (see footnotes 5~and 15).
Comparison of Soulier's unfiltered CCD magnitudes and Bryssinck's red magnitudes
suggests that after aperture correction they are essentially equal, so we apply no
color correction.  Soulier's phase-corrected absolute magnitude on August~3 is
then 18.6, constraining the projected area of the cloud to
\begin{equation}
X_{\rm frg}({\mbox{\scriptsize I}}) < 0.8 \; {\rm km}^2
\end{equation}
and for the expansion velocity
\begin{equation}
v_{\rm exp}({\mbox{\scriptsize I}}) < 0.7 \; {\rm m}\:{\rm s}^{-1}\!.
\end{equation}
This indicates that the cloud of millimeter-sized grains associated with
Outburst~I was detected {\it only\/} in the images taken on the final days
of the comet's ground-based monitoring campaign; that in the examined July~14
image the cloud contributed no more than a few percent of the signal in the
smaller aperture; that the expansion velocity was in a range that is typical
for the separation velocities of the split comets; and that Outburst I should
be classified as a {\it nonfatal event\/}.

\subsection{Results from the STEREO-A Images}
Because of the 72$^{\prime\prime}$ pixel size of the HI1 imager's detector, one
should not read too much into the information provided by the STEREO-A astrometric
positions listed in Table~4.  The very fact that they refer to the debris cloud
associated with Outburst~II comes from their photometry (Section~5), not
astrometry.

The fragmentation solution derived from the ground-based observations in
Section~8.1 (a fragmentation time of July~15.9~UT and a radial nongravitational
acceleration of 64$\,\times 10^{-8}$\,AU day$^{-2}$ at 1~AU from~the~Sun~based
on an inverse square power law of heliocentric distance) fits the 46~STEREO-A
positions, including the several clearly inferior points, with a mean
residual~of~$\pm$29$^{\prime\prime\!}$.1.  The few final positions, which left huge
negative residuals in right ascension, of up to $\sim$170$^{\prime\prime}$ from
Orbit A$_0$ (Figure~10) and up to $\sim$100$^{\prime\prime}$ from Orbit
B$_1^{\textstyle \ast}$(2.0) (Figure~14), are now accommodated with residuals of
$<$30$^{\prime\prime}$. 

No fragmentation time can at all be determined from the STEREO-A data set alone
and no meaningful results are obtained by differentially correcting more than three
fragmentation model's parameters at a time.  When the fragmentation time is fixed
at July~15.9~UT, various solutions with a mean residual of less than $\pm$15$^{
\prime\prime}$, or one fifth of the pixel size, are derived, their only~common
attribute being a nongravitational-acceleration parameter $A_1$ that exceeds the
value obtained from the ground-based observations.  The parameter's value is now
close to 100\,$\times 10^{-8}$\,AU~day$^{-2}$ (instead of 64\,$\times
10^{-8}$\,AU~day$^{-2}$), which by itself suggests that by the time the cloud
of debris left the field of view of the STEREO-A imager shortly before perihelion,
the dominant diameter of dust grains in the cloud dropped to $\sim$0.7~mm or
less,~possibly a sign of their incipient sublimation.  Because of the fashion in
which orbital data respond to sudden events (Section~7.2), this effect could
reflect the precipitous dive{\nopagebreak} of the debris cloud's intrinsic
brightness detected in the{\nopagebreak} last several STEREO-A images (Figure~3).

\subsection{Unconfirmed Post-Perihelion Ground-Based\\Observations of the Comet's
Debris}
J.\ J.\ Gonz\'alez\footnote{See the message 27461 in the Comets
Mailing List website; cf.\ footnote 2.} reported his visual detection
of the comet's debris with his 20-cm f/10 Schmidt-Cassegrain
telescope on five nights between 2018 October~16 and November~16,
estimating the total magnitude of the essentially condensation-free
cloud consistently at 9.7--10.1.  The feature's diameter was near
4$^\prime$ in October (when located low above the horizon), but
8$^\prime$--9$^\prime$ in November (when it was higher).  On the last date,
this cloud, observed at magnification of 77$\times$, was surrounded
by a ``wide and very diffused extension'' to 1$^\circ\!$.2, of complex
morphology and elongated along the orbit.

On November 20, A.\ Hale\footnote{See the message 27479 in the Comets
Mailing List website; cf.\ footnote 2.}\,inspected the region~of~the
predicted position of the comet with his 41-cm f/4 reflector, detecting no
trace of it with certainty.  On November 20.504~UT, M.\ Suzuki\footnote{See
the message 27488 in the Comets Mailing List website; cf.\ footnote 2.}
imaged a field of 49$^\prime$ by 32$^\prime$ centered on the ephemeris
position with a 43-cm f/4.5 telescope (+CCD) of the Remote Astronomical
Society Observatory of New Mexico near Mayhill. Visual inspection of the
6-minute exposure reveals no sign of any comet-related mass.  The question
we address below is whether it is possible to reconcile Gonz\'alez's
detection with Suzuki's and Hale's negative observations.  At issue is the
contrast threshold (or sensitivity) of a visual observer and a CCD device.

The most comprehensive investigation of the contrast threshold of human vision
was performed by Blackwell (1946), who employed 19 highly trained observers
with visual acuity of 20/20, each contributing tens of thousands of individual
data to the published sample.  More recent reviews and studies of the subject
(e.g., Clark 1990; Crumey 2014; Montrucchio 2015) still build on Blackwell's
results.

Blackwell examined the contrast threshold as a function of two
fundamental parameters, both highly relevant to our topic.  One was
the luminance of a uniform background and the other was the angular size
of the target disk (stimulus).  In the experiments he considered both
positive contrasts (targets brighter than the background) and negative.
The brightness of the night sky is well within the range he studied
(the work was motivated in part by the night-operations needs of the
U.S.\ military that funded it), but the largest stimulus was merely
6$^\circ$ in diameter.  Fortunately, for extended targets the
contrast threshold varied little and predictably with size, allowing
sound extrapolation.  Blackwell did not examine the dependence of the
contrast threshold on other possible variables, including the observer's
age that Montrucchio (and others) found to be important.

Gonz\'alez made the reported observations at two high-elevation locations and
provided much relevant information needed for judging the accuracy of his reports,
including the site's Sky Quality Meter's (SQM)\footnote{For a description,
see, e.g., Birriel \& Adkins (2010).} measurement.  In all five instances the
luminance of the feature, computed from its estimated total brightness and
diameter, was above the expected average contrast threshold, by a factor of 10
or more in October and~by~about~2~in November, when the feature's surface
brightness was, on the average, 67\,$\pm$\,11 S$_{10}$ units,\footnote{1
S$_{10}$ unit is a surface brightness of one magnitude~10 star per square
degree of the sky, equivalent to 27.78 mag per arcsec$^2$.} equaling
9\,$\pm$\,2 percent of the background sky brightness.  A suspicious attribute
of the reported feature is the sharp drop in its surface brightness by a factor
of $\sim$6 between October and November.

The dependence of the contrast sensitivity on a wide range of conditions
under which a visual observation is made does not rule out the possibility
that Hale's threshold was close to, or exceeded, the feature's inferred
surface brightness, in line with his failure to detect it.  Of greater
concern is the feature's apparent absence in the image taken by Suzuki.

Unlike Gonz\'alez's observations, the CCD image is accompanied by few details.
This observation was made shortly before the beginning of the astronomical
twilight, with the comet at a zenith distance of 58$^\circ$ and the Sun
19$^{\circ\!}$.5 below the horizon at midexposure.  The Moon was also below
the horizon, and the high altitude of the observatory (higher than Gonz\'alez's
observing sites) should have provided near-perfect conditions.  We do not know
the contrast threshold of the CCD device used, but it certainly was substantially
better than visually (probably $<$\,1~S$_{10}$).  We would only question whether
visual inspection of the image is enough to claim that the feature does not show
up in it.  We are unaware of any more rigorous means of examination or an
image-enhancement technique having been applied.  It is perhaps likely,
but not certain, that we have a problem.

In support of the possible conflict, we should point out that the total
magnitudes reported by Gonz\'alez imply a phase-corrected absolute magnitude
as bright as 6.7 in October and 6.1 in November 2018.\footnote{When derived
with the standard $r^{-2}$ law.}  These do not reconcile easily with the
absolute magnitudes of 10.7 for the intact comet before Outburst~I
(Section~4.2) and 9.9 for the debris cloud after Outburst~II (Section~5),
not to mention the sudden drop noted in the last \mbox{STEREO-A} images
shortly before perihelion, when the absolute magnitude slumped to 11.3
and the brightness was subsiding at a rate of 0.8 mag per day (Table~1).
The only physically plausible mechanism that we know of which could bring
about a major intrinsic brightening of a defunct comet after a period of a few
months is the process of continuing progressive dust-grain fragmentation.  If
the mass of millimeter-sized grains at the end of July should have increased
its intrinsic brightness more than thirtyfold, their dominant dimension should
over this period of time drop by the same factor, i.e., from 1~mm to 30~$\mu$m
across.  If so, however, they would occupy a strongly elongated volume of
space several degrees --- not 9$^\prime$ --- long in a position angle of
210$^\circ$.

The light curve based on the Gonz\'alez magnitudes in October--November
is also problematic.  A plot, against heliocentric distance $r$, of the
five magnitude estimates corrected for the geocentric distance and phase
effect is flat between 1.5 and 2.1~AU:\ in the standard law $r^{-n}$ the
exponent is calculated to equal \mbox{$n = 0.2\pm0.3$} instead of the
expected \mbox{$n = 2$}.

Any other avenue of explaining the Gonz\'alez feature would have
to be purely speculative.  For example, one can postulate that most
of the mass of the original nucleus was inert, not participating
in activity and not fragmenting, until after the comet left both
the fields of view of the HI1 imager on board STEREO-A and the
C3 coronagraph on board SOHO, since there is no sign of detection
of such an object by either detector.  Very impressive observational
evidence would be required to make this hypothesis look credible.

We believe that while the existence of a residual mass of the
disintegrated comet C/2017~S3 is unquestionable, it is difficult to
identify it with the unconfirmed, very bright feature reported by
Gonz\'alez on five occasions in October--November.  It is unfortunate
that no additional CCD imaging has been attempted to help resolve
this issue.

\section{Summary of Properties of the Debris Clouds}
Now that we know the total projected areas of the two debris clouds,
$X_{\rm frg}$, and the dominant diameter, $D$, of the grains that make them
up, we can readily estimate the clouds' masses at an assumed bulk density
$\delta$.  Eliminating the number of grains from the expressions for the
total projected area and total mass ${\cal M}_{\rm frg}$, we have
\begin{equation}
{\cal M}_{\rm frg} = {\textstyle \frac{2}{3}} \delta D_{\rm frg} X_{\rm frg}.
\end{equation}
Adopting again a bulk density of \mbox{$\delta = 0.53$ g cm$^{-3}$} and inserting
for Outburst II, \mbox{$D_{\rm frg}({\mbox{\scriptsize II}}) = 1$ mm} from
{\vspace{-0.04cm}}Section~8.1 and \mbox{$X_{\rm frg}({\mbox{\scriptsize II}})
= 2200$ km$^2$} from Equation~(12), we obtain for the mass of the cloud
associated with this outburst
\begin{equation}
{\cal M}_{\rm frg}({\mbox{\scriptsize II}}) = 7.8 \times \! 10^{11} \,{\rm g},
\end{equation}
which equals the mass of a spherical object 140~meters in diameter.  Taking for
{\vspace{-0.04cm}}Outburst~I, \mbox{$D_{\rm frg}({\mbox{\scriptsize I}}) = 3.8$
mm} from Section~8.2 and \mbox{$X_{\rm frg}({\mbox{\scriptsize I}}) < 0.8$
km$^2$} from the relation (29), we similarly find for the mass of this cloud
\begin{equation}
{\cal M}_{\rm frg}({\mbox{\scriptsize I}}) < 1.1 \times \! 10^{9} \,
 {\rm g},
\end{equation}
equivalent to the mass of a sphere less than 16 meters in diameter.  It would add
negligibly to the dimensions of an object 140~meters across.

Since the square of a mean outward velocity of the grains that make up a cloud
expanding with a velocity $v_{\rm exp}$ equals \mbox{$\langle v^2 \rangle =
\frac{1}{3} v_{\rm exp}^2$}, {\vspace{-0.04cm}}the kinetic energy needed~to~release
the cloud of millimeter-sized grains in the course of Outburst~II amounts to
\begin{equation}
{\cal E}_{\rm kin}({\mbox{\scriptsize II}}) = 7.5 \times \! 10^{18} \, {\rm erg},
\end{equation}
while for the cloud conceived during Outburst I the kinetic energy is constrained
by
\begin{equation}
{\cal E}_{\rm kin}({\mbox{\scriptsize I}}) < 0.9 \times \! 10^{12} \, {\rm erg}.
\end{equation}
Although an investigation of the physical processes that caused the
outbursts is not an objective of this paper, we note that the kinetic
energy of the Outburst~II cloud amounts to the energy released in the course of
the transformation of less than 10$^{10}$\,g of amorphous water ice into cubic
ice at a temperature of $\sim$140\,K, given that the relevant heat of
crystallization equals 10$^9$\,erg~g$^{-1}$ (Ghormley 1968).  At a distance of
70~meters from the center of the nucleus, this mass of ice would be contained
in a layer of less than 30~cm thick.

The mass and size of the original nucleus of the comet were of course greater than
the estimates of the large-debris clouds suggest, although probably not by much.
The losses of outgassed ices and emitted microscopic dust integrated
over one orbital period about the Sun are for a comet with a perihelion distance
of 0.2~AU typically equivalent to a layer several meters deep.  Given that
C/2017~S3 was a dust-poor comet and that its activity had terminated by the time
of Outburst~II,~prior~to perihelion and before the comet reached 0.9~AU, the
total losses in the dimensions of the nucleus by~gas~and~dust emission should
have been almost trivial.  The only significant mass contribution could
have derived from sizable undetected fragments of inert material, pebbles and
boulders, should they have survived Outburst~II.

For an Oort Cloud comet of this size and perihelion distance, it is
rather unusual to exhibit no nongravitational effects in the orbital
motion about the Sun (Marsden et al.\ 1978).  Yet, C/2017~S3 is found
to exemplify this peculiarity before Outburst~I, as demonstrated
convincingly by our first two solutions, Orbits~A$_0$ and A$_1$.  A
rapidly rotating nucleus of a low obliquity may represent a scenario
that is in line with the comet's observed motion.  The rapid rotation
may also have been a driver of the fragmentation events. 

Comparison of the class B solutions, which
span a time period that includes Outburst~I, shows the presence of
moderate nongravitational perturbations. This event did not paralyze the
nucleus, but may have acted as a precursor event to Outburst~II by
providing direct access to the nucleus' interior over a limited
area of the surface, thereby facilitating the looming breakup. 

These considerations already suggest that the two outbursts differed
from each other markedly.  The differences are documented in the
upper part of Table~13 that summarizes the information gathered from
the photometric data.  Besides the quantities measured directly from
the light curve (the time of onset, rise time, and amplitude), the
listed parameters include quantities that better describe the yield of
either outburst's active stage and provide a more accurate comparison
of the two events:\ the flux rise and its average rate (Sekanina 2017).
In particular, the average flux-rise rates confirm that Outburst~II was a
much more explosive event than Outburst~I, a finding that certainly
is not obvious from comparison of the two events' amplitudes or
{\it Af}$\rho$ values.  

\begin{table}[t]
\vspace{-4.2cm}
\hspace{4.22cm}
\centerline{
\scalebox{1}{
\includegraphics{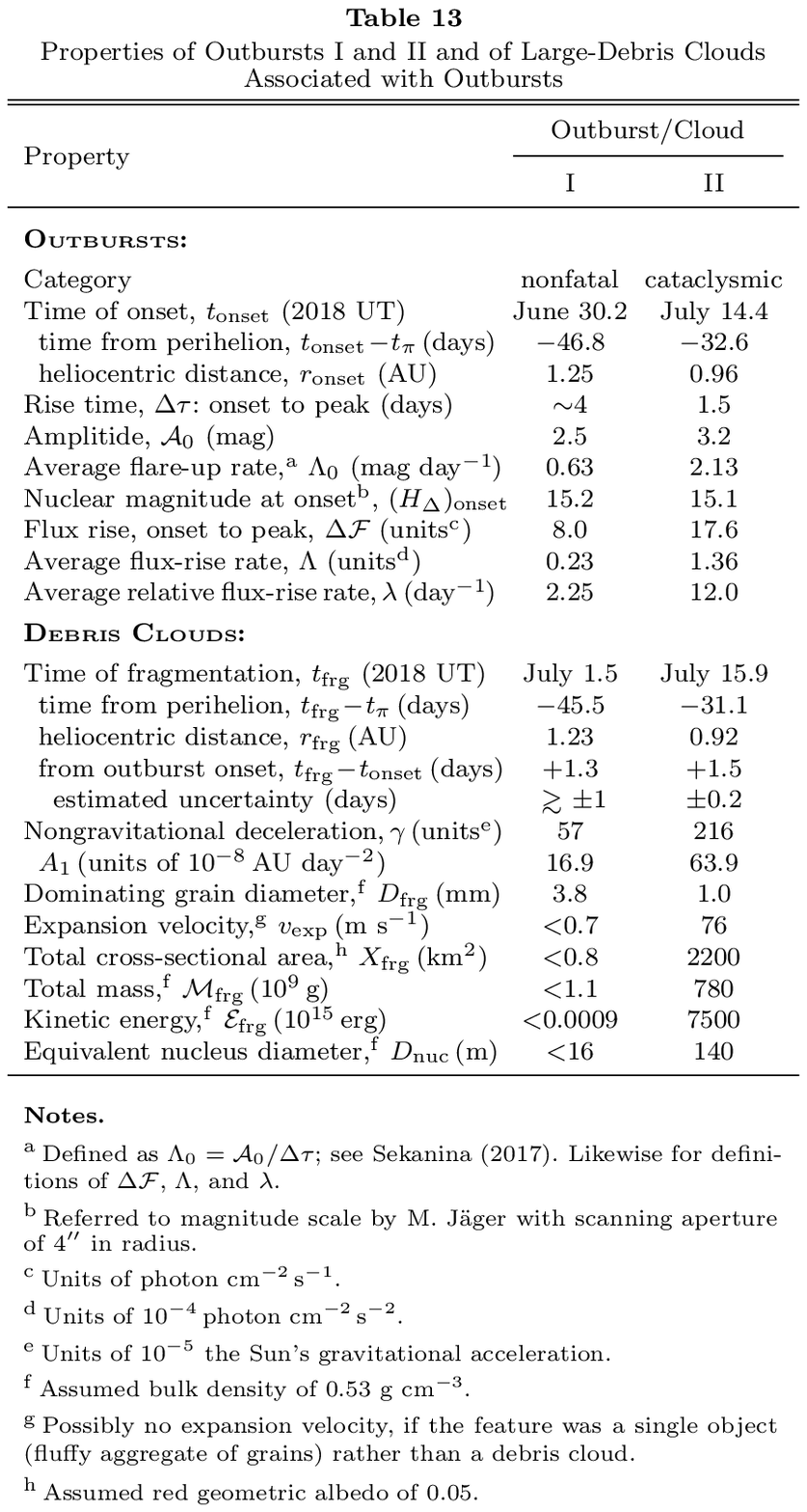}}}
\vspace{-9.05cm}
\end{table}

Information on the debris clouds is summarized in the lower part of Table~13.
The existence of the cloud associated with Outburst~II was strikingly
demonstrated by the divergence of the residuals left by the post-event
observations from the early gravitational orbits (A$_0$ and B$_0$) linking
the pre-outburst astrometric observations.  Thanks to the large-aperture
measurements of the cloud's total brightness in the
STEREO-A images taken between July~31 and August~14, it was possible
to determine the feature's expansion rate.

By contrast, we were unaware of the large-debris cloud related to Outburst~I
until we noticed that the ground-based observations from July~30 to August~3
increasingly deviated from the expected positions of~the~\mbox{Outburst}~II
cloud.  Although the Outburst~I cloud was much fainter, its rate of expansion
was extremely low (if any; see below), so that its peak surface brightness
eventually exceeded that of the Outburst~II cloud in the last days of
ground-based observation.  Only outer regions of the Outburst~II cloud,
some 100,000~km in radius by this time, were superposed on the compact
Outburst~I cloud.

For this reason, Soulier's imaging on August~3 is shown by the relation (29) to
provide only an upper limit on the projected area $X_{\rm frg}$(\scriptsize I}).
The tight expansion-velocity constraint in the relation (30) does not rule
out the limit of \mbox{$v_{\rm exp} = 0$}, which would imply that the feature
related to Outburst~I was not in fact a cloud of debris, but a {\small \bf
single sizable subfragment}.  Its estimated brightness and exposure to moderate
radiation-pressure effects are only reconciled if the subfragment was a
{\small \bf devolatilized, ``fluffy'' aggregate of
loosely-bound dust grains} (i.e., of very high porosity), with the dominant
grain diameter definitely smaller than the nominal size (3.8~mm across) and
more in line with the dominant size of the grains in the Outburst~II cloud.
Orbit~B$_1^{\textstyle \ast}$(2.0), the best B-class solution (Table~7), suggests
that the putative aggregate's motion was consistent with the motion of the measured
condensation that consisted of the debris of the original nucleus' major fragment
that was released in the course of Outburst~I and might have been still active
for a short period of time afterwards.  Late July and early August was apparently
the only time when the aggregate subfragment, the surviving part of the major
Outburst~I fragment, was picked up and monitored, when already inert, for several
days by four independent observers (Tables~11 and 12).  Survival of such a sizable
piece of fluffy nuclear debris is fascinating; major ramifications are addressed in
Section 10.

Whatever its nature, the modeling of its motion should have been carried out on
the residuals of the relevant observations from Orbit A$_0$.  However, we replaced
them with the residuals from Orbit B$_0$ to mitigate effects of extrapolation.
This decision seems justified by the above mentioned agreement between the
parameter $A_1$ from Orbit B$_1^{\textstyle \ast}$(2.0) and the deceleration
from the fragmentation model based on Orbit~B$_0$.

Table 13 shows that fragmentation followed the onset of either outburst by about
1.5~days.  While this information is burdened by a large error for Outburst~I,
the computed uncertainty for Outburst~II is only a small fraction of a day, the
time of fragmentation coinciding in fact with the time of peak brightness (i.e.,
the end of the event's active stage).  As a process, fragmentation does not
occur instantaneously, yet the data on Outburst~II suggest that the duration
was very short.  

From the information available, we classify Outburst~I as a nonfatal event and
Outburst~II as a cataclysmic event.  C/2017~S3 has a special place among comets
because enough data are available to classify it equally as a disintegrating
comet and a split comet.   In its capacity as a split comet, the comet displays
a very strong correlation between outbursts and nuclear fragmentation --- a topic
repeatedly addressed in the scientific literature over the past decades.  However,
C/2017~S3 is an atypical split comet in that it displays no primary nucleus (its
position at any time after an outburst approximated by extrapolating the comet's
pre-outburst purely-gravitational motion) and two companions that have never been
measured or, indeed, observed, at the same time.

Fitting a debris cloud's motion with a constant deceleration implies the presence
of a dominant grain size, which may be surprising.  These grain dimensions, in
the millimeter-size range, are typical for the debris detected in cometary dust
trails (e.g., Sykes et al.\ 1990; Reach et al.\ 2007) and for meteoroids in
related meteor outbursts (e.g., Jenniskens 1998).  Dust grains of this size are
important in that they are large enough not to get rapidly dispersed in space by
radiation pressure, yet small enough to have a rather high cross-section-to-mass
ratio for fairly efficiently scattering sunlight and abundant to be readily noticed
{\vspace{-0.04cm}}(unlike pebbles and boulders) in CCD images.\footnote{Also of
help is a relatively steep size-distribution function for large dust particles;
Sykes et al.\ (1990) specifically note that most mass and surface area of cometary
trails is contributed by grains $\sim$1~mm in diameter.{\vspace{-0.07cm}}}
Interestingly, the post-perihelion disintegration of C/2011~W3 resulted
in a cloud of dust grains whose dominant size (Sekanina \& Chodas 2012)
was very similar to the size we find for the cloud related{\nopagebreak}
to Outburst~II.

\section{Ramifications}
Our investigation of C/2017~S3 has ramifications primarily for short-lived
companions of the split comets.  Based on the assumption that --- as a
fragment of the original nucleus --- a companion is a miniature comet that
sublimates and releases dust, its nongravitational deceleration has been
interpreted as the product of the outgassing-driven momentum transfer
the same mechanism that was invoked by Whipple (1950) for comets in general.
With evidence that either ``companion'' of C/2017~S3 is an expanding cloud
of (inert) millimeter-sized dust grains, the nongravitational deceleration
is identified with solar radiation pressure.  Could it be that most, if not
all, {\small \bf short-lived companions of the split comets are expanding
clouds of grains\/}?  This idea is corroborated by well-known properties
of such companion nuclei, which (i)~become more diffuse with time; (ii)~grow
steadily in size; (iii)~get progressively elongated; and (iv)~disappear on
time scales equivalent at 1~AU from the Sun to days, weeks, or, at most,
\mbox{1--2}~months.  The property under (iii) would in this context imply
grain crumbling, which increases the radiation pressure.  The major
difference between the outgassing-driven deceleration and radiation
pressure --- the variation with heliocentric distance --- is mitigated
by the short lifetime of the companion nuclei and by the fact that they
are mostly observed at smaller distances from the Sun, at which the two
mechanisms imply nearly identical rates of variation.  Thus, the
heliocentric-distance law cannot readily be employed as a criterion to
discriminate between the two interpretations and the problem remains open.
For persistent companions of the split comets the outgassing-driven
deceleration appears to be preferable except perhaps shortly before the
end of their lifetime.

We conclude with speculation:\ {\small \bf Is there a lesson to be learned
from the physical behavior of C/2017~S3 that is relevant to 1I/'Oumuamua?}~Both
objects had~similar perihelion distances, 0.21~AU and 0.25~AU, respectively,
and they both arrived from interstellar space, except that from a distance of
10$^5$\,AU it took the comet 300 times longer than 1I to get to perihelion.
From a distance of 50~AU the ratio is still about 3.  This means that the comet
could have accommodated to the high temperatures more gradually and its interior
was, on the average, subjected to lower thermal stresses, yet C/2017~S3 did
disintegrate.  Should not one expect 1I to do likewise?  

\begin{figure}[t]
\vspace{0.17cm}
\hspace{-0.17cm}
\centerline{
\scalebox{0.535}{
\includegraphics{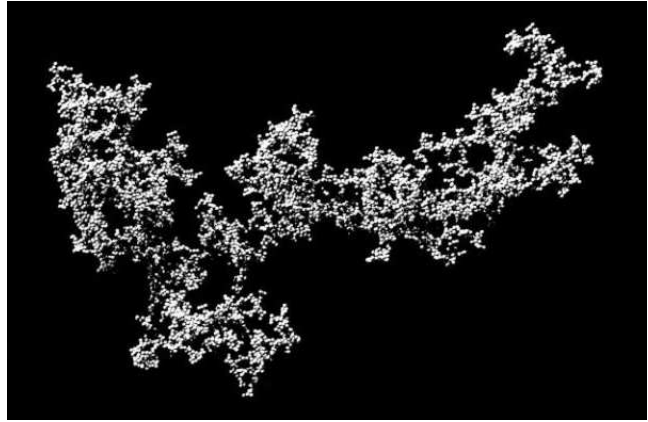}}}
\vspace{-0.04cm}
\caption{Computer-generated image of a loosely-bound fluffy aggregate of dust
grains (ballistic cluster-cluster agglomeration or BCCA; from Wada et al.\ 2008),
{\vspace{-0.05cm}}whose effective bulk density may drop below 0.0001~g~cm$^{-3}$.
Models of this kind are devised for studies of formation processes of
planetesimals in protoplanetary disks.  With the dimensions scaled up,
the model is used to simulate outburst-surviving high-porosity fragments
of a cometary nucleus.  The debris ``cloud'' associated with Outburst~I of
C/2017~S3 may have had this kind of morphology, and we propose to apply this
paradigm to the interstellar object 1I/`Oumuamua as well.  (Image credit:\
K.~Wada et al., Hokkaido University, Sapporo, Japan.){\vspace{0.45cm}}}
\end{figure}

A remarkable property of many comets is that their appearance in the same
range of heliocentric distances before and after perihelion is very different.
For C/2017~S3, the disparity was astounding.  Given that the earliest
pre-discovery observation of 1I was made five weeks after perihelion,
we have absolutely no idea on how the object looked like before perihelion,
a great disadvantage compared to C/2017~S3.  On the other hand, the
post-perihelion near-encounter of 1I with the Earth allowed close-up
examination of its physical behavior, which was unavailable for C/2017~S3.
In spite of these differences and rather limited supporting evidence, we
find it both intriguing and irresistible to suggest that, to a degree, the
{\small \bf two objects may have shared a similar recent history} in that
they both underwent preperihelion outbursts, albeit on different scales, with 1I
exhibiting features that were unobservable for C/2017~S3, and vice versa.

We propose that 1I had been approaching the Sun as a small interstellar comet that
experienced some weeks before perihelion an outburst, probably much smaller than
was either of the two outbursts of C/2017~S3.\footnote{All relevant images taken by
the HI1 camera on board the STEREO-A spacecraft should carefully be inspected for
potential evidence of 1I.  It is unclear whether such a search has ever been
undertaken; Bannister et al.\ (2017) quote K.~Battams that no trace of the
object was detected near perihelion on 2017 September~9, implying that at the
time it was fainter than magnitude $\sim$13.5.{\vspace{0.15cm}}}  The nucleus was
completely devolatilized in the course of the event, but did not disintegrate into
dust entirely.  Part~of it held together as a {\small \bf sizable, fluffy aggregate
fragment made up of loosely-bound dust grains} --- a kind of drifting skeleton of
the spent comet, which, as a survivor of the outburst, was eventually discovered.

The attractive properties of such a devolatilized fluffy aggregate fragment are
the {\small \bf absence of activity, strongly irregular shape} that could lead
to large brightness variations during rotation or tumbling, {\small \bf high
porosity}, i.e., high cross-section-to-mass ratio, and, consequently, {\small
\bf propensity to effects of the Sun's radiation pressure}, implying the presence
of nongravitational perturbations that could be confused with outgassing-driven
effects.

Consider, as a mathematical construct, an aggregate consisting of $N_{\rm grain}$
spherical dust grains of arbitrary~but equal diameter, $D_{\rm grain}$, an example
of which is displayed in Figure~20.  The objective is to constrain, as a function
of the grain diameter, the mass of the aggregate, ${\cal M}_{\rm aggr}$, which is to
be consistent with the light curve, the geometric albedo, and the nongravitational
acceleration of the object 1I.  If $\delta_{\rm grain}$ is the bulk density of the
grains, {\vspace{-0.07cm}}the aggregate's mass~is \mbox{${\cal M}_{\rm aggr} =
\frac{1}{6} \pi \delta_{\rm grain} D_{\rm grain}^3 N_{\rm grain}$},{\vspace{-0.07cm}}
whereas the sum of the projected areas of the grains{\vspace{-0.01cm}} is equal to
\mbox{$X_{\rm sum} = \frac{1}{4} \pi D_{\rm grain}^2 N_{\rm grain}$}. Eliminating
{\vspace{-0.04cm}}$N_{\rm grain}$, one derives, similarly to Equation~(31),
\begin{equation}
{\cal M}_{\rm aggr} = {\textstyle \frac{2}{3}} \delta_{\rm grain} D_{\rm grain}
X_{\rm sum}.
\end{equation}
Because of the mutual
occultations among neighboring grains in the aggregate fragment and effects of
multiple scattering, the amount of scattered sunlight in any particular
direction is attenuated.  The degree of attenuation of the signal
is measured by a logarithmic quantity $\alpha$, which normalizes the observed
projected area of the aggregate fragment, $X_{\rm aggr}$, to $X_{\rm sum}$,
\begin{equation}
X_{\rm aggr}(\alpha) = X_{\rm sum} \, 10^{-\alpha} \;\;\;\; (\alpha \!>\! 0).
\end{equation}
Variations in the projected area and brightness of the aggregate fragment depend
on its morphology and are a function of the degree of attenuation.  The radiation
pressure acceleration, to which the fluffy fragment is subjected,
exhibits significant quasi-periodic variations in time, like the signal of
scattered sunlight, but they are far too rapid to be detected by orbital
analysis, only an averaged effect being discerned.

Turning to the light curve of 1I, its peak brightness refers to the minimum
degree of attenuation, $\alpha_{\rm min}$, the minimum on the light curve to
the maximum degree of attenuation, \mbox{$\alpha_{\rm max} = \alpha_{\rm min}
+ \Delta \alpha$}, where $\Delta \alpha$ equals 0.4 times the amplitude in
magnitudes.  The mean projected area of the aggregate fragment, $\langle
X_{\rm aggr} \rangle$, taken as an average of the minimum and maximum projected
areas, is with help of Equation~(37) expressed as
\begin{equation}
\langle X_{\rm aggr} \rangle = {\textstyle \frac{1}{2}} X_{\rm peak} \left(
 1 \!+\! 10^{-\Delta \alpha} \right),
\end{equation}
where \mbox{$X_{\rm peak} = X_{\rm aggr}(\alpha_{\rm min})$} is the projected
area of the aggregate fragment when its light curve reaches the peak.  It is
the projected area $\langle X_{\rm aggr} \rangle$ that controls the magnitude
of the detected effect of the radiation pressure.

The nongravitational parameter determined by Micheli et al.\ (2018) for 1I
{\vspace{-0.04cm}}equals, in the units used in this paper, \mbox{$A_1 = (+24.55
\pm 0.80) \, \times 10^{-8}$\,AU day$^{-2}$}, which happens to fit the range of
the decelerations of the debris clouds related to the two outbursts of C/2017~S3
(Table~13)!  With the grain bulk density we assume in this paper, the nominal
grain diameter, $D_0$, fitting Micheli et al.'s value of $A_1$ equals, following
Equation~(17), \mbox{$D_0 = 2.6$ mm}.  Because of the attenuation of scattered
light, the grains in the aggregate must individually be subjected to a higher
radiation pressure acceleration and their diameter must be smaller than $D_0$.
Since the total grain mass does not change, the grain diameter $D_{\rm grain}$
is related to the nominal diameter $D_0$ by
\begin{equation}
D_{\rm grain} = D_0 \frac{\langle X_{\rm aggr} \rangle}{X_{\rm sum}}.
\end{equation}
Inserting from Equation (39) into Equation (36), we have
\begin{equation}
{\cal M}_{\rm aggr} = {\textstyle \frac{2}{3}} \delta_{\rm grain} D_0 \langle
 X_{\rm aggr} \rangle,
\end{equation}
where $\langle X_{\rm aggr} \rangle$ is given by Equation~(38).  Expressing now
$D_0$ in terms of the radiation pressure parameter $A_1$ (in AU\,day$^{-2}$) and
bulk density $\delta_{\rm grain}$ (in g\,cm$^{-3}$), the diameter (in mm) of the
grains that make up the fragment~is
\begin{equation}
D_{\rm grain} = 0.17 \, \frac{10^{-(6 + \alpha_{\rm min})}}{A_1 \delta_{\rm grain}}
 \left( 1 \!+\! 10^{-\Delta \alpha} \right)
\end{equation}
and the fragment's mass (in g) is
\begin{equation}
{\cal M}_{\rm aggr} = 113 \, \frac{X_{\rm peak}}{A_1} \left( 1 \!+\! 10^{-\Delta
 \alpha} \right),
\end{equation}
where $X_{\rm peak}$ is in km$^2$.  Equation (42) shows that the fragment's mass
is a function of only the quantities that are derived directly from the
observations and is independent of the grain size and the degree of attenuation.

We now tackle the photometric data.  Drahus et al.'s (2018) light curves
show that, on the average, the peak red absolute magnitude was 21.7 and the
amplitude equaled 2.6~mag.  Based on Trilling et al.'s (2018) considerations,
we adopt a geometric albedo of 0.1, so that the peak projected area of 1I
amounts to
\begin{equation}
X_{\rm peak} = 0.021 \; {\rm km}^2
\end{equation}
and the aggregate model's predicted mass for 1I is
\begin{equation}
{\cal M}_{\rm aggr} = 10^7 \; {\rm g},
\end{equation}
nearly four orders of magnitude lower than the value implied at an assumed bulk
density of 0.5~g~cm$^{-3}$ by the dimensions determined by Jewitt et al.\ (2017),
demonstrating the enormous difference between the two models.  The grain diameter
is merely a parameter of the aggregate model.  As expected, it decreases with
the increasing degree of attenuation; it is always smaller than 1.4~mm in the
examined model.

Although the proposed hypothesis is as speculative as any other published,
it avoids difficulties with the interpretation of the nongravitational
acceleration and does not introduce new difficulties with other data.
We feel that speculations on the object's behavior along the unobserved
preperihelion arc of the orbit should be part of the global picture and that
comparison with the fate of a small comet of similar perihelion distance
should provide some insight.  In this context we again call attention to the
so-far ignored possibility that the object's appearance and morphology
changed near perihelion, so that at least some of its observed properties
were not necessarily acquired before 1I had entered the inner Solar System. 

The results of this investigation illustrate the advantages of combining
photometric and orbital analysis in describing the attributes and impact
of explosive phenomena in comets of small perihelion distance.  A topic for
future studies of this kind is the proposed relationship in the orbital
behavior among the disintegrating comets, the short-lived companions of the
split comets, and other peculiar objects with similar orbital signatures.

\vspace{0.2cm}

We thank J.-F.\ Soulier for granting us permission to reproduce his images
of the comet.  This research was carried out in part at the Jet Propulsion
Laboratory, California Institute of Technology, under contract with the
National Aeronautics and Space Administration. \\[-0.2cm]
\begin{figure}[t]
\vspace{-4.28cm}
\hspace{2.75cm}
\centerline{
\scalebox{0.77}{
\includegraphics{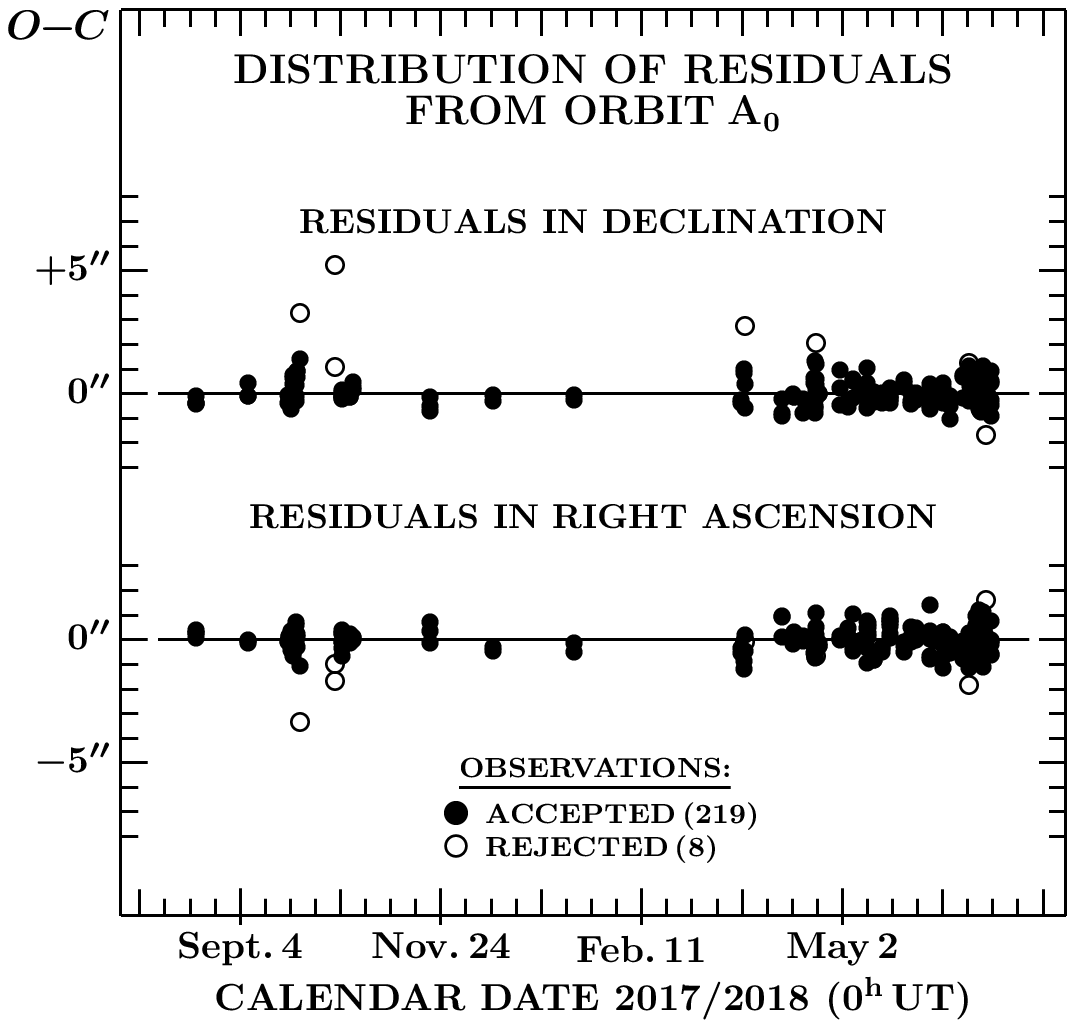}}}

\vspace{-10.5cm}
%
\parbox{8.6cm}{\footnotesize {\bf Figure A.1.} Distribution of residuals
\mbox{$O \!-\! C$} from Orbit A$_0$ (gravi\-tational) left by 227 ground-based
observations made on 2017 August 17--2018 June 30, 219 of which satisfy the
rejection threshold~of 1$^{\prime\prime\!}$.5 in either
coordinate\,(solid$\;$circles), eight do not\,(open$\;$circles).}

{\vspace{0.5cm}}
\end{figure}

\begin{center}
\bf APPENDIX\\[0.2cm]
{\rm REJECTING OBSERVATIONS THAT LEAVE UNACCEPTABLY LARGE RESIDUALS}
\end{center}
An important component of any orbit-determination effort is to subject the
collected set of astrometric observations to tests in order to extract a
subset of high-accuracy data and separate them from the data of lower accuracy.
Only the high-accuracy data are then used in the orbit-determination procedure,
the lower-accuracy data being rejected.  To ensure that this has indeed been
achieved requires the introduction of a {\it rejection cutoff\/} or {\it
rejection{\nopagebreak} threshold\/}, which eliminates all entries whose (observed
minus computed) residuals in right ascension, \mbox{$(O \!-\! C)_{\rm RA}$},
or declination, \mbox{$(O \!-\! C)_{\rm Decl}$}, exceed the limit.  Thus, the
extraction of the high-accuracy observations is always a matter of personal
choice.  In general, a residual may exceed the rejection cutoff for one of two
fundamentally different reasons:\ either it is of inferior quality {\it or\/}
the orbital motion of the comet is modeled inadequately.

The inferior quality of a positional observation has one or more causes, some
of which originate with the observer and his equipment (e.g., the comet being
at the limit of detection of the telescope; a short focal length of the
telescope; a large pixel size of the CCD array; an inferior reduction
technique; etc.), whereas others do not (e.g., a diffuse appearance of the
comet's nuclear condensation; the lack or unfavorable distribution of comparison
stars in the imaging field or their inaccurate positions; a star trail
interfering with the comet's image; poor observing conditions, such as a
low elevation of the object above the horizon or poor seeing; etc.).

The aim of the extraction procedure is to reject the data points of inferior
quality (with unacceptable residuals of the first category) to achieve a
high-quality orbital solution (thereby unacceptable residuals of the second
category be automatically eliminated).  The discrimination
between the acceptable and unacceptable data is a process that requires the
incorporation of a digital filter.  The category into which an observation
with a large residual belongs is determined by comparing, with one another,
observations over a limited orbital arc:\ if all or most of them exhibit {\it
systematic trends\/}) in the residuals, the problem is in the model of the
orbital motion; if one or a few stand out in reference to the
majority, it is the poor quality of the individual observations.  Obviously, this
filter fails when only very few observations are available widely scattered
over a long arc of the orbit.

Besides the digital filter, each converging orbital{\nopagebreak}
solution{\nopagebreak} is to be confronted with a self-sustaining
test{\pagebreak} of the rejection cutoff's enforcement.
Because of minor differences between two consecutive iterations of a solution\,---
before and after elimination of the observations with~residuals exceeding
the rejection cutoff --- the residuals of all accepted observations change
slightly and some that marginally exceeded the limit in the first solution
(and were therefore removed from the set of used data) just satisfied the
limit in the second solution (and were to be incorporated back into the
set), while the residuals of some of the other observations just satisfying
the limit in the first solution (and therefore kept in the set) subtly
exceeded the limit in the second solution (and were~now to be removed
from the set).  This process should be iterated until none of the
accepted observations has residuals exceeding the limit in the next
iteration and none of the eliminated observations has residuals complying
with it.

As an example, we show in Figure A.1 the distribution of the residuals from
Orbit A$_0$ (Table 5) by 219 accepted and eight rejected observations.

\begin{center}
{\footnotesize REFERENCES}
\end{center}
\vspace*{-0.5cm}
\begin{description}
{\footnotesize
%
%
\item[\hspace{-0.3cm}]
Bannister,\,M.\,T., Schwamb,\,M.\,E., Fraser,\,W.\,C., et al.\,2017, ApJL,{\linebreak}
 {\hspace*{-0.6cm}}851, 38
\\[-0.57cm]
\item[\hspace{-0.3cm}]
Birriel, J., \& Adkins, J.\ K.\ 2010, J.\,Amer.\,Assoc.\,Var.\,Star Obs.,~38,{\linebreak}
 {\hspace*{-0.6cm}}221
\\[-0.57cm]
\item[\hspace{-0.3cm}]
Blackwell, H. R. 1946, J. Opt. Soc. Amer., 36, 624
\\[-0.57cm]
%
%
\item[\hspace{-0.3cm}]
Bortle, J. E. 1991, Int.\ Comet Quart., 13, 89
\\[-0.57cm]
\item[\hspace{-0.3cm}]
Clark, R. N. 1990, Visual Astronomy of the Deep Sky. (Cam-{\linebreak}
 {\hspace*{-0.6cm}}bridge, UK:\ Cambridge University Press; and Cambridge,
 MA:{\linebreak}
 {\hspace*{-0.6cm}}Sky Publishing Corporation; 355pp)
\\[-0.57cm]
\item[\hspace{-0.3cm}]
Crumey, A. 2014, MNRAS, 442, 2600
\\[-0.57cm]
\item[\hspace{-0.3cm}]
Drahus, M., Guzik, P., Waniak, W., et al. 2018, Nature Astron., 2,{\linebreak}
 {\hspace*{-0.6cm}}407
\\[-0.57cm]
\item[\hspace{-0.3cm}]
Ghormley, J. A. 1968, J. Chem. Phys., 48, 503
\\[-0.57cm]
\item[\hspace{-0.3cm}]
Green, D. W. E. 2017, CBET 4432
\\[-0.57cm]
\item[\hspace{-0.3cm}]
Guigay, G. 1955, J. Obs., 38, 189
\\[-0.57cm]
\item[\hspace{-0.3cm}]
Jenniskens, P. 1998, Earth, Plan.\ \& Space, 50, 555
\\[-0.57cm]
\item[\hspace{-0.3cm}]
Jewitt, D., Luu, J., Rajagopal, J., et al.\ 2017, ApJL, 850, 36 (7pp)
\\[-0.57cm]
\item[\hspace{-0.3cm}]
Marcus, J. N. 2007, Int.\ Comet Quart., 29, 39
\\[-0.57cm]
%
%
%
\item[\hspace{-0.3cm}]
Marsden, B. G., Sekanina, Z., \& Everhart, E. 1978, AJ, 83, 64
\\[-0.57cm]
\item[\hspace{-0.3cm}]
Marsden, B.\,G., Sekanina, Z., \& Yeomans, D.\,K.\ 1973, AJ,\,78,\,211
\\[-0.57cm]
\item[\hspace{-0.3cm}]
Meech, K. J., Pittichov\'a, J., Bar-Nun, A., et al.\ 2009, Icarus, 201,{\linebreak}
 {\hspace*{-0.6cm}}719
\\[-0.57cm]
\item[\hspace{-0.3cm}]
Micheli, M., Farnocchia, D., Meech, K. J., et al.\ 2018, Nature, 559,{\linebreak}
 {\hspace*{-0.6cm}}223
\\[-0.57cm]
\item[\hspace{-0.3cm}]
Montrucchio, B. 2015, IEEE Trans.\ Hum.-Mach.\ Syst,, 45, 739
\\[-0.57cm]
%
%
\item[\hspace{-0.3cm}]
Nakano, S. 1994, MPC 23649
\\[-0.57cm]
\item[\hspace{-0.3cm}]
Nakano, S. 2018a, Nakano Note 3525
\\[-0.57cm]
\item[\hspace{-0.3cm}]
Nakano, S. 2018b, Nakano Note 3675
\\[-0.57cm]
%
%
\item[\hspace{-0.3cm}]
Reach, W. T., Kelley, M. S., \& Sykes, M. V. 2007, Icarus, 191, 298
\\[-0.57cm]
\item[\hspace{-0.3cm}]
Roemer, E. 1962, PASP, 74, 351
\\[-0.57cm]
\item[\hspace{-0.3cm}]
Roemer, E. 1963, AJ, 68, 544
\\[-0.57cm]
\item[\hspace{-0.3cm}]
Sekanina, Z. 1975, Icarus, 25, 218
\\[-0.57cm]
\item[\hspace{-0.3cm}]
Sekanina, Z. 1977, Icarus, 30, 574
\\[-0.57cm]
\item[\hspace{-0.3cm}]
Sekanina, Z. 1978, Icarus, 33, 173
\\[-0.57cm]
\item[\hspace{-0.3cm}]
Sekanina, Z. 1979, Icarus, 38, 300
\\[-0.57cm]
%
%
\item[\hspace{-0.3cm}]
Sekanina, Z. 1982, in Comets, ed.\ L. L. Wilkening (Tucson, AZ:\
 {\hspace*{-0.6cm}}University of Arizona), 251
\\[-0.57cm]
\item[\hspace{-0.3cm}]
Sekanina, Z. 2010, Int.\ Comet Quart., 32, 45
\\[-0.57cm]
%
%
\item[\hspace{-0.3cm}]
Sekanina, Z. 2017, eprint arXiv:1712.03197
\\[-0.57cm]
%
%
%
\item[\hspace{-0.3cm}]
Sekanina, Z., \& Chodas, P. W. 2012, ApJ, 757, 127 (33pp)
\\[-0.57cm]
\item[\hspace{-0.3cm}]
Sekanina, Z., \& Kracht, R. 2015, ApJ, 801, 135 (19pp)
\\[-0.57cm]
\item[\hspace{-0.3cm}]
Sykes, M. V., Lien, D. J., \& Walker, R. G. 1990, Icarus, 86, 236
\\[-0.57cm]
\item[\hspace{-0.3cm}]
Trilling, D. E., Mommert, M., Hora, J. L., et al.\ 2018, AJ, 156,{\linebreak}
 {\hspace*{-0.6cm}}261 (9pp)
\\[-0.57cm]
\item[\hspace{-0.3cm}]
Wada, K., Tanaka, H., Suyama, T., et al.\ 2008, ApJ, 677, 1296
\\[-0.57cm]
\item[\hspace{-0.3cm}]
Whipple, F. L. 1950, ApJ, 111, 375
\\[-0.57cm]
\item[\hspace{-0.3cm}]
Whipple, F. L. 1978, Moon \& Plan., 18, 343
\\[-0.57cm]
\item[\hspace{-0.3cm}]
Williams, G. V. 2017a, MPEC 2017-S160
\\[-0.57cm]
\item[\hspace{-0.3cm}]
Williams, G. V. 2017b, MPC 106348
\\[-0.65cm]
\item[\hspace{-0.3cm}]
Williams, G. V. 2018, MPC 111770; also MPEC 2018-Q62}
{\vspace{-0.42cm}}
\end{description}
\end{document}